\documentclass[a4paper,11pt]{article}
\pdfoutput=1 
\usepackage{verbatim}

\usepackage{jheppub}
\usepackage{amsmath,amsfonts,amssymb,amsthm,graphicx,color}
\usepackage[T1]{fontenc} 
\usepackage{enumerate}
\usepackage{tikz, tikz-3dplot, pgfplots}
\usepackage[utf8]{inputenc}
\usepackage{slashed}
\usepackage{booktabs}
\usepackage{caption}
\usepackage{subcaption}
\usepackage{footmisc}
\usepackage{comment}
\usepackage{ulem}

\usepackage{array}

\usepackage{changepage}
\usepackage{float}
\usepackage{tikz}
\usetikzlibrary{decorations.text}
\usepackage{enumitem}
\setlist{itemsep=0pt}
\usetikzlibrary{math}
\usepackage{physics}

\usepackage{tikz}
\usepackage{pgfplots}
\usetikzlibrary{decorations.text,intersections, pgfplots.fillbetween}
\usetikzlibrary{math}
 \usetikzlibrary{snakes,3d,shapes.geometric,shadows.blur}
\usepackage{tikz-3dplot}

\definecolor{amaranthred}{rgb}{0.83,0.13,0.18}
\definecolor{amazon}{rgb}{0.23,0.48,0.34}
\definecolor{bdazzledblue}{rgb}{0.18,0.35,0.58}
\definecolor{absolutezero}{rgb}{0.0,0.28,0.73}
\definecolor{bitterlemon}{rgb}{0.79,0.88,0.05}
\definecolor{byzantine}{rgb}{0.74,0.2,0.64}
\definecolor{turquoise}{rgb}{0.19, 0.84, 0.78}

\renewcommand{\comm}[1]{} 

\def\({\left(}
\def\){\right)}
\def\[{\left[}
\def\]{\right]}

\def\coeff#1#2{{\textstyle \frac{#1}{#2}}}

\def\One{{\hbox{ 1\kern-.8mm l}}}
\def\AdS{\mathrm{AdS}}
\def\Sph{\mathrm{S}}

\def\barray{\begin{array}}
\def\earray{\end{array}}
\def\be{\begin{equation}}
\def\ee{\end{equation}}
\def\bea{\begin{eqnarray}}
\def\eea{\end{eqnarray}}
\def\bal{\begin{align}}
\def\eal{\end{align}}
\def\nn{\nonumber}

\def\-{\,-\,}
\def\={\,=\,}
\def\+{\,+\,}
\def\equi{\,\equiv\,}

\numberwithin{equation}{section} 

\definecolor{cardinal}{rgb}{0.6,0,0}
\definecolor{darkgreen}{rgb}{0,0.4,0}
\definecolor{golden}{rgb}{0.92, 0.7, 0}
\definecolor{midnight}{rgb}{0, 0, 0.5}
\definecolor{darkblue}{rgb}{0, 0, 0.7}
\definecolor{purple}{rgb}{0.5, 0, 0.5}

\def\IR{\mathbb{R}}

\def\cC{{\cal C}}

\def\cF{{\cal F}}
\def\cK{{\cal K}}

\def\cH{{\cal H}}

\def\cL{{\cal L}}
\def\cM{{\cal M}}

\def\cV{{\cal V}}

\def\cZ{{\cal Z}}

\makeatletter
\newcommand\footnoteref[1]{\protected@xdef\@thefnmark{\ref{#1}}\@footnotemark}
\makeatother

\usetikzlibrary{positioning,calc,fadings,decorations.pathreplacing}

\tikzset{
 diffuse color/.initial = black,                       
}

\tikzset{
 linear opacity/.initial=0.5,                          
 linear stroke/.style = {                              
   preaction={                                         
     draw=\pgfkeysvalueof{/tikz/diffuse color},        
     line width = (2.0-#1)*\pgflinewidth,              
     opacity=\pgfkeysvalueof{/tikz/linear opacity},white}},  
 diffuse gradient/.style={                             
   draw = none,                                        
   linear opacity=#1,                                  
   linear stroke/.list={0.0,#1,...,1.0}},              
 diffuse gradient/.default=1,                          
}

\tikzset{
 non-linear stroke/.style = {                          
   preaction={                                         
     draw=\pgfkeysvalueof{/tikz/diffuse color},        
     line width = (2.0-#1)*\pgflinewidth,              
     opacity=#1,white}},                                     
 diffuse falloff/.style={                              
   draw = none,                                        
   non-linear stroke/.list={0.0,#1,...,1.0}},          
 diffuse falloff/.default=1,                           
}
 
\tikzset{%
  >=latex, 
  inner sep=0pt,%
  outer sep=2pt,%
  mark coordinate/.style={inner sep=0pt,outer sep=0pt,minimum size=3pt,
    fill=black,circle}%
}

\title{\boldmath Solitonic Excitations in AdS$_2$}

\author[a]{Pierre Heidmann} 
\author[b]{and Anthony Houppe} 
\affiliation[a]{Department of Physics and Astronomy,\\
Johns Hopkins University,\\
3400 North Charles Street, Baltimore, MD 21218, USA}
\affiliation[b]{
Université Paris-Saclay, CNRS, CEA, Institut de physique théorique, \\
Orme des Merisiers, 91191, Gif-sur-Yvette, France}

\emailAdd{pheidma1@jhu.edu}
\emailAdd{anthony.houppe@ipht.fr}

\abstract{We construct large families of supergravity solutions that are asymptotic to AdS$_2$ and terminate with a cap that is singular in two dimensions but smooth in higher dimensions.  These solutions break supersymmetry and conformal invariance.  We list arguments suggesting that they correspond to finite-energy excitations in empty AdS$_2$ that back-react on the geometry by inducing non-trivial bubbling topology.  They are constructed from the novel technique associated with the Ernst formalism for AdS$_D\times\cC$ solitons in supergravity \cite{Bah:2022pdn}.  The technique is applied to $D=2$ in M-theory with $\cC=\,$S$^3\times$T$^6$.  The directions of $\cC$ degenerate smoothly as a chain of bolts  which ends the spacetime in the IR and generates non-supersymmetric bubbles supported by M2-brane flux.  Some specific solutions have ``flat'' directions where the sizes of their bubbles are totally unconstrained and can be arbitrarily tuned while the asymptotics remains fixed.  The solitons should correspond to regular non-supersymmetric states of a holographically dual CFT$_1$.}

\preprint{}
\begin{document}

\maketitle
\flushbottom

\section{Introduction}
\label{sec:Intro}

Despite the numerous gravity and string theory solutions that asymptote to two-dimensional Anti-de Sitter spacetime, such as the near-horizon limit of extremal four-dimensional black holes \cite{Strominger:1998yg,Spradlin:1999bn} or supersymmetric smooth horizonless geometries \cite{Bena:2007ju,Lunin:2015hma,Bena:2018bbd,Heidmann:2018vky},  the AdS$_2$/CFT$_1$ duality is less understood than its higher-dimensional counterparts.  There are several essential characteristics that distinguish it from other dualities and have prevented a fully satisfactory realization of the correspondence in the field theory side.  

First,  an immediate puzzle is that empty global AdS$_2$ has two disconnected boundaries.  Hence,  its holographic dual seems to consist of two copies of an one-dimensional conformal quantum mechanics. This appears to be in contradiction with the derivation of black hole entropy in string theory from bound states of a (single) brane system \cite{Sen:1995in,Strominger:1996sh,Maldacena:1997de,Benini:2016rke,Azzurli:2017kxo}, so that the entropy derivation in AdS$_2$ should involve counting the ground states of a single CFT$_1$ \cite{Sen:2008vm,Gupta:2008ki,Sen:2011cn}.

Second,  it has been shown in \cite{Maldacena:1998uz,Almheiri:2014cka} that empty AdS$_2$ fibered over a sphere does not allow finite-energy excitations: the backreaction necessarily makes the sphere blow up at one of the two asymptotic boundaries.  As a consequence,  most of the attempts to understand quantum gravity in AdS$_2$ has been made by modifying the UV for ``nearly-AdS$_2$'' geometries, with the addition of a running dilaton.  This has led to interest in a NAdS$_2$/NCFT$_1$ correspondence,  including the focus on the SYK
model and Jackiw-Teitelboim gravity (see for example \cite{Kitaev:2015talks,Maldacena:2016hyu,Cotler:2016fpe,Balasubramanian:2016ids,Kitaev:2017awl,Jackiw:1984je,Teitelboim:1983ux}).  

\begin{figure}[t]
  \centering
  \includegraphics[width=.75\textwidth]{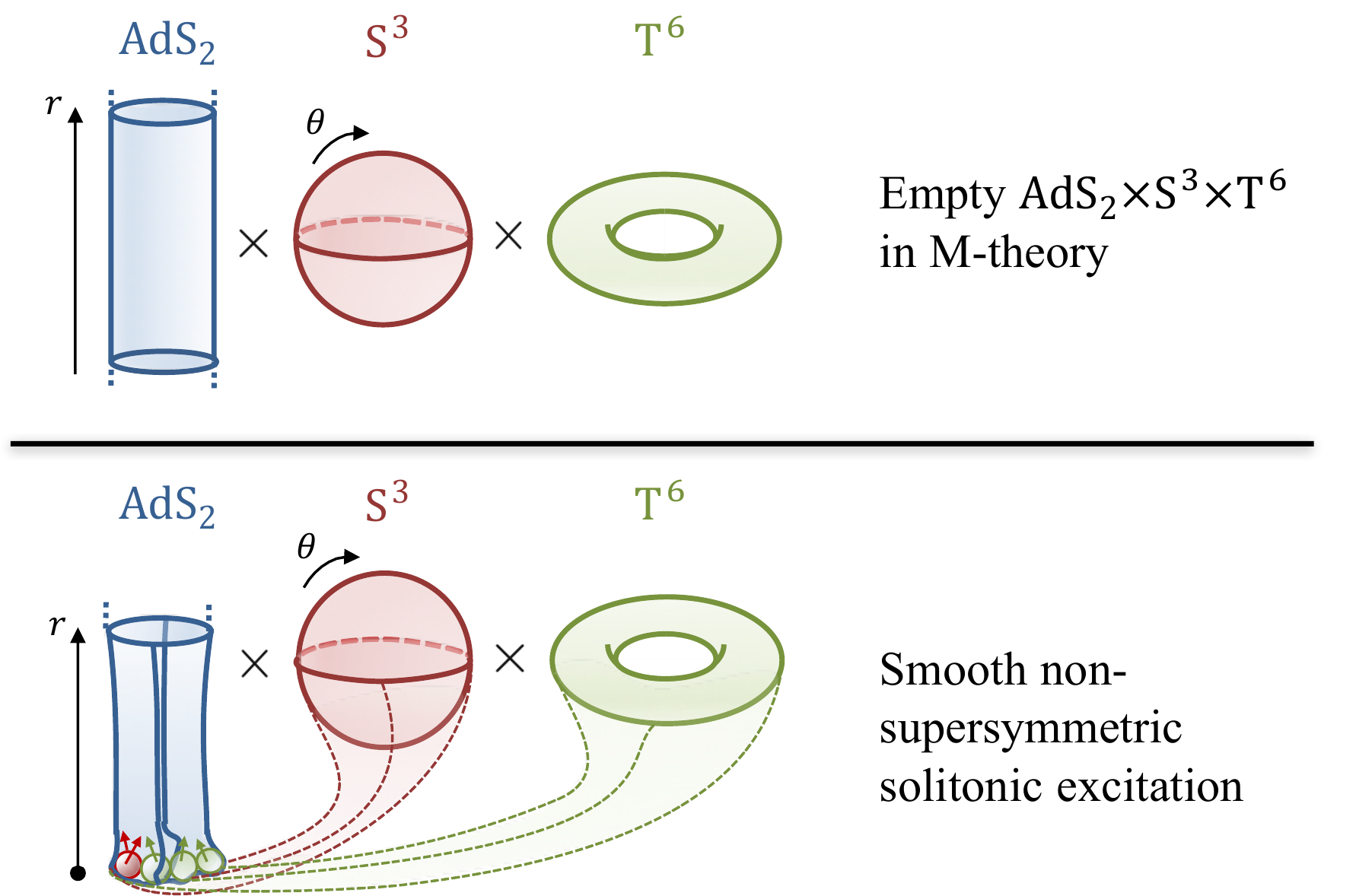}
  \caption{The topology of empty AdS$_2\times$S$^3\times$T$^6$ in M-theory and the non-supersymmetric excitations constructed in this paper.  They are produced by the smooth degeneracies of some S$^3$ and T$^6$ coordinates that end the spacetime as a chain of bolts supported by M2-brane flux.}
  \label{fig:Intro}
\end{figure}

The main goal of this paper is to  use string-theory degrees of freedom to construct smooth backreacted non-supersymmetric solutions in AdS$_2$. We will give arguments suggesting that these solutions must correspond to finite-energy excitations of empty AdS$_2$.  This requires to go beyond Jackiw-Teitelboim  gravity,  and to include a much richer field content.  More precisely,  we explicitly construct smooth non-supersymmetric geometries asymptotic to AdS$_2\times$S$^3\times$T$^6$,  that correspond to non-perturbative and regular excitations of AdS$_2$  in M-theory.  The solutions are asymptotic to AdS$_2$ in the UV,  and the compact space S$^3\times$T$^6$ has a finite size everywhere.  Moreover,  they only have one asymptotic boundary. This is made possible by non-trivial geometric transitions of M2 branes in the IR: the spacetime caps off smoothly where some directions in S$^3\times$T$^6$ degenerate,  forming bubbles supported by M2 brane flux, and breaking conformal invariance. We have schematically represented the construction in Fig.\ref{fig:Intro}. Since these solutions are generated by the dynamics of extra compact dimensions, they would appear geometrically singular if one reduces them to a two-dimensional theory of gravity.  Our results show that this kind of singularity can be resolved by  embedding them in M-theory.

Preserving the AdS$_2$ region in the UV while modifying the IR by brane polarization and smooth bubbling geometries has been one of the main proposal of \cite{Bena:2018bbd}. Because these solutions admit a single AdS$_2$ boundary, this allows us to study the bulk side of the AdS$_2$/CFT$_1$ correspondence using  usual holographic techniques similar to those existing in higher-dimensional AdS spaces.  Indeed,  for AdS$_5$ and AdS$_3$,  many CFT$_4$ and CFT$_2$ supersymmetric ground states correspond to regular capped geometries in supergravity,  as the LLM geometries in AdS$_5\times$S$^5$ \cite{Lin:2004nb} or the microstate geometries in AdS$_3\times$S$^3\times$T$^4$ \cite{Lunin:2002iz,Kanitscheider:2007wq,Giusto:2012yz,Giusto2015,Bena:2015bea,Bena:2016agb,Heidmann:2019zws,Heidmann:2019xrd,Shigemori:2020yuo}.  In \cite{Bena:2018bbd},  it has been proposed similarly that supersymmetric ground states of  (a single copy of) a CFT$_1$ are dual to smooth bulk geometries in AdS$_2\times\cC$ that break conformal invariance by capping off smoothly in the IR.  Large families of such supersymmetric solutions have been constructed in various supergravity frameworks \cite{Bena:2018bbd,Heidmann:2018vky}. 

Moreover,  the presence of a smooth IR cap had allowed the asymptotically-AdS$_2$ supersymmetric solutions of \cite{Bena:2018bbd} to support an infinite tower of non-supersymmetric linearized excitations. These excitations are localized near the cap, and are normalizable.  An important question that was still opened is whether  the backreaction of these excitations preserves the AdS$_2$ asymptotics or destroys the asymptotics as suggested by the derivation in lower dimensions in \cite{Maldacena:1998uz,Almheiri:2014cka}.  The present paper tends to validate the first scenario by showing the existence of non-supersymmetric fully-backreacted geometries in M-theory that modify the topology in the IR but preserve the asymptotics. 

To construct these geometries,  we adapt the solution-generating technique of \cite{Bah:2022pdn} that has allowed to systematically construct non-supersymmetric solitons in AdS$_3\times$S$^3\times$T$^4$ in type IIB.  The technique can be generically applied to backgrounds with $d-2$ commuting Killing vectors and suitable electromagnetic flux,  where $d$ is the dimension of the spacetime.   This reduces the Einstein-Maxwell equations into a set of Ernst equations, thereby admitting an integrable structure associated to the action of the Geroch group \cite{Geroch:1970nt,Geroch:1972yt}.  Then,  solutions in AdS$_D\times\cC$ can be extracted using methods associated to the Ernst formalism: the curvature of the AdS spacetime can be balanced off with that of the internal space $\cC$.  The solutions end in the IR as a chain of charged bolts that are inherently non-supersymmetric,  and where the spacelike compact directions in $\cC$ or AdS$_D$ smoothly degenerate. 

We apply this formalism to M-theory on T$^6=\,$T$^2\times$T$^2\times$T$^2$,  with three stacks of M2 branes wrapping the orthogonal two-tori.  The geometries are asymptotic to AdS$_2\times$S$^3\times$T$^6$ and depend on two variables which we choose to be the radial distance in AdS$_2$, and the angular position on the S$^3$.  The spacetimes end at a certain radius as a chain of bolts where directions of the T$^6$ and the S$^3$ alternatively degenerate.  This induces non-trivial bubbling topologies supported by M2-brane fluxes as depicted in Fig.\ref{fig:Intro}.

The regularity of the bolts leads to algebraic constraints on the M2-brane charges,  asymptotic sizes of the T$^6$ and sizes of the bubbles in the IR.  We discuss in which regions of the moduli space our solitons can exist.  We show that,  in very specific points of the moduli space,  there are solutions that have a totally unconstrained IR topology.  Indeed, the size of their internal bubbles can be set arbitrarily while the asymptotics are fixed, corresponding to certain flat directions in the phase space of solutions.  Moreover,  we show that generic solutions with a large number of bubbles are typically regular in large regions of the moduli space,  such that our non-supersymmetric smooth solitons describe an incredibly large family of solitonic excitations of AdS$_2$ in M-theory.

The organization of this paper is as follows.  In section \ref{sec:EOM}, we detail our solution generating technique for systematically constructing asymptotically-AdS$_2$ non-supersymmetric solitons in M-theory.  In section \ref{sec:BuAdS2T2}, we build and discuss the physics of the most primitive solution.  It consists of three bolts on a line where one direction in each orthogonal T$^2$ degenerates alternatively.  Then,  we derive the most generic solutions in section \ref{sec:BuAdS2T2s}.  Finally, we summarize the results and give some perspectives in section \ref{sec:conclusion}.  The paper also includes three appendices that provide additional details on the derivations.

\section{Integrable structure in \texorpdfstring{AdS$_2$}{AdS2}}
\label{sec:EOM}


In \cite{Bah:2020pdz,Bah:2021owp,Bah:2021rki,Heidmann:2021cms,Bah:2022pdn},  it has been shown that Einstein-Maxwell equations for static $d$-dimensional backgrounds with $d-2$ commuting Killing vectors and suitable electromagnetic flux decomposes into a set of Ernst equations.  This reduces the equations of motion for the $d$-dimensional spacetime to an integrable system on a two-dimensional plane.  Generic solutions are generated by rod sources, that are segments on a symmetry axis.  For regular solutions,  the rod sources must induce spacelike or timelike coordinate degeneracies in the spacetime.

The integrable structure allows to restrict to a class of solutions that relies on a linear structure of the cylindrical axially-symmetric Laplace problem, thereby generalizing the Weyl formalism to a charged Weyl formalism.  If the system is still linear,  it remarkably goes beyond the supersymmetric equations of motion allowing for non-supersymmetric sources.

In this section,  we present this class of solutions in the context of M-theory backgrounds on T$^6$ with M2-M2-M2 flux.  The decomposition of the Einstein-Maxwell equations in terms of Ernst equations has been initially derived in \cite{Heidmann:2021cms},  and the interested reader can found a summary in the Appendix \ref{App:EOM}.  Then,  we compute the internal and asymptotic boundary conditions to construct regular geometries that are asymptotic to AdS$_2\times$S$^3\times$T$^6$.

\subsection{Static M2-M2-M2 solutions on \texorpdfstring{T$^6$}{T6}}

We consider static M-theory solutions with 9 commuting Killing vectors,  thereby depending on two coordinates.  In the Weyl formalism, we can freely choose these coordinates, denoted as $(\rho,z)$, such that the induced metric on the two-dimensional space is conformally flat,  and the induced metric on the remaining nine-dimensional spacetime satisfies $\det h \= -\rho^2$ \cite{Weyl:book,Emparan:2001wk}. Moreover, one can consider one of the isometries, denoted as $\phi$, to have a metric coefficient proportional to $\rho^2$,  so that the $(\rho,z,\phi)$ space defines a three-dimensional base in Weyl cylindrical coordinates, and $z$ plays the role of an axis of symmetry \cite{Emparan:2001wk,Bah:2020pdz,Heidmann:2021cms}.

The solutions are constructed on T$^6$ and are supported by M2-M2-M2 flux.  The three stacks of M2 branes are wrapping three orthogonal 2-tori inside the T$^6$ that we parametrized by $(y_1,y_2)$,  $(y_3,y_4)$ and $(y_5,y_6)$.  Finally,  we consider the remaining S$^1$,  written in terms of an angle $\psi$,  as a Hopf fibration over the $(\rho,z,\phi)$ base with a KK vector along $\phi$.

An ansatz of metric and fields that suits the spacetime symmetries and flux is given by
\begin{align}
ds_{11}^2 = &- \frac{dt^2}{\left(W_0 Z_1 Z_2 Z_3 \right)^{\frac{2}{3}}} + \left(\frac{Z_1 Z_2 Z_3}{W_0^2}\right)^{\frac{1}{3}} \left[\frac{1}{Z_0} \left(d\psi +H_0 d\phi\right)^2 + Z_0 \left( e^{2\nu} \left(d\rho^2 + dz^2 \right) +\rho^2 d\phi^2\right)\right] \nonumber\\
&\hspace{-0.5cm} + \left(W_0Z_1 Z_2 Z_3\right)^{\frac{1}{3}} \left[\frac{1}{Z_1} \left(\frac{dy_1^2}{W_1} + W_1 \,dy_2^2 \right) + \frac{1}{Z_2} \left(\frac{dy_3^2}{W_2} + W_2 \,dy_4^2 \right) + \frac{1}{Z_3} \left(\frac{dy_5^2}{W_3} + W_3 \,dy_6^2 \right)\right], \nonumber\\
F_4 = &d\left[ T_1 \,dt \wedge dy_1 \wedge dy_2 \+T_2 \,dt \wedge dy_3 \wedge dy_4 \+T_3\,dt \wedge dy_5 \wedge dy_6 \right]\,. \label{eq:MtheoryAnsatz}
\end{align}
We will generically use capital Greek letters, $\Lambda,\Sigma,...$,  for indices from $0$ to $3$ and the capital Latin letters,  $I,J,...$,  for indices from $1$ to $3$.  The warp factors $(Z_\Lambda,W_\Lambda,\nu)$ and gauge potentials $(T_I,H_0)$ are functions of $(\rho,z)$.  

There are three electric gauge potentials, $T_I$,  induced by the M2 branes, and a magnetic gauge potential, $H_0$.  We have introduced four warp factors, $\{Z_\Lambda\}_{\Lambda=0,1,2,3}$, which couple naturally with the four gauge potentials.  In addition, we have four warp factors, $\{W_\Lambda\}_{\Lambda=0,1,2,3}$,  which are associated with T$^6$ deformations.  Finally, $e^{2\nu}$ determines the nature of the three-dimensional base. 

Remarkably,  the Einstein-Maxwell equations decompose into 8 sectors (see  Appendix \ref{App:EOM}).  Four sectors concern the T$^6$ deformations,  $W_\Lambda$,  and give linear equations as for vacuum Weyl solutions \cite{Weyl:book,Emparan:2001wk}. The four remaining sectors involve the pairs of warp factors and their associated gauge fields $(Z_I,T_I)$ and $(Z_0,H_0)$.  They are non-linear differential equations that can be mapped to Ernst equations for four-dimensional electrostatic backgrounds \cite{Bah:2022pdn}.  Thus, solutions to these non-linear equations can be  built from the numerous solution-generating techniques in the Ernst formalism.  The charged Weyl formalism \cite{Bah:2020ogh,Bah:2020pdz,Bah:2021owp,Bah:2021rki,Heidmann:2021cms,Bah:2022pdn} is one of them and has the advantage to have a linear structure without restricting to supersymmetric equations.  Finally,  the base factor $e^{2\nu}$ takes contribution from each sector into a first-order differential equation that can be easily integrated.

\subsection{Charged Weyl formalism}

The solutions are given in terms of \textit{eight functions} for which their logarithms are harmonic:
\begin{equation}
\Delta \log L_\Lambda \= \Delta \log W_\Lambda = 0 \,,\qquad \Lambda=0,1,2,3,
\end{equation}
where $\Delta\equiv \frac{1}{\rho}\,\partial_\rho \left( \rho \,\partial_\rho \right) \+ \partial_z^2$ is the flat Laplacian in cylindrical coordinates.  Then,  the M-theory fields of \eqref{eq:MtheoryAnsatz} are given by 
\begin{equation}
Z_\Lambda \= \frac{e^{b_\Lambda} \, L_\Lambda - e^{-b_\Lambda} \,L_\Lambda^{-1}}{2 a_\Lambda} \,,\qquad T_I \= -\frac{\sqrt{1+a_I^2 Z_I^2}}{Z_I}\,,\qquad \star_2 dH_0 \= \frac{\rho}{a_0}\, d(\log L_0)\,,
\label{eq:LinearSolGen}
\end{equation}
where $a_\Lambda$ and $b_\Lambda$ are positive arbitrary constants, and the base warp factor $\nu$ can be obtained by integrating 
\begin{align}
\frac{2}{\rho}\, \partial_z \nu &\=\sum_{\Lambda=0}^4  \left[\partial_\rho \log L_\Lambda \,\partial_z \log L_\Lambda+ \partial_\rho \log W_\Lambda \,\partial_z \log W_\Lambda\right]\,, \nn \\
 \frac{4}{\rho}\, \partial_\rho \nu  &\=\sum_{\Lambda=0}^4  \left[\left( \partial_\rho \log L_\Lambda\right)^2 - \left(\partial_z \log L_\Lambda\right)^2 +\left( \partial_\rho \log W_\Lambda\right)^2 - \left(\partial_z \log W_\Lambda\right)^2 \right].
\end{align}

\begin{figure}[ht]
\centering
\includegraphics[height=63mm]{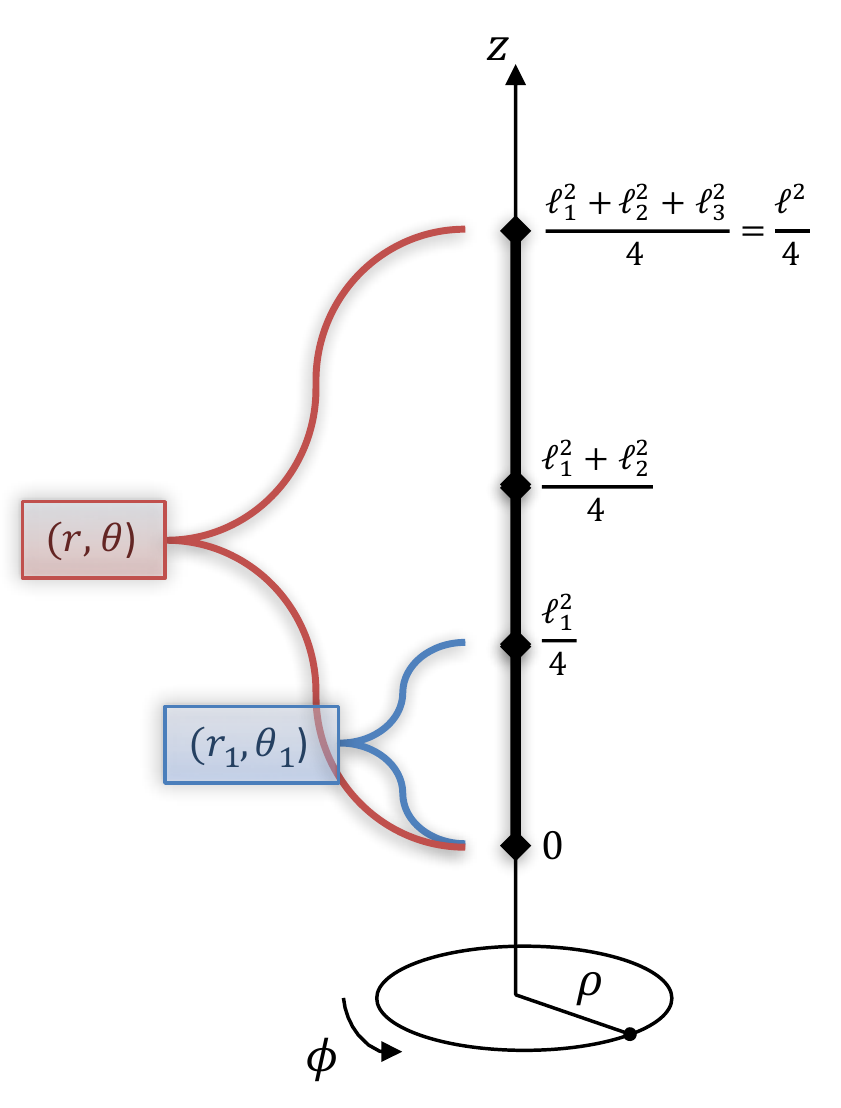}
\caption{Schematic description of connected rod sources on the $z$-axis. We depict the local spherical coordinates on the first rod,  $(r_1,\theta_1)$ \eqref{eq:DefDistance}, and the global spherical coordinates for the whole configuration, $(r,\theta)$ \eqref{eq:DefDistanceglobal}.}
\label{fig:rodsources}
\end{figure}

The harmonic functions are sourced on the $z$-axis by \textit{an arbitrary number of rods}.  We assume for the scope of this paper that they have a \textit{finite length} and are \textit{connected}.\footnote{Inspired by the results of \cite{Elvang:2002br,Bah:2020pdz,Bah:2021owp,Bah:2021rki,Heidmann:2021cms,Bah:2022yji},  the rod sources are connected to prevent struts in between two disconnected rods.  A strut is a string with negative tension that manifests itself as a conical excess along a segment where a compact coordinate degenerates and that cannot be resolved classically in supergravity.} Thus,  we consider $n$ connected rod sources such that the origin of the $z$-axis is located at the extremity of the first rod.  We depicted a generic rod configuration in Fig.\ref{fig:rodsources}.  The rod lengths are given by $\ell_i^2/4$,  $i=1,\ldots,n$,  while the overall length is $\ell^2/4$ such that
\begin{equation}
\ell^2 \equi \sum_{i=1}^n \ell_i^2\,.
\end{equation}

We introduce the \textit{local spherical coordinates on the $i^\text{th}$ rod}, $(r_i,\theta_i)$,  given by
\begin{equation}
\begin{split}
r_i^2 &\equi 2\left[\sqrt{\rho^{2}+\left(z-\frac{1}{4} \sum_{j=1}^i \ell_j^2\right)^{2}}+\sqrt{\rho^{2}+\left(z-\frac{1}{4} \sum_{j=1}^{i-1} \ell_j^2\right)^{2}} -\frac{\ell_i^2}{4}\right]\,, \\
\cos 2\theta_i &\equi \frac{4}{\ell_i^2} \left[\sqrt{\rho^{2}+\left(z-\frac{1}{4} \sum_{j=1}^{i-1} \ell_j^2\right)^{2}}- \sqrt{\rho^{2}+\left(z-\frac{1}{4} \sum_{j=1}^{i} \ell_j^2\right)^{2}} \right]\,,
\end{split}
\label{eq:DefDistance}
\end{equation}
where $0\leq \theta_i\leq \frac{\pi}{2}$ and $r_i \geq 0$.  The coordinate $r_i$ measures the radial distance to the rod.  Indeed,  taking $r_i=0$ and varying $\theta_i$ from $0$ to $\pi/2$ is equivalent to a shift along the $i^\text{th}$ rod such that $\rho=0$ with $z$ varying from $\coeff{1}{4}\sum_{j=1}^{i-1}\ell_j^2$ to $\coeff{1}{4}\sum_{j=1}^{i}\ell_j^2$.

Moreover,  we also define the \textit{global spherical coordinates} $(r,\theta)$,
\begin{equation}
\begin{split}
r^2 &\equi 2\left[\sqrt{\rho^{2}+\left(z-\frac{\ell^2}{4} \right)^{2}}+\sqrt{\rho^{2}+z^{2}} -\frac{\ell^2}{4}\right]\,, \\
\cos 2\theta &\equi \frac{4}{\ell^2} \left[\sqrt{\rho^{2}+z^{2}}-\sqrt{\rho^{2}+\left(z-\frac{\ell^2}{4} \right)^{2}}\right]\,,
\end{split}
\label{eq:DefDistanceglobal}
\end{equation}
They are the spherical coordinates centered on the whole rod configuration,  such that the rod sources are located at $r^2=0$, and varying $\theta$ from $0$ to $\pi/2$ moves from the first rod to the very last.  In this paper,  we will prefer to describe the solutions in terms of these global spherical coordinates since they are the physical coordinates centered around the gravitational sources.  The solutions will also depend on the local spherical coordinates at each source $(r_i,\theta_i)$,  which can be considered as function of $(r,\theta)$ by simply replacing $(\rho,z)$ in \eqref{eq:DefDistance} by
\begin{equation}
\begin{split}
\rho= \frac{r\sqrt{r+\ell^2}}{4} \sin 2\theta \,,\qquad z = \frac{2 r^2+\ell^2}{8} \cos 2\theta+  \frac{\ell^2}{8}\,.
\end{split}
\label{eq:ri_thetaidef}
\end{equation}

The eight functions $(L_\Lambda,W_\Lambda)$ are \textit{sourced at the rods with specific weights} $(P_i^{(\Lambda)},G_i^{(\Lambda)})$:
\begin{equation}
L_\Lambda \= \prod_{i=1}^{n} \left(1+ \frac{\ell_i^2}{r_i^2} \right)^{P_i^{(\Lambda)}}\,,\quad W_\Lambda \=  \prod_{i=1}^{n} \left(1+ \frac{\ell_i^2}{r_i^2} \right)^{G_i^{(\Lambda)}}\,,\qquad \Lambda=0,1,..,3.
\label{eq:HarmFunc2}
\end{equation}
The warp factors $Z_\Lambda$ and the gauge potentials $T_I$ can be directly derived from \eqref{eq:LinearSolGen}, and we have in addition
\begin{equation}
\begin{split}
H_0  &\= \frac{1}{4a_0}\sum_{i=1}^{n} \ell_i^2 P_i^{(0)}\,\cos 2\theta_i\,,\\
e^{2\nu} &\=   \prod_{i, j=1}^{n} \left( \frac{\left( \left(r_i^2+\ell_i^2 \right) \cos^2\theta_i +  \left(r_j^2+\ell_j^2\right) \sin^2\theta_j \right)\left(r_i^2 \cos^2\theta_i +  r_j^2 \sin^2\theta_j \right)}{\left( \left(r_i^2+\ell_i^2 \right) \cos^2\theta_i + r_j^2 \sin^2\theta_j \right)\left(r_i^2 \cos^2\theta_i +  \left(r_j^2+\ell_j^2\right) \sin^2\theta_j \right)}\right)^{\alpha_{ij}}\,, 
\end{split}
\end{equation}
where we have defined
\begin{equation}
\alpha_{ij}\equi \sum_{\Lambda=0}^3 \left[ P_i^{(\Lambda)} P_j^{(\Lambda)} +G_i^{(\Lambda)} G_j^{(\Lambda)} \right]\,.
\label{eq:AlphaDef}
\end{equation}
The magnetic duals of the electric M2 gauge potentials have exactly the same form as $H_0$ by replacing $a_0 \to a_I$ and $P_i^{(0)} \to P_i^{(I)}$ with $I=1,2,3$.  At large distance $r\to \infty$,  we have $\cos 2\theta_i \sim \cos 2\theta$.  Therefore,  the solutions correspond to M2-M2-M2 geometries with potential KKm charges along $\psi$, $k$,  such that the supergravity charges are given by
\begin{equation}
k \=  \frac{1}{4a_0}\sum_{i=1}^{n} \ell_i^2 P_i^{(0)}\,,\qquad Q_I \=  \frac{1}{a_I}\sum_{i=1}^{n} \ell_i^2 P_i^{(I)}\,,\quad I=1,2,3\,,
\label{eq:M2M2M2chargesGen}
\end{equation}
where $Q_1$, $Q_2$ and $Q_3$ are the M2 brane charges.  One can use these expressions to directly fix the constants $a_I$ in terms of the net charges and the rod parameters.

In \cite{Bah:2020ogh,Bah:2020pdz,Bah:2021owp,Bah:2021rki,Heidmann:2021cms},  asymptotically-flat geometries have been extracted from these solutions.  We will adapt the construction to obtain geometries asymptotic to AdS$_2\times$S$^3/\mathbb{Z}_k\times$T$^6$ in M-theory.  As their asymptotically-flat cousins,  they will be internally sourced by rods leading to a chain of regular bolts.  The difference in the asymptotic constraints however will change the constants $b_\Lambda$ and modify the geometries globally.  

Before doing so, we point out some useful expressions between the local and global spherical coordinates:
\begin{align}
&r_{i}^2 \cos^2 \theta_i \= (r_{i+1}^2+\ell_{i+1}^2) \cos^2\theta_{i+1}\,,\qquad  r_{i+1}^2 \sin^2 \theta_{i+1} \= (r_i^2+\ell_i^2) \sin^2\theta_i\,, \nn \\
&\prod_{i=1}^{n} \left(1+ \frac{\ell_i^2}{r_i^2} \right) \= 1+ \frac{\ell^2}{r^2}\,,\qquad \sum_{i=1}^{n} \ell_i^2 \cos 2\theta_i \= \ell^2 \cos 2\theta\,,\nn \\
& \prod_{i, j=1}^{n}\frac{\left( \left(r_i^2+\ell_i^2 \right) \cos^2\theta_i +  \left(r_j^2+\ell_j^2\right) \sin^2\theta_j \right)\left(r_i^2 \cos^2\theta_i +  r_j^2 \sin^2\theta_j \right)}{\left( \left(r_i^2+\ell_i^2 \right) \cos^2\theta_i + r_j^2 \sin^2\theta_j \right)\left(r_i^2 \cos^2\theta_i +  \left(r_j^2+\ell_j^2\right) \sin^2\theta_j \right)} \label{eq:SimplRelations2}\\
& \hspace{8.5cm} \= \frac{r^2(r^2+\ell^2)}{(r^2+\ell^2\sin^2\theta)(r^2+\ell^2\cos^2\theta)}\,. \nn
\end{align}

\subsection{Regular asymptotically-\texorpdfstring{AdS$_2$}{AdS2} solutions}

We first derive the constraints on the asymptotics before discussing internal boundary conditions at the rod sources.

\subsubsection{Asymptotic boundary conditions}

We expand the solutions at large distance $r\to \infty$:
\begin{align}
&W_\Lambda,\, e^{2\nu} \,\sim \,1\,,\qquad H_0   \,\sim \,k\cos 2\theta\,, \qquad  Z_0 \,\sim\, \frac{4k \sinh b_0}{\sum_{i=1}^{n} \ell_i^2 P_i^{(0)}} + \frac{4k \cosh b_0}{r^2}\,,\nn \\
& Z_{I} \,\sim\, \frac{Q_{I} \sinh b_{I}}{\sum_{i=1}^n \ell_i^2 P_i^{(I)}} + \frac{Q_{I} \cosh b_{I}}{r^2}\,, \qquad I=1,2,3.
\label{eq:AsympBehav}
\end{align}
As argued in \cite{Bena:2018bbd},  four-charge solutions are asymptotic to AdS$_2$ if the warp factors associated to the gauge potentials vanish asymptotically.  Thus, we impose
\begin{equation}
b_0 \= b_1 \= b_2 \= b_3 \= 0\,.
\label{eq:AsympBehav_afterb}
\end{equation}
The metric and fields \eqref{eq:MtheoryAnsatz} are asymptotic to
\begin{align}
ds_{11}^2 &\,\sim - \frac{r^4\, dt^2}{Q^2} +kQ\,\left[\frac{dr^2}{r^2} +d\theta^2+\cos^2 \theta\,d\varphi_1^2+\sin^2 \theta\,d\varphi_2^2 \right]+ Q\, \sum_{I=1}^3 \frac{dy_{2I-1}^2+dy_{2I}^2}{Q_I} , \nonumber\\
F_4 &\,\sim \,-2 r dr \wedge dt \wedge \sum_{I=1}^3 \frac{dy_{2I-1}\wedge dy_{2I}}{Q_I}\,,\qquad Q\equi (Q_1 Q_2 Q_3)^{\frac{1}{3}}\,,\label{eq:AdS2Asymp}
\end{align}
where we have defined \textit{the hyperspherical angles} of the S$^3$ from the Hopf fibration angles:
\begin{equation}
\varphi_1 \equi \frac{1}{2}\left( \phi +\frac{\psi}{k}\right)\,,\qquad \varphi_2 \equi \frac{1}{2}\left( \phi -\frac{\psi}{k}\right)\quad \Leftrightarrow \quad \phi \= \varphi_1 +\varphi_2 \,,\qquad \psi= k \,(\varphi_1 -\varphi_2)\,.
\label{eq:DefHyperspher}
\end{equation}
We define the periodicity of the compact directions such as
\begin{equation}
\begin{split}
(\psi,\phi) &\= (\psi,\phi) \+ (4\pi,0)\,,\qquad (\psi,\phi) \= (\psi,\phi) \+ (2\pi,2\pi)\,, \qquad \theta \= \theta \+ \frac{\pi}{2} \,,\\
y_a &\= y_a \+ 2\pi R_{y_a}\,, \quad a=1,\ldots,6\,,
\end{split}
\label{eq:psi_phiPerio}
\end{equation}
where $R_{y_a}$ are the intrinsic radii of the T$^6$ directions.  

Thus,  the geometries are manifestly asymptotic to AdS$_2\times$S$^3/\mathbb{Z}_k\times$T$^6$ in the coordinates $\hat{t} \equi 2t/\sqrt{kQ} $ and $\hat{r} = r^2/Q$.  Moreover,  one can restrict to solutions without orbifold asymptotically by simply considering $k=1$ in the above expressions.

\subsubsection{Internal boundary conditions}
\label{sec:InternalBCLinear}

The locus of each rod,  $\rho=0$ and $\coeff{1}{4}\sum_{j=1}^{i-1}\ell_j^2\leq z \leq \coeff{1}{4}\sum_{j=1}^{i}\ell_j^2$,  corresponds to $r_i=0$ and $0\leq \theta_i \leq \pi/2$ in the local spherical coordinates \eqref{eq:DefDistance}.  Thus,  the eight functions $(L_\Lambda, W_\Lambda)$  \eqref{eq:HarmFunc2} are either blowing or vanishing if their weights are non-zero.  More precisely,  we have
\begin{equation}
Z_\Lambda \propto r_i^{-2|P_i^{(\Lambda)}|}\,,\qquad W_\Lambda \propto r_i^{-2G_i^{(\Lambda)}}\,,\qquad e^{2\nu} \propto r_i^{2\alpha_{ii}}\,,\qquad r_i\to 0\,.
\end{equation}

Therefore,  there are only 8 non-trivial combinations for which the rod corresponds to a regular coordinate degeneracy on the $z$-axis such that the local metric behaves as
\begin{equation}
ds_{11}^2 \,\propto\, dr_i^2 - \frac{r_i^2}{\kappa_t^2} dt^2 +ds(\text{horizon})^2\,,\quad \text{or}\quad ds_{11}^2 \,\propto\, dr_i^2 + \frac{r_i^2}{\kappa_x^2}  dx^2 - g_{tt} dt^2 +ds(\text{bubble})^2\,,
\label{eq:boltDef}
\end{equation}
where $x$ is one of the compact direction $(\psi,y_1,y_2,y_3,y_4,y_5,y_6)$, and $\kappa$ is a constant. Moreover, $ds(\text{horizon})$ or $ds(\text{bubble})$ are the line elements of the compact space that correspond to either a horizon if the rod induces the degeneracy of the timelike direction or a bubble if it is a spacelike direction.  In addition,  $\kappa$ must be fixed by regularity in terms of the periodicity of the compact direction,  generically denoted as $x\to x+2\pi R_x$ here, or the black hole temperature, $T$:
\begin{equation}
\kappa_t \= \frac{1}{2\pi T}\,,\qquad \text{or}\qquad \kappa_x \= R_x\,.
\label{eq:ConstraintInternal}
\end{equation}

The 8 values of weights,  $(P_i^{(\Lambda)},G_i^{(\Lambda)})$,  that lead to these local geometries are summarized in Table \ref{tab:internalBC}.
\begin{table}[ht]
  \centering
  \begin{tabular}{|c||c|c|c|c!{\vrule width 1.4pt}c|c|c|c|}
\hline 
 &  $P_i^{(0)}$ & $P_i^{(1)}$& $P_i^{(2)}$&  $P_i^{(3)}$&$G_i^{(0)}$& $G_i^{(1)}$& $G_i^{(2)}$& $G_i^{(3)}$ \\ \hline \hline
Horizon & $\frac{1}{2}$ &$\frac{1}{2}$ &$\frac{1}{2}$ &$\frac{1}{2}$ &$0$ &$0$ &$0$ &$0$ \\ \noalign{\hrule height 0.8pt}
$\psi$ degeneracy & $1$ &$0$ &$0$ &$0$ &$0$ &$0$ &$0$ &$0$ \\ \hline
$y_1$ degeneracy & $\frac{1}{2}$ &$\frac{1}{2}$ &$0$  &$0$ &$-\frac{1}{2}$&$\frac{1}{2}$ &$0$ &$0$ \\ \hline
$y_2$ degeneracy  &$\frac{1}{2}$ &$\frac{1}{2}$ &$0$  &$0$ &$-\frac{1}{2}$&$-\frac{1}{2}$ &$0$ &$0$  \\ \hline
$y_3$ degeneracy &$\frac{1}{2}$ &$0$ &$\frac{1}{2}$  &$0$ &$-\frac{1}{2}$&$0$ &$\frac{1}{2}$ &$0$  \\ \hline
$y_4$ degeneracy &$\frac{1}{2}$ &$0$ &$\frac{1}{2}$  &$0$ &$-\frac{1}{2}$&$0$ &$-\frac{1}{2}$ &$0$ \\ \hline
$y_5$ degeneracy &$\frac{1}{2}$ &$0$ &$0$  &$\frac{1}{2}$ &$-\frac{1}{2}$&$0$ &$0$ &$\frac{1}{2}$  \\ \hline
$y_6$ degeneracy &$\frac{1}{2}$ &$0$ &$0$  &$\frac{1}{2}$ &$-\frac{1}{2}$&$0$ &$0$ &$-\frac{1}{2}$ \\ \hline
  \end{tabular}
  \caption{\label{tab:internalBC} The eight possible weights at the $i^\text{th}$ rod source leading to a regular coordinate degeneracy of the timelike direction or a compact spacelike direction.}
\end{table}

Moreover,  the regularity conditions \eqref{eq:ConstraintInternal} give a set of $n$ algebraic equations that constrain the rod lengths and their charges in terms of the total charges, temperature and the radii of the extra dimensions. 

Interestingly,  for the regular sources in Table \ref{tab:internalBC},  the exponents $\alpha_{ij}$ \eqref{eq:AlphaDef} simplify to
\begin{equation}
\alpha_{ij} \= \begin{cases} 1 \quad \text{if the $i^\text{th}$ and $j^\text{th}$ rods are of the same nature,} \\
\frac{1}{2} \quad \text{otherwise,}
\end{cases}
\label{eq:AlphaSimple}
\end{equation}
where by ``same nature'' we meant that the same coordinate degenerates at both rods.

Finally,  a necessary condition for having asymptotically-AdS$_2$ solutions is to have non-zero charges $Q_1 , Q_2,  Q_3 \neq 0$ \eqref{eq:AsympBehav}.  From \eqref{eq:M2M2M2chargesGen},  we notice that each M2-brane charge $Q_I$ is induced internally by the rods that have a non-zero weight $P_i^{(I)}$. Thus, this requires that $P_i^{(1)} \neq 0$ for at least one rod in the configuration,  and similarly for $P_i^{(2)}$ and $P_i^{(3)}$.  Let us first consider the charge $Q_1$.  From Table \ref{tab:internalBC},  only two types of smooth rods have $P_i^{(1)}\neq 0$: the rods that force either $y_1$ or $y_2$ to degenerate,  that is a direction of the first T$^2$.  Therefore,  the solutions require a direction of the first T$^2$ to degenerate on one rod at least.  By using the same argument for $P_i^{(2)}$ and $P_i^{(3)}$, \textit{asymptotically-AdS$_2$ horizonless bubbling geometries require each internal T$^{\,2}$ inside the T$^{\,6}$ to degenerate in the spacetime.}

\subsection{Quantized charges and moduli}

In supergravity,  quantized charges are expressed in terms of an integral of the field strength or its Hodge dual. The Hodge dual of $F_4=\star F_7$ have a simple form for our solutions:
$$
F_7 = \sum_{I=1}^3 F_7^{(I)} \,,\quad
F_7^{(I)} \equiv d\left[ \sum_{J,K=1}^3 \frac{|\epsilon_{IJK}|}{2} \, H_I \,d\phi \wedge d\psi \wedge dy_{2J-1} \wedge dy_{2J} \wedge dy_{2K-1} \wedge dy_{2K}\right],
$$
where $\epsilon_{IJK}$ is the rank 3 Levi-Civita tensor, and we have
\begin{equation}
H_I \= \frac{Q_I}{4\sum_{i=1}^n \ell_i^2 \,P_i^{(I)}}\,\,\sum_{i=1}^{n}\ell_i^2 P_i^{(I)}\,\cos 2\theta_i\,,\qquad I=1,2,3\,.
\label{eq:H_i}
\end{equation}

The quantized charges $N_I$, corresponding to the total number of M2 branes wrapping the $I$-th 2-torus, are given by
\begin{equation}
 N_I \= \frac{1}{ (2 \pi l_p)^6}\int F_7^{(I)} \,,
\end{equation}
where $l_p$ is the Planck length in eleven dimensions.
For our solutions, using \eqref{eq:H_i} and the periodicities of the angles \eqref{eq:psi_phiPerio}, we find
\begin{equation}
  N_1 ~=~ \frac{R_{y_3}R_{y_4}R_{y_5}R_{y_6}}{(l_p)^6} Q_1 \,,\qquad
  N_2 ~=~ \frac{R_{y_1}R_{y_2}R_{y_5}R_{y_6}}{(l_p)^6} Q_2 \,,\qquad
  N_3 ~=~ \frac{R_{y_1}R_{y_2}R_{y_3}R_{y_4}}{(l_p)^6} Q_3 \,.
   \label{eq:quantized_charges}
\end{equation}

Since our solutions decompose into a chain of charged bubbles,  one can compute quantized charges associated to each rod, corresponding to the numbers of M2 branes wrapping each individual bubble.  In practice,  they are given by integrating the field strength at the locus of each rod:
\begin{equation}
  n_i^{(I)} \=\frac{1}{(2 \pi l_p)^6 } \int_{ \coeff{1}{4}\sum_{j=1}^{i-1}\ell_j^2\leq z \leq \coeff{1}{4}\sum_{j=1}^{i}\ell_j^2}\, F_7^{(I)} \bigl|_{\rho=0}\,, \quad I = 1,2,3 \,,\quad  1 \leq i \leq n \,.
\end{equation}
On these intervals,  one can check that $\theta_i$ varies from $0$ to $\pi/2$ \eqref{eq:DefDistance}, while the other angles $\theta_j$, $j \neq i$,  are all constant, equal to $\pi/2$ if $j < i$ and $0$ otherwise. Thus, the local quantized charges at the $i^\text{th}$ rod are given by:
\begin{equation}
\begin{aligned}
  n_i^{(1)} &= \frac{R_{y_3}R_{y_4}R_{y_5}R_{y_6}}{(l_p)^6} \frac{l_i^2 P_i^{(1)}}{\sum_{j=1}^{n} l_j^2 P_j^{(1)}} Q_1 \,,\\
  n_i^{(2)} &= \frac{R_{y_1}R_{y_2}R_{y_5}R_{y_6}}{(l_p)^6} \frac{l_i^2 P_i^{(2)}}{\sum_{j=1}^{n} l_j^2 P_j^{(2)}} Q_2 \,,\\
  n_i^{(3)} &=  \frac{R_{y_1}R_{y_2}R_{y_3}R_{y_4}}{(l_p)^6} \frac{l_i^2 P_i^{(3)}}{\sum_{j=1}^{n} l_j^2 P_j^{(3)}} Q_3 \,.
\end{aligned}
   \label{eq:local_quantized_charges}
\end{equation}

For a solution to be physical,  all local charges $n_i^{(I)}$ needs to be integers. 

The moduli of the solutions is also given by the asymptotic sizes of the internal space,  S$^3\times$T$^6$.  Using the asymptotic expansion of the fields \eqref{eq:AdS2Asymp},  we find
\begin{align}
\mathrm{Vol}_{T^6} &= (2\pi)^6\,\cV_6\,, \quad  \mathrm{Vol}_{T^2(1)} = (2\pi)^2\, \cV_6^\frac{1}{3} \,\left(\frac{N_2 N_3}{N_1^2} \right)^\frac{1}{3}\,,\quad  \mathrm{Vol}_{T^2(2)} = (2\pi)^2\, \cV_6^\frac{1}{3} \,\left(\frac{N_1 N_3}{N_2^2} \right)^\frac{1}{3}\,,\nonumber \\  \mathrm{Vol}_{T^2(3)} & = (2\pi)^2\, \cV_6^\frac{1}{3} \,\left(\frac{N_1 N_2}{N_3^2} \right)^\frac{1}{3}\,,\qquad   \mathrm{Vol}_{S^3} = 2\pi^2 \sqrt{ k N_1 N_2 N_3} \,\frac{(l_p)^9}{\cV_6}  \,,\label{eq:AsymVol}
\end{align}
where we have defined the product of radii
\begin{equation}
\cV_6 \equiv \prod_{a=1}^6 R_{y_a}\,.
\label{eq:RadiusProduct}
\end{equation}
Note that the radius of the S$^3$ is fixed in terms of the AdS$_2$ radius such that the curvature of the AdS spacetime is balanced off with that of the sphere.

\subsection{Profile in five dimensions and supersymmetry breaking}
\label{sub:profile5D_susy}

The ansatz \eqref{eq:MtheoryAnsatz} can be written in five dimensions by reducing on the T$^6$,  using the standard rules \cite{Cremmer:1997ct}.  As detailed in \cite{Heidmann:2021cms},  this leads to the STU model in five dimensions.  The resulting metric is
\begin{equation}
ds_{5}^2 = - \frac{dt^2}{(Z_1Z_2Z_3)^{\frac23}} + (Z_1Z_2Z_3)^{\frac13} \left[\frac{1}{Z_0} \left(d\psi +H_0 d\phi\right)^2 + Z_0 \left( e^{2\nu} \left(d\rho^2 + dz^2 \right) +\rho^2 d\phi^2\right)\right] \,.
\label{eq:ds5met}
\end{equation}
\\
Remarkably, the torus warp factors $W_I$ do not appear in the five-dimensional metric: from the point of view of the five-dimensional theory they are decoupled scalar fields.
\\
Using the asymptotic behavior of the fields,  \eqref{eq:AsympBehav} with\eqref{eq:AsympBehav_afterb}, the solutions are asymptotic to $\AdS_2\times\Sph^3/\mathbb{Z}_k$.  However, they will look singular in the IR since they are principally sourced by regular T$^6$ coordinate singularities.  These singularities are resolved by considering the whole string theory description and the uplift in M-theory.

We now turn to the potential supersymmetries preserved by the solutions. Our aim is to prove that the class of solutions described in this paper are generically non-supersymmetric. This is not as simple a task as in asymptotically flat space, where one can simply compare the total mass (computed from an asymptotic expansion) with the sum of the charges: one needs to turn to the BPS equations.

It will be sufficient to prove that a single BPS constraint is not satisfied. While these constraints are gauge invariant, they are often expressed more easily in very specific gauges.  The BPS equations require a Killing vector to be written as a bilinear of the Killing spinor: $V^\mu = i \bar\epsilon_i \gamma^\mu \epsilon^i$. The gauge used to describe such supersymmetric solutions then differs whether this vector $V$ is time-like or null.  Focusing on the solutions presented in this paper, we  derive conditions when the Killing vector is time-like. 

In addition,  the metric \eqref{eq:ds5met} has two $U(1)$ isometries along the $\phi$ and $\psi$ directions. Following \cite{Bellorin:2006yr},  this allows for a refinement and a simplification of the BPS constraints. With $V \equiv\partial_\tau$,  the metric can be cast into the form
\begin{equation}
  ds_5^2 ~=~ -f^2 (d\tau + \omega)^2 + f^{-1} \qty(H^{-1} (d\hat\psi + \chi)^2 + H ds_3^2) \,,
  \label{eq:ansatz_susy_5D}
\end{equation}
where $\hat\psi$ is a coordinate associated with one isometry.  Among other constraints, supersymmetry implies that the function $H$ is harmonic with respect to the covariant derivative associated to $ds_3^2$ (see section 4.2.1 of \cite{Bellorin:2006yr}).

Thus, to prove that a given solution is not supersymmetric, it is sufficient to cast it in the form of \eqref{eq:ansatz_susy_5D}, and show that the function $H$ is not harmonic. Without prior knowledge of what the Killing vector $V$ might be, one needs to do this test for any time-like Killing vectors that preserve the form of the metric. This can be done by considering the change of variables:
\begin{equation}
  \tau ~\equiv~ t \,,\qquad \hat\psi ~\equiv~ \psi + \alpha t \,, \qquad \hat\phi ~\equiv~ \phi + \beta t \,,
\end{equation}
where $\alpha$ and $\beta$ are arbitrary real numbers verifying $g_{tt} + \alpha^2 g_{\psi\psi} + \beta^2 g_{\phi\phi} + 2\alpha \beta g_{\psi\phi} < 0$. 

Applying this to our solutions \eqref{eq:ds5met},  one finds
\begin{align}
  H^2 ~&=~ \frac{Z_0^2 - Z_0 Z^3 (\alpha + H_0 \beta)^2 - Z_0^3\, Z^3 \,\beta^2 \,\rho^2}{\qty(1 - Z_0 Z^3 \,\beta^2 \,\rho^2)^2} \,, \label{eq:warp_harmonic}
  \\
  ds_3^2 ~&=~ e^{2\nu} \qty(1- Z_0 \, Z^3 \, \beta^2 \rho^2)\qty(dz^2 + d\rho^2) + \rho^2 d\hat\phi^2 \,. \nn
\end{align}
Moreover, for fields that do not depend on $\hat\phi$, the Laplacian operator in the metric $ds_3^2$ is proportional to the flat Laplacian $\Delta\equiv \frac{1}{\rho}\,\partial_\rho \left( \rho \,\partial_\rho \right) \+ \partial_z^2$.  We will prove that the solutions are non-supersymmetric by showing that
\begin{equation}
\Delta H \,\neq\, 0 \,,\qquad \forall( \alpha,\beta) \in \mathbb{R}.
\end{equation}
Since proving it in full generality is rather tedious and not specifically enlightening,  we will do it in a case by case manner when dealing with explicit solutions.  In a word,  we found that the solutions break supersymmetry as soon as one rod source has a non-zero size:
\begin{equation}
\exists \ell_i \,,\qquad \ell_i >0\,.
\end{equation}
Note that when all $\ell_i=0$,  the sources become point-like.  As explained in \cite{Heidmann:2021cms},  the ansatz reduces to the static four-charge BPS multicenter solutions,  such that $W_\Lambda =1$ \eqref{eq:HarmFunc2} and $Z_\Lambda$ are harmonic functions.

While this method shows clearly when a solution is not supersymmetric, it does not quantify the ``amount of supersymmetry breaking.'' In asymptotically flat spacetime,  the difference between the mass and the sum of charges is a good indicator of the supersymmetry breaking.  In the present configurations,  this amount is somehow a function of the rod sizes.  However,  solutions in AdS$_2$ are necessarily scale invariant, so that their size,  which is here given by the $\ell_i^2$,  can be arbitrarily dialed by a coordinate change.  Therefore,  the amplitude of the rod sizes cannot indicate the amount of supersymmetry breaking of the geometries and how far from empty AdS$_2$ they are in the phase space of solutions.

\subsection{Final form of the solutions}
\label{sec:linearbranchSum}

We summarize here the family of solutions we have constructed so far.  The M-theory fields are given by \eqref{eq:MtheoryAnsatz}
\begin{align}
ds_{11}^2 = &- \frac{dt^2}{\left(W_0 Z_1 Z_2 Z_3 \right)^{\frac{2}{3}}} + \left(\frac{Z_1 Z_2 Z_3}{W_0^2}\right)^{\frac{1}{3}} \left[\frac{1}{Z_0} \left(d\psi +H_0 d\phi\right)^2 + Z_0 \left( e^{2\nu} \left(d\rho^2 + dz^2 \right) +\rho^2 d\phi^2\right)\right] \nonumber\\
&\hspace{-0.5cm} + \left(W_0Z_1 Z_2 Z_3\right)^{\frac{1}{3}} \left[\frac{1}{Z_1} \left(\frac{dy_1^2}{W_1} + W_1 \,dy_2^2 \right) + \frac{1}{Z_2} \left(\frac{dy_3^2}{W_2} + W_2 \,dy_4^2 \right) + \frac{1}{Z_3} \left(\frac{dy_5^2}{W_3} + W_3 \,dy_6^2 \right)\right], \nonumber\\
F_4 = &d\left[ T_1 \,dt \wedge dy_1 \wedge dy_2 \+T_2 \,dt \wedge dy_3 \wedge dy_4 \+T_3\,dt \wedge dy_5 \wedge dy_6 \right]\,. \label{eq:MtheoryAnsatz2}
\end{align}
The geometries obtained from the charged Weyl formalism that are asymptotic to AdS$_2\times$S$^3/\mathbb{Z}_k$ $\times$T$^6$ are sourced by $n$ connected rods on the $z$-axis of length $\ell_i^2/4$.  The fields are given by eight functions for which their logarithms are harmonic functions sourced at the rods
\begin{equation}
L_\Lambda \= \prod_{i=1}^{n} \left(1+ \frac{\ell_i^2}{r_i^2} \right)^{P_i^{(\Lambda)}}\,,\quad W_\Lambda \=  \prod_{i=1}^{n} \left(1+ \frac{\ell_i^2}{r_i^2} \right)^{G_i^{(\Lambda)}}\,,\quad \Lambda=0,1,..,3,
\label{eq:HarmFunc}
\end{equation}
such that
\begin{align}
Z_I &\= Q_I\,\frac{L_I-L_I^{-1}}{2 \sum_{i=1}^{n} \ell_i^2 P_i^{(I)}}\,,\qquad T_I \= -\frac{\sum_{i=1}^{n} \ell_i^2 P_i^{(I)}}{Q_I}\,\frac{L_I^2+1}{L_I^2-1}\,,\qquad I=1,2,3\,, \nn \\
Z_0 &\= 2k\,\frac{L_0 -L_0^{-1}}{\sum_{i=1}^{n} \ell_i^2 \,P_i^{(0)}}\,, \qquad H_0 \= \frac{k}{\sum_{i=1}^n \ell_i^2 \,P_i^{(0)}}\,\,\sum_{i=1}^{n}\ell_i^2 P_i^{(0)}\,\cos 2\theta_i\,,  \label{eq:LinearAdS2} \\
e^{2\nu} &\=   \prod_{i, j=1}^{n} \left( \frac{\left( \left(r_i^2+\ell_i^2 \right) \cos^2\theta_i +  \left(r_j^2+\ell_j^2\right) \sin^2\theta_j \right)\left(r_i^2 \cos^2\theta_i +  r_j^2 \sin^2\theta_j \right)}{\left( \left(r_i^2+\ell_i^2 \right) \cos^2\theta_i + r_j^2 \sin^2\theta_j \right)\left(r_i^2 \cos^2\theta_i +  \left(r_j^2+\ell_j^2\right) \sin^2\theta_j \right)}\right)^{\alpha_{ij}}\,, \nn
\end{align}
where the local spherical coordinates at each rod $(r_i,\theta_i)$ are given in terms of Weyl cylindrical coordinates,  $(\rho,z)$, in \eqref{eq:DefDistance} and in terms of the global spherical coordinates, $(r,\theta)$,  using \eqref{eq:ri_thetaidef},  the exponents $\alpha_{ij}$ are given in \eqref{eq:AlphaSimple}.

The $i^\text{th}$ rod can carry three M2 brane charges given in terms of the total charge by 
\begin{equation}
q_{\text{M2}\,\,i}^{(I)} \=  \frac{\ell_i^2 \,P_i^{(I)}}{\sum_{j=1}^n \ell_j^2 \,P_j^{(I)}} \, Q_I \,, 
\qquad  I=1,2,3. 
\label{eq:chargesRodGen}
\end{equation}
and the local numbers of M2 branes associated to these are given in \eqref{eq:local_quantized_charges}.

Moreover,  the weights $(P_i^{(\Lambda)},G_i^{(\Lambda)})$ at each rod take one of the eight possible values in Table \ref{tab:internalBC} depending on the coordinate that degenerates there.  Note that the rods that force a T$^6$ direction to shrink carry only one M2 charge which is associated to the stack of M2 branes that is wrapping the T$^2$ of the shrinking direction.  The rods that force the degeneracy of the S$^3$ Hopf angle, $\psi$,  do not carry any M2-brane charges.

Moreover,  the solutions are constrained by $n$ regularity equations that must be derived in a case by case manner \eqref{eq:ConstraintInternal}.  They will give constraints on all rod lengths, $\ell_i^2$,  and asymptotic quantities that are the M2-brane charges,  the radii of the T$^6$,  and the asymptotic orbifold parameter $k$.  

Note that we have a priori $n$ rod lengths and $n$ regularity conditions.  However, since asymptotically-AdS$_2$ geometries are scale-invariant, at least one rod length must remain a free parameter to give a scale to the geometries.  One has $n-1$ internal parameters for $n$ constraints at best.  A regularity condition will necessarily constrain the asymptotic quantities.  Thus, a given configuration cannot exist for arbitrary values of total charges and T$^6$ radii,  and exists only for specific corners of the moduli space.  We will see through specific examples what kind of constraints exists.  However,  there is a wide variety of solutions at hand, so that the whole moduli space is still densely populated by these types of solutions.

\section{\texorpdfstring{Three smooth T$^2$ deformations in AdS$_2\times$S$^3\times$T$^6$}{Three smooth T2 deformations in AdS2 x S3 x T6}}
\label{sec:BuAdS2T2}

In this section,  we construct and discuss the physics of the simplest non-supersymmetric solitons one can build.

As previously argued in section \ref{sec:InternalBCLinear},  one needs at least three rods that force the degeneracy of a direction in each internal T$^2$.  This generates the necessary three M2 charges to be asymptotic to AdS$_2$ in M-theory.  Thus,  we first construct the solutions obtained from such a three-rod configuration.  They correspond to asymptotically-AdS$_2$ geometries that cap off smoothly in the interior at $r=0$, where $r$ is the radius of the global spherical coordinates \eqref{eq:DefDistanceglobal}.  At this locus,  the spacetime ends as a chain of three bolts where a T$^2$ direction shrink depending on the position along the S$^3$.
Moreover,  we will show that the sizes of the bolts are unconstrained such that they can be dialed independently and arbitrarily while the total M2-brane charges are fixed in terms of the torus radii.

Finally,  we will prove that these bubbling geometries indeed break supersymmetry and correspond to non-supersymmetric smooth excitations of empty AdS$_2$ in M-theory.

\subsection{Rod profile}
\label{sec:3rod}

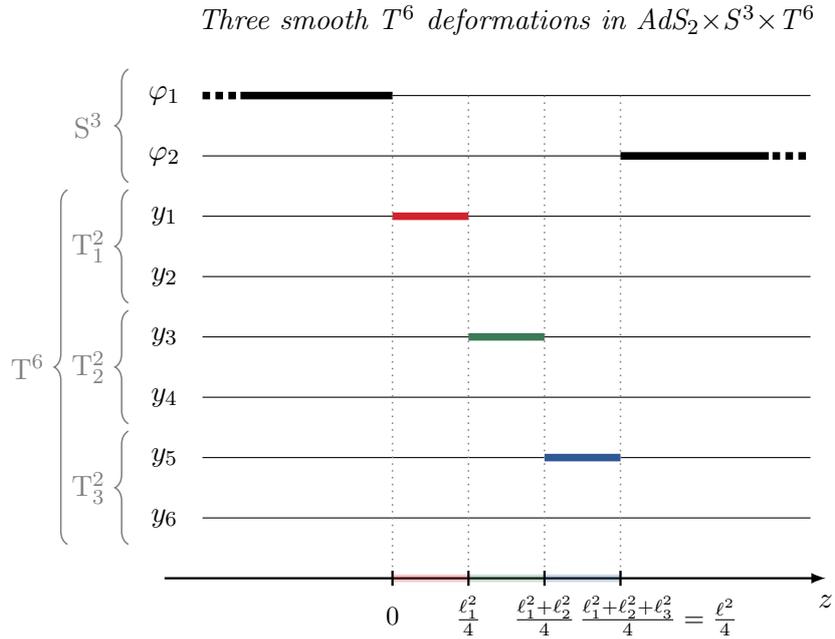
\begin{figure}[ht]
\centering
    \begin{tikzpicture}

\def\deb{-10} 
\def\inter{0.8} 
\def\ha{2.8} 
\def\zaxisline{8} 
\def\rodsize{2} 
\def\numrod{1.5} 

\def\fin{\deb+1+2*\rodsize+\numrod*\rodsize} 




\draw (\deb+0.5+\rodsize+0.5*\numrod*\rodsize,\ha+1) node{{{\it Three smooth T$^{\,6}$ deformations in AdS$_2\times$S$^{\,3}\times$T$^{\,6}$}}}; 


\draw[black,thin] (\deb+1,\ha) -- (\fin,\ha);
\draw[black,thin] (\deb,\ha-\inter) -- (\fin-1,\ha-\inter);
\draw[black,thin] (\deb,\ha-2*\inter) -- (\fin,\ha-2*\inter);
\draw[black,thin] (\deb,\ha-3*\inter) -- (\fin,\ha-3*\inter);
\draw[black,thin] (\deb,\ha-4*\inter) -- (\fin,\ha-4*\inter);
\draw[black,thin] (\deb,\ha-5*\inter) -- (\fin,\ha-5*\inter);
\draw[black,thin] (\deb,\ha-6*\inter) -- (\fin,\ha-6*\inter);
\draw[black,thin] (\deb,\ha-7*\inter) -- (\fin,\ha-7*\inter);
\draw[black,->, line width=0.3mm] (\deb-0.5,\ha-\zaxisline*\inter) -- (\fin+0.2,\ha-\zaxisline*\inter);

\draw [decorate, 
    decoration = {brace,
        raise=5pt,
        amplitude=5pt},line width=0.2mm,gray] (\deb-0.8,\ha-1.5*\inter+0.05) --  (\deb-0.8,\ha+0.5*\inter-0.05);
\draw [decorate, 
    decoration = {brace,
        raise=5pt,
        amplitude=5pt},line width=0.2mm,gray] (\deb-0.8,\ha-3.5*\inter+0.05) --  (\deb-0.8,\ha-1.5*\inter-0.05);
\draw [decorate, 
    decoration = {brace,
        raise=5pt,
        amplitude=5pt},line width=0.2mm,gray] (\deb-0.8,\ha-5.5*\inter+0.05) --  (\deb-0.8,\ha-3.5*\inter-0.05);
\draw [decorate, 
    decoration = {brace,
        raise=5pt,
        amplitude=5pt},line width=0.2mm,gray] (\deb-0.8,\ha-7.5*\inter+0.05) --  (\deb-0.8,\ha-5.5*\inter-0.05);
\draw [decorate, 
    decoration = {brace,
        raise=5pt,
        amplitude=5pt},line width=0.2mm,gray] (\deb-1.6,\ha-7.5*\inter+0.05) --  (\deb-1.6,\ha-1.5*\inter-0.05);

\draw[gray] (\deb-1.5,\ha-0.5*\inter) node{S$^3$};
\draw[gray] (\deb-1.5,\ha-2.5*\inter) node{T$^2_1$};
\draw[gray] (\deb-1.5,\ha-4.5*\inter) node{T$^2_2$};
\draw[gray] (\deb-1.5,\ha-6.5*\inter) node{T$^2_3$};
\draw[gray] (\deb-2.3,\ha-4.5*\inter) node{T$^6$};

\draw (\deb-0.5,\ha) node{$\varphi_1$};
\draw (\deb-0.5,\ha-\inter) node{$\varphi_2$};
\draw (\deb-0.5,\ha-2*\inter) node{$y_1$};
\draw (\deb-0.5,\ha-3*\inter) node{$y_2$};
\draw (\deb-0.5,\ha-4*\inter) node{$y_3$};
\draw (\deb-0.5,\ha-5*\inter) node{$y_4$};
\draw (\deb-0.5,\ha-6*\inter) node{$y_5$};
\draw (\deb-0.5,\ha-7*\inter) node{$y_6$};

\draw (\fin+0.2,\ha-\zaxisline*\inter-0.3) node{$z$};


\draw[black, dotted, line width=1mm] (\deb,\ha) -- (\deb+0.5,\ha);
\draw[black,line width=1mm] (\deb+0.5,\ha) -- (\deb+0.5+\rodsize,\ha);
\draw[black,line width=1mm] (\fin-0.5-\rodsize,\ha-\inter) -- (\fin-0.55,\ha-\inter);
\draw[black, dotted,line width=1mm] (\fin-0.5,\ha-\inter) -- (\fin,\ha-\inter);


\draw[amaranthred,line width=1mm] (\deb+0.5+\rodsize,\ha-2*\inter) -- (\deb+0.5+1.5*\rodsize,\ha-2*\inter);
\draw[amazon,line width=1mm] (\deb+0.5+1.5*\rodsize,\ha-4*\inter) -- (\deb+0.5+2*\rodsize,\ha-4*\inter);
\draw[bdazzledblue,line width=1mm] (\deb+0.5+2*\rodsize,\ha-6*\inter) -- (\deb+0.5+2.5*\rodsize,\ha-6*\inter);

\draw[amaranthred,line width=1mm,opacity=0.25] (\deb+0.5+\rodsize,\ha-\zaxisline*\inter) -- (\deb+0.5+1.5*\rodsize,\ha-\zaxisline*\inter);
\draw[amazon,line width=1mm,opacity=0.25] (\deb+0.5+1.5*\rodsize,\ha-\zaxisline*\inter) -- (\deb+0.5+2*\rodsize,\ha-\zaxisline*\inter);
\draw[bdazzledblue,line width=1mm,opacity=0.25] (\deb+0.5+2*\rodsize,\ha-\zaxisline*\inter) -- (\deb+0.5+2.5*\rodsize,\ha-\zaxisline*\inter);


\draw[gray,dotted,line width=0.2mm] (\deb+0.5+\rodsize,\ha) -- (\deb+0.5+\rodsize,\ha-\zaxisline*\inter);
\draw[gray,dotted,line width=0.2mm] (\deb+0.5+1.5*\rodsize,\ha) -- (\deb+0.5+1.5*\rodsize,\ha-\zaxisline*\inter);
\draw[gray,dotted,line width=0.2mm] (\deb+0.5+2*\rodsize,\ha) -- (\deb+0.5+2*\rodsize,\ha-\zaxisline*\inter);
\draw[gray,dotted,line width=0.2mm] (\deb+0.5+2.5*\rodsize,\ha) -- (\deb+0.5+2.5*\rodsize,\ha-\zaxisline*\inter);

\draw[line width=0.3mm] (\deb+0.5+1.5*\rodsize,\ha-\zaxisline*\inter+0.1) -- (\deb+0.5+1.5*\rodsize,\ha-\zaxisline*\inter-0.1);
\draw[line width=0.3mm] (\deb+0.5+\rodsize,\ha-\zaxisline*\inter+0.1) -- (\deb+0.5+\rodsize,\ha-\zaxisline*\inter-0.1);
\draw[line width=0.3mm] (\deb+0.5+2*\rodsize,\ha-\zaxisline*\inter+0.1) -- (\deb+0.5+2*\rodsize,\ha-\zaxisline*\inter-0.1);
\draw[line width=0.3mm] (\deb+0.5+2.5*\rodsize,\ha-\zaxisline*\inter+0.1) -- (\deb+0.5+2.5*\rodsize,\ha-\zaxisline*\inter-0.1);

\draw (\deb+0.5+\rodsize,\ha-\zaxisline*\inter-0.5) node{{\small $0$}};
\draw (\deb+0.5+1.5*\rodsize,\ha-\zaxisline*\inter-0.5) node{{\small $\frac{\ell_1^2}{4}$}};
\draw (\deb+0.5+2*\rodsize,\ha-\zaxisline*\inter-0.5) node{{\small $\frac{\ell_1^2+\ell_2^2}{4}$}};
\draw (\deb+0.5+2.75*\rodsize,\ha-\zaxisline*\inter-0.5) node{{\small $\frac{\ell_1^2+\ell_2^2+\ell_3^2}{4}=\frac{\ell^2}{4}$}};

\end{tikzpicture}
\caption{Rod diagram of the shrinking directions on the $z$-axis after sourcing the solutions with three connected rods that force the degeneracy of the $y_1$, $y_3$ and $y_5$ circles respectively. }
\label{fig:3rodsources}
\end{figure}  

We consider three connected rod sources of length parameter $\ell_i^2$ and total length $\ell^2$ (see Fig.\ref{fig:3rodsources}).  They force the $y_1$,  $y_3$ and $y_5$ circle to degenerate.  From Table \ref{tab:internalBC},  the weights at the rods are
\begin{equation}
\begin{split}
P_1^{(0)} &\= P_1^{(1)} \= -G_1^{(0)} \= G_1^{(1)} \= \frac{1}{2}\,,\qquad P_1^{(2)} \= P_1^{(3)} \= G_1^{(2)} \= G_1^{(3)} \= 0\,, \\
P_2^{(0)} &\= P_2^{(2)} \= -G_2^{(0)} \= G_2^{(2)} \=  \frac{1}{2}\,,\qquad P_2^{(1)} \= P_2^{(3)} \= G_2^{(1)} \= G_2^{(3)} \= 0\,, \\
P_3^{(0)} &\= P_3^{(3)} \= -G_3^{(0)} \= G_3^{(3)} \=  \frac{1}{2}\,,\qquad P_3^{(1)} \= P_3^{(2)} \= G_3^{(1)} \= G_3^{(2)} \= 0\,.
\end{split}
\end{equation}
For simplicity,  we consider that $k=1$ such that the S$^3$ has no conical defect asymptotically.

\subsection{The solution}

We simplify \eqref{eq:LinearAdS2} for the present rod configuration using  \eqref{eq:SimplRelations2}.  We find that
\begin{align}
Z_0 &\= \frac{4}{r \sqrt{r^2+\ell^2}}\,,\qquad H_0 \= \cos 2\theta \,,  \qquad W_0 \= \left(1+\frac{\ell^2}{r^2} \right)^{-\frac{1}{2}} ,\nn\\
 Z_I &\= \frac{Q_I}{r_I \sqrt{r_I^2+\ell_I^2}}\,,\qquad T_I \= - \frac{r_I^2+\frac{\ell_I^2}{2}}{Q_I}\,,\qquad  W_I \= \sqrt{1+\frac{\ell_I^2}{r_I^2}}\,,\quad I=1,2,3\,, \qquad \qquad  \nn 
 \end{align}
 \begin{align}
 e^{2\nu} &\= \frac{r^2(r^2+\ell^2)}{(r^2+\ell^2 \cos^2\theta)(r^2+\ell^2 \sin^2\theta)}\\
 & \hspace{1cm} \times \frac{\prod_{I=1}^3 (r_I^2+\ell_I^2)}{(r_2^2+\ell_2^2 \cos^2\theta_2)(r_2^2+\ell_2^2 \sin^2\theta_2)\(\(r_1^2+\ell_1^2\)\cos^2\theta_1+\(r_3^2+\ell_3^2\)\sin^2\theta_3\)}\,, \nn
\end{align}
where we remind that $(r,\theta)$ are the global spherical coordinates of the three-rod configuration \eqref{eq:DefDistanceglobal}, while $(r_i,\theta_i)$ are the local spherical coordinates centered at the $i^\text{th}$ rod,  given by \eqref{eq:DefDistance} and \eqref{eq:ri_thetaidef}.

Thus,  the M-theory solution \eqref{eq:MtheoryAnsatz2} is\footnote{One can also write the metric in the Weyl cylindrical coordinate system by replacing $$ \frac{dr^2}{r^2+\ell^2}+d\theta^2= \frac{4}{\left( r^2+\ell^2\cos^2\theta\right)\left( r^2+\ell^2\sin^2\theta\right)} \left(d\rho^2+dz^2\right) .$$}
\begin{align}
ds_{11}^2 \= & \frac{1}{\cZ^2} \,\left[  -\frac{(r^2+\ell^2)^2}{Q^2}\,dt^2+ \frac{Q \cF}{r^2+\ell^2}\, dr^2 \right] + Q\, \cZ \left[ \frac{\cF}{\cZ^3}\, d\theta^2+ \cos^2 \theta \,d\varphi_1^2 + \sin^2 \theta\, d\varphi_2^2 \right]  \nn \\
&+\frac{Q\, \cZ}{r^2+\ell^2}\,\sum_{I=1}^3\left( \frac{r_I^2 \,dy_{2I-1}^2+(r_I^2+\ell_I^2) \,dy_{2I}^2}{Q_I}\right), \nn \\
A_3 \= &-\sum_{I=1}^3 \frac{r_I^2}{Q_I} \,dt\wedge dy_{2I-1} \wedge dy_{2I}\,, \label{eq:met3AdS2Sph} 
\end{align}
where we have defined the deformation factors and the charge product
\begin{equation}
\begin{split}
Q&\equi (Q_1 Q_2 Q_3)^\frac{1}{3}\,,\quad \cZ \equi \frac{r^2+\ell^2}{\prod_{I=1}^3 (r_I^2+\ell_I^2)^\frac{1}{3}}\,,\\
 \cF &\equi \frac{(r^2+\ell^2)^3}{(r_2^2+\ell_2^2 \cos^2\theta_2)(r_2^2+\ell_2^2 \sin^2\theta_2)\(\(r_1^2+\ell_1^2\)\cos^2\theta_1+\(r_3^2+\ell_3^2\)\sin^2\theta_3\)}.
\end{split}
\label{eq:DefDefFactors}
\end{equation}

The solutions are asymptotic to AdS$_2\times$S$^3\times$T$^6$ as in \eqref{eq:AdS2Asymp} with $k=1$ since $\cZ$ and $\cF$ go to $1$ at large $r$. The rod sources have not only broken the rigidity of the T$^6$ but have also deformed the S$^3$ and AdS$_2$ spaces non-trivially.  

\subsection{Regularity and topology}

The rod sources are located at $\rho=0$ and $0\leq z \leq \ell^2/4$.  In the $(r,\theta)$ coordinate system \eqref{eq:DefDistanceglobal},  they are at $r=0$ and
\begin{center}
\begin{tabular}{c|c|c}
 $1^\text{st}$ rod: $~\theta_c^{(1)} \leq \theta \leq \frac{\pi}{2}\quad$ &
 $\quad2^\text{nd}$ rod: $~\theta_c^{(2)} \leq \theta \leq \theta_c^{(1)}\quad$ &
 $\quad3^\text{rd}$ rod: $~0 \leq \theta \leq \theta_c^{(2)}$,
\end{tabular} \\[2ex]
\end{center}
where we have defined the angles delimiting the rods as
\begin{equation}
\cos^2 \theta_c^{(1)} \equi \frac{\ell_1^2}{\ell^2}\,,\qquad \cos^2 \theta_c^{(2)} \equi \frac{\ell_1^2+\ell_2^2}{\ell^2}\,.
\label{eq:DefThetaCri}
\end{equation}
First,  at $r>0$,  one can check that $r_I>0$,  and $\cF$ and $\cZ$ are finite and positive.  Therefore,  all metric components \eqref{eq:met3AdS2Sph} are finite and the geometries are regular there for $\theta\neq 0,\pi/2$.  The loci $r>0$ and $\theta=0,\pi/2$ correspond to the two semi-infinite segments above and below the rod sources on the $z$-axis depicted in Fig.\ref{fig:rodsources}.  They define the North and South poles of the S$^3$ where $\varphi_2$ and $\varphi_1$ degenerate respectively.  One can check that $\cF/\cZ^3=1$ and the angles degenerate smoothly without conical singularities at the poles: $ds(S^3) \sim d\theta^2 + \cos^2 \theta d\varphi_1^2 +\sin^2 \theta d\varphi_2^2$.  The spacetime is therefore regular outside the rod sources at $r>0$ and have a S$^3\times$T$^6$ topology.

At the sources, $r=0$,  the $y_1$-circle degenerates at the first rod, $\theta_c^{(1)} \leq \theta \leq \pi/2$ where $r_1=0$ and $r_2,r_3>0$,  the $y_3$-circle degenerates at the second rod,  $\theta_c^{(2)} \leq \theta \leq \theta_c^{(1)}$ where $r_2=0$ and $r_1,r_3>0$,  and finally the $y_5$-circle shrinks at the third rod $0 \leq \theta \leq \theta_c^{(2)}$ where $r_3=0$ and $r_1,r_2>0$.  Thus,  the S$^3$ splits into three regions at $r=0$  such that a different T$^6$ direction shrinks there.  The local geometries are better described in terms of the local spherical coordinates,  $ i=1,2$ or $3$, 
\begin{equation}
\rho = \frac{r_i \sqrt{r_i^2+\ell_i^2}}{4}\,\sin 2\theta_i \,,\qquad z \= \frac{2r_i^2+\ell_i^2}{8} \,\cos 2\theta_i + \frac{1}{4} \sum_{j=1}^i \ell_j^2 - \frac{\ell_i^2}{8}\,, 
\label{eq:LocalSpher1}
\end{equation}
which implies
\begin{equation}
d\rho^2 +dz^2 \= \frac{\left( r_i^2+\ell_i^2\cos^2\theta_i\right)\left( r_i^2+\ell_i^2\sin^2\theta_i\right)}{4} \, \left(\frac{dr_i^2}{r_i^2+\ell_i^2}+ d\theta_i^2 \right)\,.
\label{eq:LocalSpher2}
\end{equation}
Therefore, at the first rod $r_1 \to 0$,  the time slices of the metric \eqref{eq:met3AdS2Sph} give
\begin{align}
ds_{11} |_{dt=0} &\,\propto \,   dr_1^2 +\frac{r_1^2}{Q_1} dy_1^2 +\ell_1^2 \left(d\theta_1^2 +\cos^2\theta_1\,d\varphi_1^2 +\frac{\sin^2\theta_1}{Q_2}\,dy_3^2 \right)\label{eq:Met1stRod} \\
&\hspace{0.5cm} +(\ell^2-\ell_1^2\cos^2\theta_1) \left( d\varphi_2^2+\frac{dy_6^2}{Q_3} \right) +\frac{\ell_1^2}{Q_1} dy_2^2 +(\ell_2^2+\ell_1^2\sin^2\theta_1) \left(\frac{dy_4^2}{Q_2}+\frac{dy_5^2}{Q_3} \right)\,.\nn
\end{align}
At the second rod,  $r_2 \to 0$,  we have
\begin{align}
ds_{11} |_{dt=0} &\,\propto \,   dr_2^2 +\frac{r_2^2}{Q_2} dy_3^2 +\ell_2^2 \left(d\theta_2^2 +\frac{\cos^2\theta_2}{Q_1}\,dy_1^2 +\frac{\sin^2\theta_2}{Q_3}\,dy_5^2 \right)\label{eq:Met2ndRod} \\
&\hspace{0.5cm} +(\ell_1^2+\ell_2^2\cos^2\theta_2) \left( d\varphi_1^2+\frac{dy_2^2}{Q_1} \right) +\frac{\ell_2^2}{Q_2} dy_4^2 +(\ell_3^2+\ell_2^2\sin^2\theta_2) \left(d\varphi_2^2+\frac{dy_6^2}{Q_3} \right)\,.\nn
\end{align}
And finally,  at the third rod, $r_3 \to 0$, 
\begin{align}
ds_{11} |_{dt=0} &\,\propto \,   dr_3^2 +\frac{r_3^2}{Q_1} dy_5^2 +\ell_3^2 \left(d\theta_3^2 +\frac{\cos^2\theta_3}{Q_2}\,dy_3^2 +\sin^2\theta_3\,d\varphi_2^2 \right)\label{eq:Met3rdRod} \\
&\hspace{0.5cm} +(\ell^2-\ell_3^2\sin^2\theta_3) \left( d\varphi_1^2+\frac{dy_2^2}{Q_1} \right) +\frac{\ell_3^2}{Q_3} dy_6^2 +(\ell_2^2+\ell_3^2\cos^2\theta_3) \left(\frac{dy_1^2}{Q_1}+\frac{dy_4^2}{Q_2} \right)\,.\nn
\end{align}
The geometries correspond to regular S$^3\times$T$^5$ fibrations over an origin of a $\IR^2$ space if
\begin{equation}
R_{y_1} \= \sqrt{Q_1}\,,\qquad R_{y_3} \= \sqrt{Q_2} \,,\qquad R_{y_5} \= \sqrt{Q_3} \,.
\label{eq:Regularity3rods}
\end{equation}

Moreover,  one can check that the four-form flux is regular at the rods such that the components along the shrinking directions vanish. Using  \eqref{eq:chargesRodGen}, we see that each bolt carries an M2 brane charge such that
\begin{equation}
\begin{split}
q_{\text{M2}\,\,1}^{(1)}  &\= Q_1 \= R_{y_1}^2 \,,\qquad q_{\text{M2}\,\,1}^{(2)}  \=q_{\text{M2}\,\,1}^{(3)}  \=0\,,\\
q_{\text{M2}\,\,2}^{(2)}  &\= Q_2 \= R_{y_3}^2 \,,\qquad q_{\text{M2}\,\,2}^{(1)}  \=q_{\text{M2}\,\,2}^{(3)}  \=0\,,\\
q_{\text{M2}\,\,3}^{(3)} & \= Q_3 \= R_{y_5}^2 \,,\qquad q_{\text{M2}\,\,3}^{(1)}  \=q_{\text{M2}\,\,3}^{(2)}  \=0\,.
\end{split}
\end{equation}
More precisely,  the bolt that force $y_1$ to shrink carries a charge of the M2 branes that are wrapping the first T$^2$,  and similarly for the other bolts with the second and third T$^2$.

\begin{figure}[ht]
\centering
\includegraphics[width=0.95 \columnwidth]{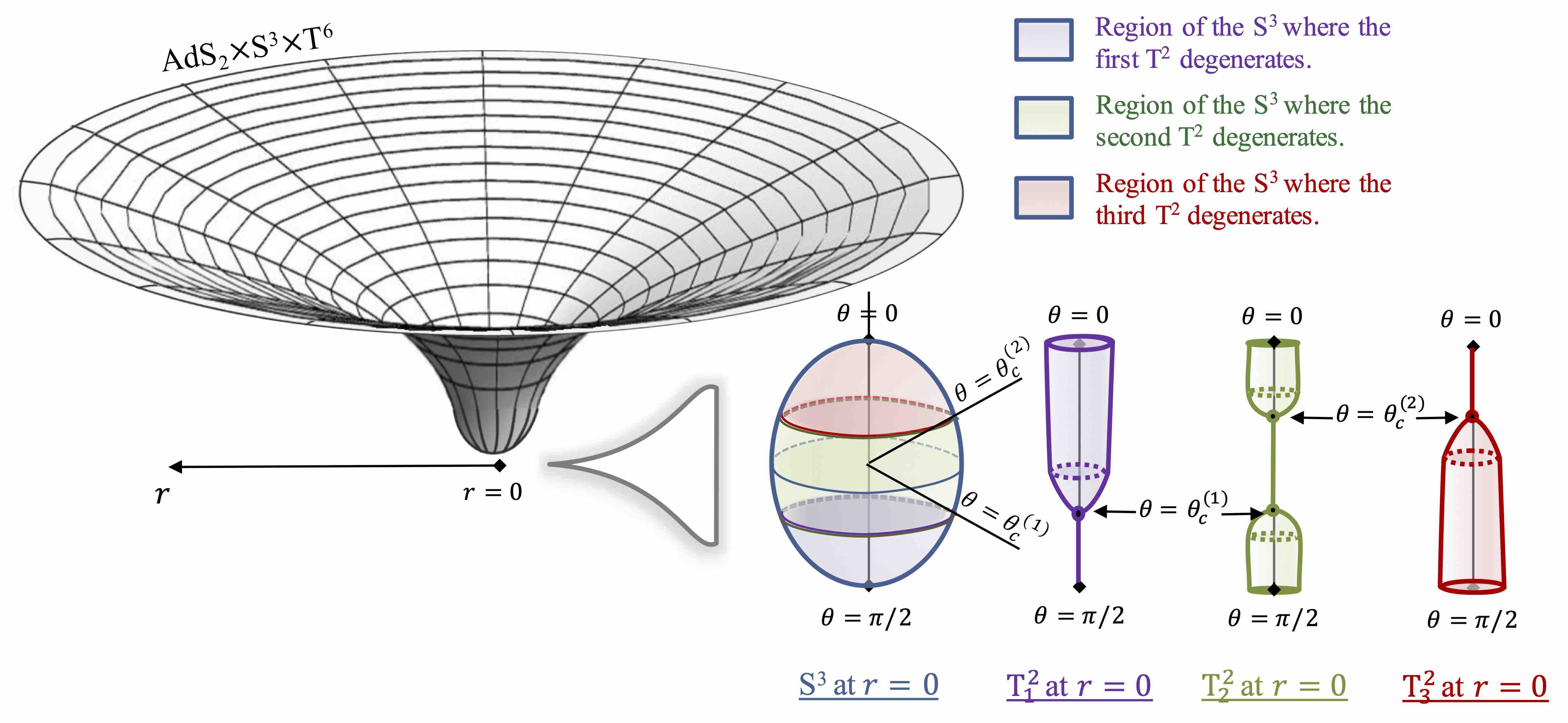}
\caption{Schematic description of the spacetime induced by three connected rods inducing the degeneracy of the $y_1$,  $y_3$ and $y_5$ circles.  On the left hand-side,  we depict the overall geometry in terms of the radius $r$.  On the right hand-side,  we describe the behavior of the S$^3$ and the three T$^2$ inside T$^6$ at $r=0$ and as a function of $\theta$,  the S$^3$ coordinate.  The spacetime ends smoothly at $r=0$ as a chain of three bolts where a T$^2$ direction degenerates.}
\label{fig:AdS2+3T6pic}
\end{figure}

We finally obtained regular geometries which are asymptotic to AdS$_2\times$S$^3\times$T$^6$.  The spacetime ends smoothly at $r=0$ as a chain of three bolts where a direction of each T$^2$ inside T$^6$ shrinks.  We have represented the profile of the smooth geometries,  the behavior of the S$^3$,  and each T$^2$ at the end of the spacetime in Fig.\ref{fig:AdS2+3T6pic}.  At this locus, the regions where the T$^2$ degenerate depend on the position on the S$^3$,  and are delimited by the critical angles \eqref{eq:DefThetaCri}.   Since all $\ell_i^2$ are free parameters,  one can make one or two regions very small relative to the other(s) by imposing a hierarchy of scales between the $\ell_i^2$.  The present solitonic excitations in AdS$_2$ have therefore a large variety of IR topology possible and have interesting limits that we will analyze in a moment.

\subsection{Supersymmetry breaking}
\label{sec:SUSYBreaking}

Supersymmetry requires the existence of a harmonic warp factor,  $H$ \eqref{eq:warp_harmonic}.  For the three-rods solution, one finds
\begin{equation}
\begin{aligned}
  H^2 ~=~ &\frac{(r_1 r_2 r_3)^2 (r^2+\ell^2)}{\qty[(r_1 r_2r_3)^2 (r^2 + \ell^2) - 4 \beta^2 Q^3 \rho^2]^2} \Bigg[16 \frac{(r_1 r_2 r_3)^2}{r^2}
  \\
  &- 4 Q^3  \qty((\alpha-\beta)^2 + \frac{16 \beta^2 \rho^2}{r^2(r^2+\ell^2)} + 4 (\alpha-\beta) \beta \cos^2\theta + 4 \beta^2 \cos^4\theta )\Bigg] \,.
\end{aligned}
\end{equation}
If we consider the illustrative example with $\alpha = \beta = 0$,  then $$ H = \frac{4}{r \sqrt{r^2 + \ell^2}} $$ is harmonic if and only if $\ell = 0$. This is the known BPS limit where the sizes of the rods shrink to zero size,  and one recovers the near-horizon limit of the three-charge BPS black-hole in M-theory,  that is a S$^3\times$T$^6$ fibration over a pure AdS$_2$ spacetime \cite{Heidmann:2021cms}.

More generally, using \eqref{eq:DefDistance} and \eqref{eq:DefDistanceglobal}, one can express $H$ as a function of $\rho$ and $z$, and compute its Laplacian. We find that when $\ell >0$, $H$ is not harmonic for any value of $\alpha, \beta$. This proves that the three-rod configuration with at least one rod of finite size does indeed break supersymmetry. 

\subsection{Interpretation in terms of moduli}

For the three-rod solution to be regular,  we have found that the supergravity charges are completely fixed in terms of the periodicities of the tori \eqref{eq:Regularity3rods},  which fix the solution at a certain region of the moduli space.  Here, we show that this result has a simple interpretation in terms of the quantized charges \eqref{eq:quantized_charges} and moduli space.

For simplicity, we assume that each of the two-torus have sides of the same length: $R_{y_{2I}} \= R_{y_{2I+1}},\ I=1,2,3$. Then,  the regularity constraint  \eqref{eq:Regularity3rods},  requires that
\begin{equation}
  N_1 ~=~ N_2 ~=~ N_3 ~=~ \frac{\cV_6}{(l_p)^6} \,,
  \label{eq:QuantChargCons3rods}
\end{equation} 
where the product of radii,  $\cV_6$,  is defined in \eqref{eq:RadiusProduct}.
In other words, \textit{the solutions exist when the numbers of M2 branes wrapping each torus are equal,  and are given by the volume of the T$^6$ in Planck unit}.  Moreover,  this constrains also the asymptotic volumes of the internal space \eqref{eq:AsymVol}. We find
\begin{equation}
 \mathrm{Vol}_{S^3}  \=  2\pi^2 \sqrt{\cV_6}\,,\qquad  \mathrm{Vol}_{T^2(1)}  \=  \mathrm{Vol}_{T^2(2)} \= \mathrm{Vol}_{T^2(3)}  \= (2\pi)^2 \,\cV_6^\frac{1}{3}.
\end{equation}
Thus,  the asymptotic volumes of the two-tori are also equal and fixed in terms of the T$^6$ size. 
 Note that these constraints are compatible with the regime where the supergravity solution gives a valid description of the brane bound state.  Indeed,  the supergravity picture requires the internal directions wrapped by the branes are much larger than the Planck scale to neglect quantum effects.  This requires $\cV_6 \gg (l_p)^6$,  or in other word that the periodicities of the torus direction to be much larger than the Planck length.  As a consequence,  the number of M2 branes \eqref{eq:QuantChargCons3rods} is large and can indeed be integer-valued.

These strong conditions on the quantized flux and the size of the internal spaces indicate that the solitons exist at a very specific point in moduli space.  The upside is that the lengths of the rods are not constrained at all.  Therefore, at this specific point, we have an infinite family of non-supersymmetric solitons with fixed AdS$_2\times$S$^3\times$T$^6$ in the UV but with an arbitrarily-changeable topology in IR.  However,  this is specific to solutions with only three rod sources: as we will see, more generic configurations will have fewer constraints on charges and moduli.

\subsection{Interesting limits}
\label{sec:InterestingLim}

A remarkable property of the present solutions is that they have three scale invariants associated to the lengths of each bolt,  $\ell_i^2$.  

First,  as previously argued,  turning off all $\ell_i^2$ corresponds to the supersymmetric limit where the solitons \eqref{eq:met3AdS2Sph} become a S$^3\times$T$^6$ fibration over an empty AdS$_2$ spacetime:
\begin{align}
ds_{11}^2 &\= - \frac{r^4\, dt^2}{Q^2} +Q\,\left[\frac{dr^2}{r^2} +d\theta^2+\cos^2 \theta\,d\varphi_1^2+\sin^2 \theta\,d\varphi_2^2 \right]+ Q\, \sum_{I=1}^3 \frac{dy_{2I-1}^2+dy_{2I}^2}{Q_I} , \nonumber\\
F_4 &\= \,-2 r dr \wedge dt \wedge \sum_{I=1}^3 \frac{dy_{2I-1}\wedge dy_{2I}}{Q_I}\,,\qquad Q\equi (Q_1 Q_2 Q_3)^{\frac{1}{3}}\,,
\end{align}
The $\ell_i^2$ define the non-supersymmetric backreaction of the solitons that deform smoothly AdS$_2$ in M-theory.  They ``cap off'' AdS$_2$ in the IR, at $r=0$,  as a chain of three bubbles where the T$^6$ smoothly degenerate.

However,  as soon as the $\ell_i^2$ are turned on,  one scale is nonphysical since it can be absorbed by a rescaling of $r$ and $t$.  Therefore,  the solutions with infinitesimal $\ell_i^2$ cannot be considered as small perturbations on AdS$_2$.  The solitons correspond to non-perturbative back-reaction on the geometry.

With one unphysical scale,   one has an infinite family of geometries given by two scale parameters at the specific point in the moduli space detailed in the precious section.  By imposing a hierarchy of scales in between the lengths $\ell_i^2$,  one can dial independently the size of the regions where the 2-tori degenerate.  This drastically modifies the geometries in the interior while keeping the asymptotic fixed.

\begin{figure}[t]
\centering
\includegraphics[width=0.95 \columnwidth]{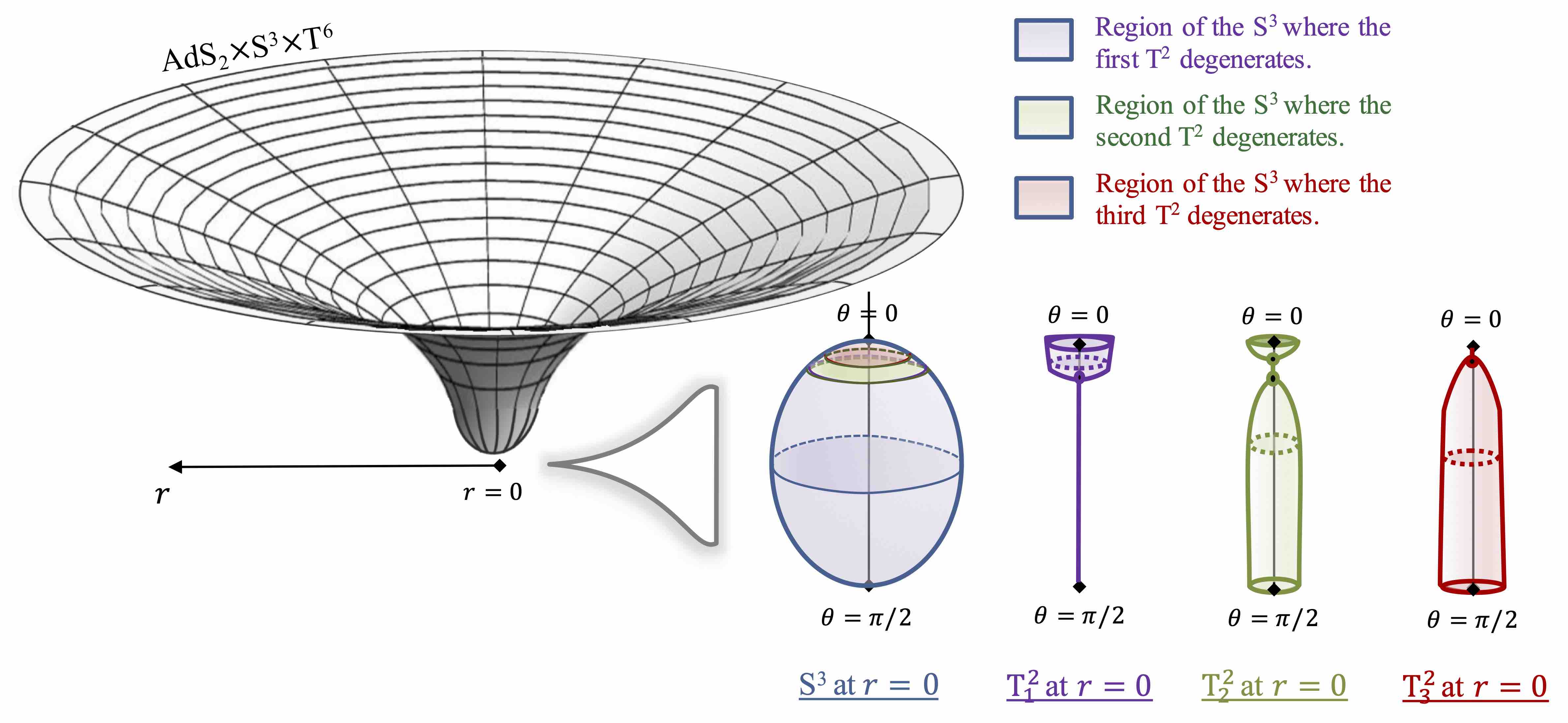}
\caption{Schematic description of the spacetime induced by three connected rods inducing the degeneracy of the $y_1$,  $y_3$ and $y_5$ circles when two rods are infinitesimal compared to the last one.  The two small rods are infinitesimally close to a point source that corresponds to singular BPS M2-brane sources.  The singularity is resolved by blowing up small non-BPS M2 bolts where the associated T$^2$ degenerates.}
\label{fig:AdS2+3T6Limpic}
\end{figure}

For instance,  we first assume that $\ell_2^2 ,\ell_3^2 \ll \ell_1^2 \sim \ell^2$.  In this regime,  the bolts where $y_3$ and $y_5$ degenerate are infinitesimally small compared to the bolt where $y_1$ degenerates.  In other words,  the regions where the second and third T$^2$ shrink are now localized at the South pole of the S$^3$ at $r=0$ (see Fig.\ref{fig:AdS2+3T6pic}),  and the S$^3$ is mostly spanned by the region where the first T$^2$ degenerates.  At leading order in the expansion $\ell_2^2 ,\ell_3^2 \to 0$,  the solutions are given by 
\begin{align}
ds_{11}^2 \= &Q \left(1+\frac{\ell^2 \cos^2 \theta}{r^2+\ell^2\sin^2\theta} \right)^\frac{2}{3}  \Biggl[-\frac{(r^2+\ell^2 \sin^2\theta )^2}{Q^3}\,dt^2 +\frac{dr^2}{r^2+\ell^2}+d\Omega_3^2 \\
&  +\frac{1}{Q_1} \left(\frac{r^2}{r^2+\ell^2}dy_1^2+dy_2^2 \right)  +\left(1+\frac{\ell^2 \cos^2 \theta}{r^2+\ell^2\sin^2\theta} \right)^{-1} \left[\frac{dy_3^2+dy_4^2}{Q_2} + \frac{dy_5^2+dy_6^2}{Q_3} \right] \Biggr] , \nn \\
A_3 \= &- \frac{r^2}{Q_1} \,dt\wedge dy_{1} \wedge dy_{2}- (r^2+\ell^2\sin^2\theta) \left( \frac{dt\wedge dy_{3} \wedge dy_{4}}{Q_2}+\frac{dt\wedge dy_{5} \wedge dy_{6}}{Q_3}\right)\,,\nn
\end{align}
where $d\Omega_3^2$ is the line element of a round three sphere.  The geometry is still regular for $r>0$ with a S$^3\times$T$^6$ topology, and $r=0$, $\theta\neq 0$ corresponds to a bolt where the $y_1$ circle smoothly degenerates.  However,  the loci of the second and third rods have degenerated to a point localized at $r^2+\ell^2\sin^2\theta=0$,  that is $r=0$ and $\theta=0$.  From the form of the gauge field,  the point source carries two M2-brane charges,  thereby corresponding to a singular locus of BPS M2 branes.\footnote{As argued in section \ref{sec:SUSYBreaking},  the sources saturate their BPS bound when they are point-like $\ell_i^2=0$.} This singularity is resolved by considering $\ell_2^2 , \ell_3^2 \ll \ell^2$ but non-zero.  The North pole of the S$^3$ at $r=0$ undergoes a geometric transition such that the singular BPS M2 branes become a chain of two infinitesimal and non-supersymmetric M2 bolts where a direction of the two 2-tori associated to the M2 branes degenerates smoothly (see Fig.\ref{fig:AdS2+3T6Limpic}).

We can obtain similar solutions by considering that a single rod is infinitely smaller than the others.  The geometries will be infinitesimally close to a solution where the small rod is replaced by a singular BPS M2-brane point source.  However,  at the vicinity of the point source,  the singularity is resolved by a small bolt that induces the smooth degeneracy of a T$^2$ direction.

\section{\texorpdfstring{Arbitrary T$^2$ and S$^3$ deformations in AdS$_2\times$S$^3\times$T$^6$}{Arbitrary T2 and S3 deformations in AdS2 x S3 x T6}}
\label{sec:BuAdS2T2s}

In this section,  we construct generic non-supersymmetric solitons in AdS$_2$ obtained from our solution-generating technique.  They are given by an arbitrary number of rod sources which deform the S$^3$ and T$^6$ in the IR such that some of their components degenerate smoothly as a chain of bolts.  More precisely,  we will first consider bubbling solutions where only the T$^6$ directions degenerate at the end-to-spacetime locus, $r=0$.  Then,  we will construct the most generic geometries where the Hopf angle of the S$^3$ can also degenerate.

For the three-rod solutions, there were no constraints on the internal parameters of the geometry,  while three constraints fixed the solutions at a specific point in the moduli space.  We will see that adding more rods and considering more generic geometries will release some constraints on the asymptotic quantities and constrain more the internal topology. 

\subsection{Generic \texorpdfstring{T$^2$}{T2} deformations}
\label{sec:ArbitraryT2s}

\subsubsection{Rod profile and weights}

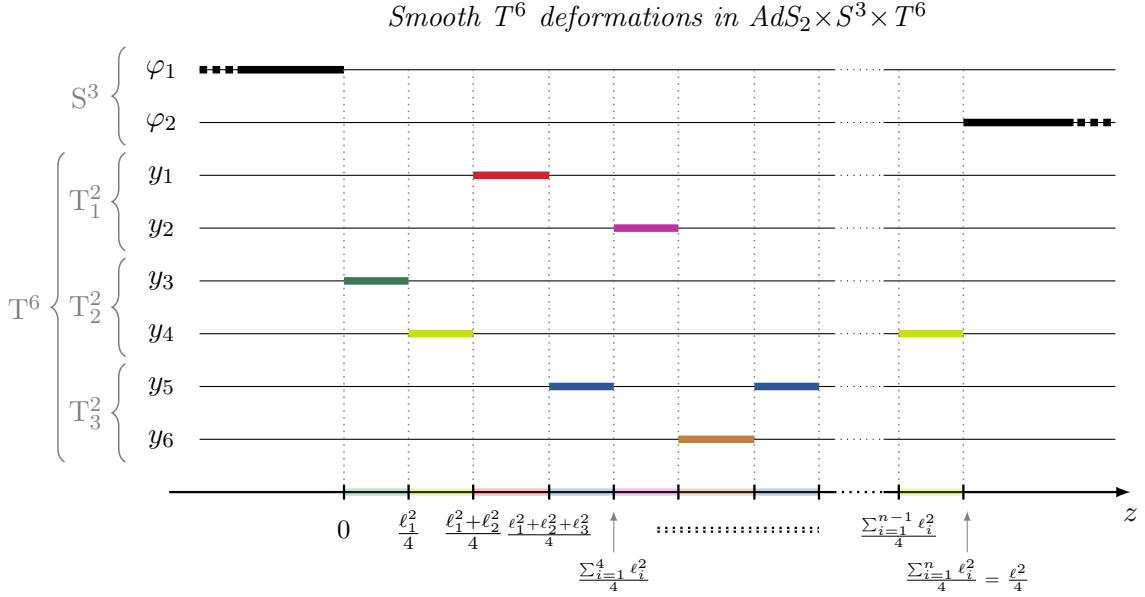
\begin{figure}[ht]
\centering
    \begin{tikzpicture}

\def\deb{-10} 
\def\inter{0.7} 
\def\ha{2.8} 
\def\zaxisline{9} 
\def\rodsize{1.7} 
\def\numrod{4.5} 

\def\fin{\deb+1+2*\rodsize+\numrod*\rodsize} 



\draw (\deb+0.5+\rodsize+0.5*\numrod*\rodsize,\ha) node{{{\it Smooth T$^{\,6}$ deformations in AdS$_2\times$S$^{\,3}\times$T$^{\,6}$}}}; 

\draw [decorate, 
    decoration = {brace,
        raise=5pt,
        amplitude=5pt},line width=0.2mm,gray] (\deb-0.8,\ha-2.5*\inter+0.05) --  (\deb-0.8,\ha-0.5*\inter-0.05);
\draw [decorate, 
    decoration = {brace,
        raise=5pt,
        amplitude=5pt},line width=0.2mm,gray] (\deb-0.8,\ha-4.5*\inter+0.05) --  (\deb-0.8,\ha-2.5*\inter-0.05);
\draw [decorate, 
    decoration = {brace,
        raise=5pt,
        amplitude=5pt},line width=0.2mm,gray] (\deb-0.8,\ha-6.5*\inter+0.05) --  (\deb-0.8,\ha-4.5*\inter-0.05);
\draw [decorate, 
    decoration = {brace,
        raise=5pt,
        amplitude=5pt},line width=0.2mm,gray] (\deb-0.8,\ha-8.5*\inter+0.05) --  (\deb-0.8,\ha-6.5*\inter-0.05);
\draw [decorate, 
    decoration = {brace,
        raise=5pt,
        amplitude=5pt},line width=0.2mm,gray] (\deb-1.6,\ha-8.5*\inter+0.05) --  (\deb-1.6,\ha-2.5*\inter-0.05);

\draw[gray] (\deb-1.5,\ha-1.5*\inter) node{S$^3$};
\draw[gray] (\deb-1.5,\ha-3.5*\inter) node{T$^2_1$};
\draw[gray] (\deb-1.5,\ha-5.5*\inter) node{T$^2_2$};
\draw[gray] (\deb-1.5,\ha-7.5*\inter) node{T$^2_3$};
\draw[gray] (\deb-2.3,\ha-5.5*\inter) node{T$^6$};

\draw (\deb-0.5,\ha-\inter) node{$\varphi_1$};
\draw (\deb-0.5,\ha-2*\inter) node{$\varphi_2$};
\draw (\deb-0.5,\ha-3*\inter) node{$y_1$};
\draw (\deb-0.5,\ha-4*\inter) node{$y_2$};
\draw (\deb-0.5,\ha-5*\inter) node{$y_3$};
\draw (\deb-0.5,\ha-6*\inter) node{$y_4$};
\draw (\deb-0.5,\ha-7*\inter) node{$y_5$};
\draw (\deb-0.5,\ha-8*\inter) node{$y_6$};

\draw (\fin+0.2,\ha-\zaxisline*\inter-0.3) node{$z$};


\draw[black,thin] (\deb,\ha-\inter) -- (\deb+0.5+4.5*\rodsize+0.2,\ha-\inter);\draw[black,thin,dotted] (\deb+0.5+4.5*\rodsize+0.2,\ha-\inter) -- (\deb+0.5+5*\rodsize,\ha-\inter);
\draw[black,thin] (\deb+0.5+5*\rodsize,\ha-\inter) -- (\fin,\ha-\inter);

\draw[black,thin] (\deb,\ha-2*\inter) -- (\deb+0.5+4.5*\rodsize+0.2,\ha-2*\inter);\draw[black,thin,dotted] (\deb+0.5+4.5*\rodsize+0.2,\ha-2*\inter) -- (\deb+0.5+5*\rodsize,\ha-2*\inter);
\draw[black,thin] (\deb+0.5+5*\rodsize,\ha-2*\inter) -- (\fin,\ha-2*\inter);

\draw[black,thin] (\deb,\ha-3*\inter) -- (\deb+0.5+4.5*\rodsize+0.2,\ha-3*\inter);\draw[black,thin,dotted] (\deb+0.5+4.5*\rodsize+0.2,\ha-3*\inter) -- (\deb+0.5+5*\rodsize,\ha-3*\inter);
\draw[black,thin] (\deb+0.5+5*\rodsize,\ha-3*\inter) -- (\fin,\ha-3*\inter);

\draw[black,thin] (\deb,\ha-4*\inter) -- (\deb+0.5+4.5*\rodsize+0.2,\ha-4*\inter);\draw[black,thin,dotted] (\deb+0.5+4.5*\rodsize+0.2,\ha-4*\inter) -- (\deb+0.5+5*\rodsize,\ha-4*\inter);
\draw[black,thin] (\deb+0.5+5*\rodsize,\ha-4*\inter) -- (\fin,\ha-4*\inter);

\draw[black,thin] (\deb,\ha-5*\inter) -- (\deb+0.5+4.5*\rodsize+0.2,\ha-5*\inter);\draw[black,thin,dotted] (\deb+0.5+4.5*\rodsize+0.2,\ha-5*\inter) -- (\deb+0.5+5*\rodsize,\ha-5*\inter);
\draw[black,thin] (\deb+0.5+5*\rodsize,\ha-5*\inter) -- (\fin,\ha-5*\inter);

\draw[black,thin] (\deb,\ha-6*\inter) -- (\deb+0.5+4.5*\rodsize+0.2,\ha-6*\inter);\draw[black,thin,dotted] (\deb+0.5+4.5*\rodsize+0.2,\ha-6*\inter) -- (\deb+0.5+5*\rodsize,\ha-6*\inter);
\draw[black,thin] (\deb+0.5+5*\rodsize,\ha-6*\inter) -- (\fin,\ha-6*\inter);

\draw[black,thin] (\deb,\ha-7*\inter) -- (\deb+0.5+4.5*\rodsize+0.2,\ha-7*\inter);\draw[black,thin,dotted] (\deb+0.5+4.5*\rodsize+0.2,\ha-7*\inter) -- (\deb+0.5+5*\rodsize,\ha-7*\inter);
\draw[black,thin] (\deb+0.5+5*\rodsize,\ha-7*\inter) -- (\fin,\ha-7*\inter);

\draw[black,thin] (\deb,\ha-8*\inter) -- (\deb+0.5+4.5*\rodsize+0.2,\ha-8*\inter);\draw[black,thin,dotted] (\deb+0.5+4.5*\rodsize+0.2,\ha-8*\inter) -- (\deb+0.5+5*\rodsize,\ha-8*\inter);
\draw[black,thin] (\deb+0.5+5*\rodsize,\ha-8*\inter) -- (\fin,\ha-8*\inter);

\draw[black,line width=0.3mm] (\deb-0.4,\ha-\zaxisline*\inter) -- (\deb+0.5+4.5*\rodsize+0.2,\ha-\zaxisline*\inter);\draw[black,line width=0.3mm,dotted] (\deb+0.5+4.5*\rodsize+0.2,\ha-\zaxisline*\inter) -- (\deb+0.5+5*\rodsize,\ha-\zaxisline*\inter);
\draw[black,->, line width=0.3mm] (\deb+0.5+5*\rodsize,\ha-\zaxisline*\inter) -- (\fin+0.2,\ha-\zaxisline*\inter);


\draw[black, dotted, line width=1mm] (\deb,\ha-\inter) -- (\deb+0.5,\ha-\inter);
\draw[black,line width=1mm] (\deb+0.5,\ha-\inter) -- (\deb+0.5+\rodsize-0.3,\ha-\inter);
\draw[black,line width=1mm] (\fin-0.5-\rodsize+0.2,\ha-2*\inter) -- (\fin-0.55,\ha-2*\inter);
\draw[black, dotted,line width=1mm] (\fin-0.5,\ha-2*\inter) -- (\fin,\ha-2*\inter);


\draw[amazon,line width=1mm] (\deb+0.5+\rodsize-0.3,\ha-5*\inter) -- (\deb+0.5+1.5*\rodsize-0.3,\ha-5*\inter);

\draw[bitterlemon,line width=1mm] (\deb+0.5+1.5*\rodsize-0.3,\ha-6*\inter) -- (\deb+0.5+2*\rodsize-0.3,\ha-6*\inter);

\draw[amaranthred,line width=1mm] (\deb+0.5+2*\rodsize-0.3,\ha-3*\inter) -- (\deb+0.5+2.5*\rodsize-0.15,\ha-3*\inter);

\draw[bdazzledblue,line width=1mm] (\deb+0.5+2.5*\rodsize-0.15,\ha-7*\inter) -- (\deb+0.5+3*\rodsize-0.15,\ha-7*\inter);

\draw[byzantine,line width=1mm] (\deb+0.5+3*\rodsize-0.15,\ha-4*\inter) -- (\deb+0.5+3.5*\rodsize-0.15,\ha-4*\inter);

\draw[brown,line width=1mm] (\deb+0.5+3.5*\rodsize-0.15,\ha-8*\inter) -- (\deb+0.5+4*\rodsize,\ha-8*\inter);

\draw[bdazzledblue,line width=1mm] (\deb+0.5+4*\rodsize,\ha-7*\inter) -- (\deb+0.5+4.5*\rodsize,\ha-7*\inter);

\draw[bitterlemon,line width=1mm] (\deb+0.5+5*\rodsize+0.2,\ha-6*\inter) -- (\deb+0.5+5.5*\rodsize+0.2,\ha-6*\inter);


\draw[amazon,line width=1mm,opacity=0.25] (\deb+0.5+\rodsize-0.3,\ha-\zaxisline*\inter) -- (\deb+0.5+1.5*\rodsize-0.3,\ha-\zaxisline*\inter);

\draw[bitterlemon,line width=1mm,opacity=0.25] (\deb+0.5+1.5*\rodsize-0.3,\ha-\zaxisline*\inter) -- (\deb+0.5+2*\rodsize-0.3,\ha-\zaxisline*\inter);
\draw[amaranthred,line width=1mm,opacity=0.25] (\deb+0.5+2*\rodsize-0.3,\ha-\zaxisline*\inter) -- (\deb+0.5+2.5*\rodsize-0.15,\ha-\zaxisline*\inter);

\draw[bdazzledblue,line width=1mm,opacity=0.25] (\deb+0.5+2.5*\rodsize-0.15,\ha-\zaxisline*\inter) -- (\deb+0.5+3*\rodsize-0.15,\ha-\zaxisline*\inter);

\draw[byzantine,line width=1mm,opacity=0.25] (\deb+0.5+3*\rodsize-0.15,\ha-\zaxisline*\inter) -- (\deb+0.5+3.5*\rodsize-0.15,\ha-\zaxisline*\inter);

\draw[brown,line width=1mm,opacity=0.25] (\deb+0.5+3.5*\rodsize-0.15,\ha-\zaxisline*\inter) -- (\deb+0.5+4*\rodsize,\ha-\zaxisline*\inter);
\draw[bdazzledblue,line width=1mm,opacity=0.25] (\deb+0.5+4*\rodsize,\ha-\zaxisline*\inter) -- (\deb+0.5+4.5*\rodsize,\ha-\zaxisline*\inter);

\draw[bitterlemon,line width=1mm,opacity=0.25] (\deb+0.5+5*\rodsize+0.2,\ha-\zaxisline*\inter) -- (\deb+0.5+5.5*\rodsize+0.2,\ha-\zaxisline*\inter);


\draw[gray,dotted,line width=0.2mm] (\deb+0.5+\rodsize-0.3,\ha-\inter) -- (\deb+0.5+\rodsize-0.3,\ha-\zaxisline*\inter);
\draw[gray,dotted,line width=0.2mm] (\deb+0.5+1.5*\rodsize-0.3,\ha-\inter) -- (\deb+0.5+1.5*\rodsize-0.3,\ha-\zaxisline*\inter);
\draw[gray,dotted,line width=0.2mm] (\deb+0.5+2*\rodsize-0.3,\ha-\inter) -- (\deb+0.5+2*\rodsize-0.3,\ha-\zaxisline*\inter);
\draw[gray,dotted,line width=0.2mm] (\deb+0.5+2.5*\rodsize-0.15,\ha-\inter) -- (\deb+0.5+2.5*\rodsize-0.15,\ha-\zaxisline*\inter);
\draw[gray,dotted,line width=0.2mm] (\deb+0.5+3*\rodsize-0.15,\ha-\inter) -- (\deb+0.5+3*\rodsize-0.15,\ha-\zaxisline*\inter);
\draw[gray,dotted,line width=0.2mm] (\deb+0.5+3.5*\rodsize-0.15,\ha-\inter) -- (\deb+0.5+3.5*\rodsize-0.15,\ha-\zaxisline*\inter);
\draw[gray,dotted,line width=0.2mm] (\deb+0.5+4*\rodsize,\ha-\inter) -- (\deb+0.5+4*\rodsize,\ha-\zaxisline*\inter);
\draw[gray,dotted,line width=0.2mm] (\deb+0.5+4.5*\rodsize,\ha-\inter) -- (\deb+0.5+4.5*\rodsize,\ha-\zaxisline*\inter);
\draw[gray,dotted,line width=0.2mm] (\deb+0.5+5*\rodsize+0.2,\ha-\inter) -- (\deb+0.5+5*\rodsize+0.2,\ha-\zaxisline*\inter);
\draw[gray,dotted,line width=0.2mm] (\deb+0.5+5.5*\rodsize+0.2,\ha-\inter) -- (\deb+0.5+5.5*\rodsize+0.2,\ha-\zaxisline*\inter);

\draw[line width=0.3mm] (\deb+0.5+\rodsize-0.3,\ha-\zaxisline*\inter+0.1) -- (\deb+0.5+\rodsize-0.3,\ha-\zaxisline*\inter-0.1);
\draw[line width=0.3mm] (\deb+0.5+1.5*\rodsize-0.3,\ha-\zaxisline*\inter+0.1) -- (\deb+0.5+1.5*\rodsize-0.3,\ha-\zaxisline*\inter-0.1);
\draw[line width=0.3mm] (\deb+0.5+2*\rodsize-0.3,\ha-\zaxisline*\inter+0.1) -- (\deb+0.5+2*\rodsize-0.3,\ha-\zaxisline*\inter-0.1);
\draw[line width=0.3mm] (\deb+0.5+2.5*\rodsize-0.15,\ha-\zaxisline*\inter+0.1) -- (\deb+0.5+2.5*\rodsize-0.15,\ha-\zaxisline*\inter-0.1);
\draw[line width=0.3mm] (\deb+0.5+3*\rodsize-0.15,\ha-\zaxisline*\inter+0.1) -- (\deb+0.5+3*\rodsize-0.15,\ha-\zaxisline*\inter-0.1);
\draw[line width=0.3mm] (\deb+0.5+3.5*\rodsize-0.15,\ha-\zaxisline*\inter+0.1) -- (\deb+0.5+3.5*\rodsize-0.15,\ha-\zaxisline*\inter-0.1);
\draw[line width=0.3mm] (\deb+0.5+4*\rodsize,\ha-\zaxisline*\inter+0.1) -- (\deb+0.5+4*\rodsize,\ha-\zaxisline*\inter-0.1);
\draw[line width=0.3mm] (\deb+0.5+4.5*\rodsize,\ha-\zaxisline*\inter+0.1) -- (\deb+0.5+4.5*\rodsize,\ha-\zaxisline*\inter-0.1);
\draw[line width=0.3mm] (\deb+0.5+5*\rodsize+0.2,\ha-\zaxisline*\inter+0.1) -- (\deb+0.5+5*\rodsize+0.2,\ha-\zaxisline*\inter-0.1);
\draw[line width=0.3mm] (\deb+0.5+5.5*\rodsize+0.2,\ha-\zaxisline*\inter+0.1) -- (\deb+0.5+5.5*\rodsize+0.2,\ha-\zaxisline*\inter-0.1);

\draw (\deb+0.5+1*\rodsize-0.3,\ha-\zaxisline*\inter-0.5) node{{\small $0$}};
\draw (\deb+0.5+1.5*\rodsize-0.3,\ha-\zaxisline*\inter-0.5) node{{\small $\frac{\ell_1^2}{4}$}};
\draw (\deb+0.5+2*\rodsize-0.3,\ha-\zaxisline*\inter-0.5) node{{\small $\frac{\ell_1^2+\ell_2^2}{4}$}};

\draw (\deb+0.5+2.5*\rodsize-0.15,\ha-\zaxisline*\inter-0.5) node{{\tiny $\frac{\ell_1^2+\ell_{2}^2+\ell_{3}^2}{4}$}};
\draw[gray,->,line width=0.1mm] (\deb+0.5+3*\rodsize-0.15,\ha-\zaxisline*\inter-0.8) -- (\deb+0.5+3*\rodsize-0.15,\ha-\zaxisline*\inter-0.25);
\draw (\deb+0.5+3*\rodsize-0.15,\ha-\zaxisline*\inter-1.1) node{{\tiny $\frac{\sum_{i=1}^4\ell_{i}^2}{4}$}};

\draw[black,line width=0.3mm,dotted,double] (\deb+0.5+3.25*\rodsize,\ha-\zaxisline*\inter-0.5) -- (\deb+0.5+4.5*\rodsize,\ha-\zaxisline*\inter-0.5);

\draw (\deb+0.5+5*\rodsize+0.2,\ha-\zaxisline*\inter-0.5) node{{\tiny $\frac{\sum_{i=1}^{n-1}\ell_{i}^2}{4}$}};
\draw[gray,->,line width=0.1mm] (\deb+0.5+5.5*\rodsize+0.2+0.05,\ha-\zaxisline*\inter-0.8) -- (\deb+0.5+5.5*\rodsize+0.2+0.05,\ha-\zaxisline*\inter-0.25);
\draw (\deb+0.5+5.5*\rodsize+0.2+0.05,\ha-\zaxisline*\inter-1.1) node{{\tiny $\frac{\sum_{i=1}^n\ell_{i}^2}{4}=\frac{\ell^2}{4}$}};

\end{tikzpicture}
\caption{Rod diagram of the shrinking directions on the $z$-axis after sourcing the solutions with $n$ connected rods that force the degeneracy of a T$^6$ direction.}
\label{fig:rodsourceAdS2+T6s}
\end{figure}

We consider a generic configurations of $n$ connected rod sources,  each one forcing the degeneracy of a T$^6$ direction.  We have depicted a generic configuration in Fig.\ref{fig:rodsourceAdS2+T6s} where we have summarized the conventions on the $z$-axis.

The eight weights at each rod,  $(P_i^{(\Lambda)},G_i^{(\Lambda)})$,  are fixed depending on which coordinate shrinks at the rod following the Table \ref{tab:internalBC}.  We partition the set of rod labels between $1$ and $n$ into six sets of labels,  $U_{y_a}$ with $a=1,\ldots,6$ such that
\begin{equation}
i \in U_{w} \quad \Leftrightarrow \quad \text{the }w\text{ direction shrinks smoothly at the }i^\text{th}\text{ rod.}
\label{eq:DefUx}
\end{equation}
For instance,  the example in Fig.\ref{fig:rodsourceAdS2+T6s} corresponds to
\begin{align}
U_{y_1} &\= \{3,\ldots\}\,, \qquad U_{y_2}=\{5,\ldots\} \,,\qquad U_{y_3}\=\{1,\ldots\} \,, \nn\\
 U_{y_4}& \=\{2,\ldots,n\} \,,\qquad U_{y_5}=\{4,7,\ldots\} \,,\qquad U_{y_6} = \{6,\ldots  \}\,.
\end{align}
The weights at the rods can then be read from Table \ref{tab:internalBC}. For instance,
\begin{equation}
i \in U_{y_4}  \quad \Rightarrow \quad P_i^{(0)} = P_i^{(2)} = -G_i^{(0)} = -G_i^{(2)} =  \frac{1}{2}\,,\quad P_i^{(1)} = P_i^{(3)} = G_i^{(1)} = G_i^{(3)} = 0. \nn
\end{equation}
Note that $\alpha_{ij}$,  the exponent in the base warp factor $e^{2\nu}$ \eqref{eq:AlphaDef},  takes simple values for regular rod sources \eqref{eq:AlphaSimple}.  Therefore,  we define the following exponent for the present configurations,  
\begin{equation}
\bar{\alpha}_{ij} = \begin{cases} 
\,\,1 \,,\qquad i,j \in U_{w}\,,\\
\,\,0\,,\qquad i\in U_w\,,\quad j\in U_{w'} \,,\quad w\neq w'.
\label{eq:DefAlphaBar}
\end{cases}
\end{equation}
We also consider that $k=1$ such that the S$^3$ has no conical defect asymptotically.

\subsubsection{The solutions}

We derive the fields from the linear branch of solutions \eqref{eq:LinearAdS2} and use the identities \eqref{eq:SimplRelations2} to simplify their form.  We find that the metric and gauge field in M-theory are generically given by\footnote{The metric in the Weyl cylindrical coordinate system is obtained by replacing $$ \frac{dr^2}{r^2+\ell^2}+d\theta^2= \frac{4}{\left( r^2+\ell^2\cos^2\theta\right)\left( r^2+\ell^2\sin^2\theta\right)} \left(d\rho^2+dz^2\right).$$}
\begin{align}
ds_{11}^2  \= & \frac{1}{\cZ^2} \,\left[  -\frac{(r^2+\ell^2)^2}{Q^2}\,dt^2+ \frac{Q \cZ^3\,\cF}{r^2+\ell^2}\, dr^2 \right] + Q\, \cZ \left[\cF\, d\theta^2+ \cos^2 \theta \,d\varphi_1^2 + \sin^2 \theta\, d\varphi_2^2 \right]  \nn \\
&+Q\, \cZ\,\sum_{I=1}^3\left( \frac{\cK_{y_{2I-1}} \,dy_{2I-1}^2+\cK_{y_{2I}}\, dy_{2I}^2}{Q_I\,\cZ_I}\right),   \label{eq:metAdS2+T6s}\\ 
A_3 \= &-\sum_{I=1}^3 \frac{\cK_{y_{2I-1}} \cK_{y_{2I}}\,(r^2+\ell^2)}{Q_I\,\cZ_I} \,dt\wedge dy_{2I-1} \wedge dy_{2I}\,, \nn 
\end{align}
where the deformation factors are given by
\begin{align}
\cK_{a} &\equi \prod_{i\in U_{y_a}} \frac{r_i^2}{r_i^2+\ell_i^2}\,,\qquad \cZ_I \equi \frac{r^2+\ell^2}{\sum_{i\in U_{y_{2I-1}}\cup U_{y_{2I}}} \ell_i^2 }\left(1-\prod_{i\in U_{y_{2I-1}}\cup U_{y_{2I}}} \left(1+\frac{\ell_i^2}{r_i^2} \right)^{-1}\right) \,,\nn\\
\cF &\equi \prod\limits_{\substack{i,j=1\\j>i}}^n\left(\frac{\left( \left(r_i^2+\ell_i^2 \right) \cos^2\theta_i +  \left(r_j^2+\ell_j^2\right) \sin^2\theta_j \right)\left(r_i^2 \cos^2\theta_i +  r_j^2 \sin^2\theta_j \right)}{\left( \left(r_i^2+\ell_i^2 \right) \cos^2\theta_i + r_j^2 \sin^2\theta_j \right)\left(r_i^2 \cos^2\theta_i +  \left(r_j^2+\ell_j^2\right) \sin^2\theta_j \right)} \right)^{\bar{\alpha}_{ij}-1} \,,\nn\\
\cZ& \equi \left(\cZ_1 \cZ_2 \cZ_3 \right)^\frac{1}{3}\,,\qquad Q \equi (Q_1 Q_2 Q_3)^\frac{1}{3}\,. \label{eq:DefDefFactorsGen}
\end{align}

One can check that,  the three-rod solution is obtained by using \eqref{eq:SimplRelations2} and by considering $n=3$,  $U_{y_1}=\{ 1\}$,  $U_{y_3}=\{ 2\}$ and $U_{y_5}=\{ 3\}$.

\subsubsection{Regularity and topology}
\label{sec:RegGen}

First,  at large distance $r\to \infty$,  all deformation factors converge to 1 and the metric is asymptotic to AdS$_2\times$S$^3\times$T$^6$ as in \eqref{eq:AdS2Asymp} with $k=1$.

At $r>0$ and $\theta \neq 0,\pi/2$,  the metric components are finite and non-zero so the solutions are regular there.  Moreover, at $r>0$ and $\theta=0$ or $\theta=\pi/2$,  the angle $\varphi_2$ or $\varphi_1$ degenerates respectively as the North and South poles of the S$^3$.  In the Weyl coordinate system $(\rho,z)$,  these loci correspond to the two semi-infinite segments above and below the rod sources as depicted in Fig.\ref{fig:rodsourceAdS2+T6s}.  One can check from \eqref{eq:DefDistance} and \eqref{eq:ri_thetaidef} that all $\theta_i=0$ when $\theta=0$ and all $\theta_i=\pi/2$ when $\theta=\pi/2$ for $r>0$. Thus, $\cF=1$ there,  and the metric of the S$^3$ is smooth,  such that $ds(S^3)^2 \sim d\theta^2 +\cos^2 \theta \,d\varphi_1^2+\sin^2 \theta \,d\varphi_2^2$ at its poles.

The rods are all located at $r=0$. The coordinate $\theta$ parametrizes the position on the rods. At a given angle $\theta$ on the $i$-th rod, the distance to the rod $r_i$ vanishes while the distance to all the other rods $r_j$ are non-zero.  More precisely, we have
\begin{equation}
\qty(r=0\,,\quad \theta_c^{(i)} < \theta < \theta_c^{(i-1)}) \quad \Leftrightarrow \quad (r_i = 0 \,,\quad r_j >0, \quad j\neq i)\,,
\label{eq:ThetaSection}
\end{equation}
where we have defined the critical angles
\begin{equation}
\cos^2 \theta_c^{(i)} \= \frac{1}{\ell^2} \sum_{j=1}^{i} \ell_j^2\,,\qquad \theta_c^{(n)}=0\,,\qquad \theta_c^{(0)}=\frac{\pi}{2}\,.
\label{eq:DefThetaCrii}
\end{equation}
Thus, we are moving along the chain of rod sources by varying $\theta$ from $0$ to $\pi/2$ at $r=0$.

We consider a segment such that $i\in U_{y_1}$.  Since $r=0$ and $\cK_{y_1}=0$,   one can check from \eqref{eq:metAdS2+T6s} that the $y_1$ coordinate degenerates.  To derive the local geometry at this segment,  one needs to rewrite the metric in the coordinate system $(r_i,\theta_i)$, take $r_i\to 0$, and expand the metric and fields.\footnote{This can be achieved by going first in the Weyl cylindrical coordinates $(\rho,z)$,  using \eqref{eq:DefDistance} and \eqref{eq:DefDistanceglobal}, and then by changing coordinates with  \eqref{eq:LocalSpher1} and \eqref{eq:LocalSpher2}.} We refer the interested reader to previous work \cite{Bah:2020pdz,Bah:2021owp,Heidmann:2021cms} for more details about this derivation.  We find that the time slices of the metric \eqref{eq:metAdS2+T6s} converge towards 
\begin{equation}
ds_{11}^2|_{dt=0} \propto dr_i^2 + \frac{r_i^2}{C_i^2}\,dy_1^2 + ds(\cC^{(i)}_\text{Bubble})^2\,,
\label{eq:reganalysis1}
\end{equation}
with\footnote{By convention, $\prod_{i=a}^b\ldots =1$ if $a>b$. \label{footnote2}}
\begin{equation}
\begin{split}
C_i^2 \=&\, \frac{Q_1 \,\ell_i^2}{\sum_{p\in U_{y_1}\cup U_{y_2}} \ell_p^2}  \,\prod_{p=1}^{i-1} \prod_{q=i+1}^n \left[\frac{1+ \frac{\ell_q^2}{\sum_{k=p}^{q-1} \ell_k^2}}{1+ \frac{\ell_q^2}{\sum_{k=p+1}^{q-1} \ell_k^2}} \right]^{\bar{\alpha}_{pq}}\\
&\times \prod_{p=1}^{i-1}\left(1+ \frac{\ell_p^2}{\sum_{k=p+1}^{i} \ell_k^2} \right)^{\bar{\alpha}_{ip}} \, \prod_{p=i+1}^{n}\left(1+ \frac{\ell_p^2}{\sum_{k=i}^{p-1} \ell_k^2} \right)^{\bar{\alpha}_{ip}}\,.
\end{split}
\label{eq:reganalysis2}
\end{equation}
The $(r_i,y_1)$ subspace describes a smooth origin of a $\IR^2$,  i.e.  a bolt,  if we impose
\begin{equation}
R_{y_1}=C_i.
\label{eq:reganalysis3}
\end{equation}

The line element,  $ds(\cC^{(i)}_\text{Bubble})$,  describes the topology of the bubble at the bolt.  As discussed in \cite{Heidmann:2021cms},  it can be either a S$^3\times$T$^5$ or a S$^2\times$T$^6$ depending on the near environment of the rod.  More precisely,  if the adjacent rods are of the same category, for example they correspond to the degeneracy of the $y_2$ coordinate,  then the topology is S$^2\times$T$^6$ where the S$^2$ and T$^6$ are respectively described by the coordinates $(\theta_i,y_2)$ and $(\varphi_1,\varphi_2,y_3,y_4,y_5,y_6)$ respectively.  If they are of different nature,  for example they correspond to the degeneracy of the $y_2$ and $y_3$ coordinates, then the topology is S$^3\times$T$^5$ where the S$^3$ and T$^5$ are respectively described by the coordinates $(\theta_i,y_2,y_3)$ and $(\varphi_1,\varphi_2,y_4,x_5,y_6)$.  The rod endpoints,  $\theta_i=0$ or $\pi/2$,  correspond to the poles of either the S$^2$ or the S$^3$.  The regularity at these poles is guaranteed by the regularity of the adjacent rods.

Moreover,  one can show that the field strength,  $F_4$,  is regular such that the component along $y_1$ vanishes.  Moreover, the rod carries a M2 charge given by \eqref{eq:chargesRodGen}
\begin{equation}
q_{M2\,\,i}^{(1)} \= \frac{\ell_i^2}{\sum_{j\in U_{y_1}\cup U_{y_2}}\ell_j^2}\,Q_1\,,\qquad q_{M2\,\,i}^{(2)} \= q_{M2\,\,i}^{(3)} \= 0\,.
\label{eq:localcharges}
\end{equation}
and the local numbers of M2 branes associated to these are given in\eqref{eq:local_quantized_charges}.

The analysis is identical if one considers another T$^6$ circle that is shrinking at the rod.  For the $i^\text{th}$ rod such that $i \in U_{y_{2I}}$ or $U_{y_{2I-1}}$ with $I=1,2,3$,  the expressions of the local metric and regularity constraint \eqref{eq:reganalysis1},  \eqref{eq:reganalysis2} and \eqref{eq:reganalysis3} are simply modified by replacing $y_1 \to y_{2I}$ or $y_{2I-1}$,  $Q_1 \to Q_I$ and $U_{y_1}\cup U_{y_2} \to U_{y_{2I-1}}\cup U_{y_{2I}}$.

\vspace{1ex}

To summarize,  $r=0$ corresponds to a smooth locus where the spacetime ends as a chain of $n$ bolts.  Each bolt makes one of the T$^6$ coordinates degenerate smoothly provided that $n$ algebraic equations are satisfied
\begin{equation}
\begin{split}
R_{y_{2I-1}} &\= \frac{\ell_i\,d_i\,\sqrt{Q_I}}{\sqrt{\sum_{p\in U_{y_{2I-1}}\cup U_{y_{2I}}} \ell_p^2}} \= \sqrt{q_{M2\,\,i}^{(I)} }\,d_i\,,\qquad \text{if }i\in U_{2I-1}\,,\quad I=1,2,3\,,\\
R_{y_{2I}} &\= \frac{\ell_i\,d_i\,\sqrt{Q_I}}{\sqrt{\sum_{p\in U_{y_{2I-1}}\cup U_{y_{2I}}} \ell_p^2}} \= \sqrt{q_{M2\,\,i}^{(I)} }\,d_i\,,\qquad \text{if }i\in U_{2I}\,,\quad I=1,2,3\,,
\end{split}
\label{eq:BuEqT6s}
\end{equation}
where we have defined the aspect ratios
\begin{equation}
d_i \equi \prod_{p=1}^{i-1} \prod_{q=i+1}^n \left[\frac{1+ \frac{\ell_q^2}{\sum_{k=p}^{q-1} \ell_k^2}}{1+ \frac{\ell_q^2}{\sum_{k=p+1}^{q-1} \ell_k^2}} \right]^{\frac{\bar{\alpha}_{pq}}{2}} \,\prod_{p=1}^{i-1}\left(1+ \frac{\ell_p^2}{\sum_{k=p+1}^{i} \ell_k^2} \right)^{\frac{\bar{\alpha}_{ip}}{2}} \, \prod_{p=i+1}^{n}\left(1+ \frac{\ell_p^2}{\sum_{k=i}^{p-1} \ell_k^2} \right)^{\frac{\bar{\alpha}_{ip}}{2}}\,. \label{eq:DefdiAspect}
\end{equation}
\begin{figure}[ht]
\centering
\includegraphics[width=0.95 \columnwidth]{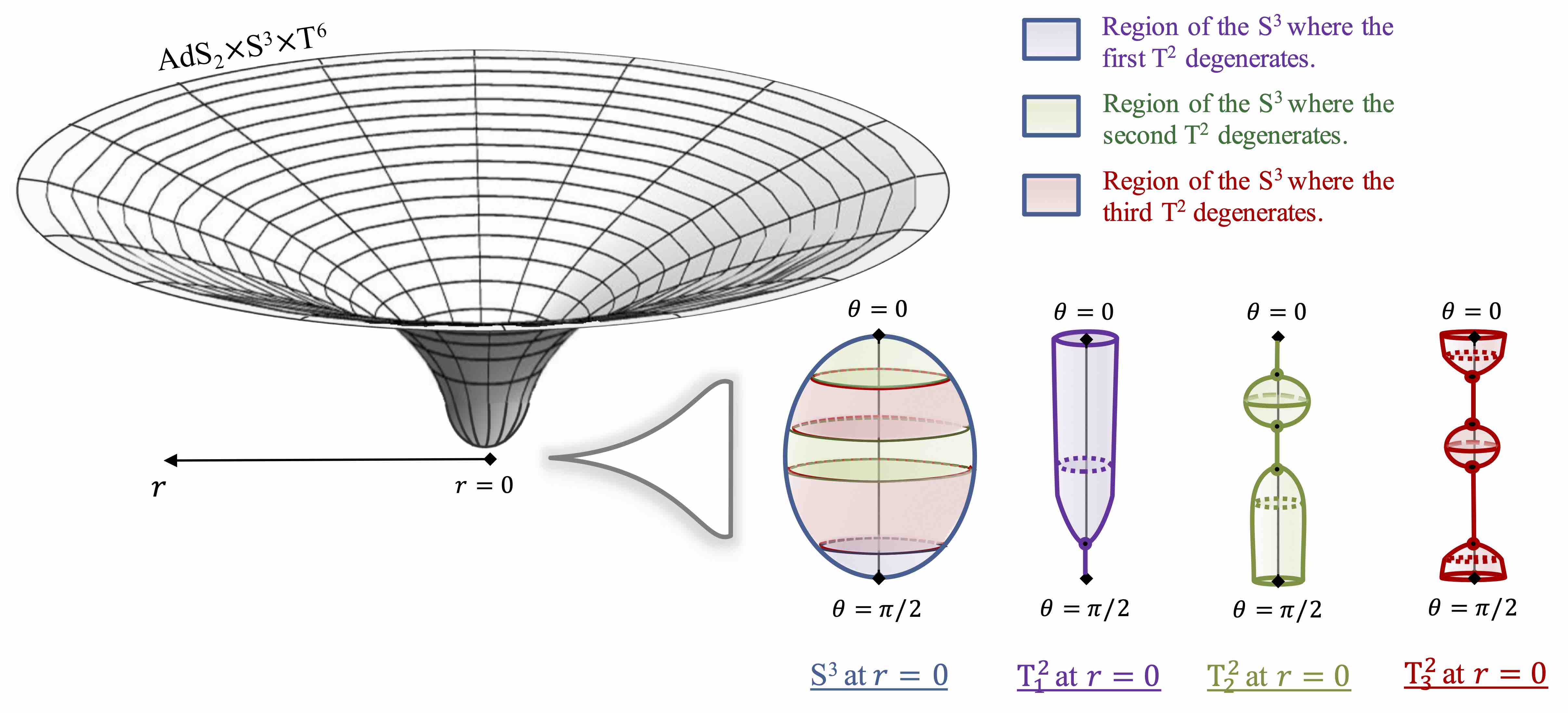}
\caption{Schematic description of the spacetime induced by an arbitrary number of connected rods inducing the degeneracy of the T$^6$ coordinates.  On the left hand-side,  we depict the overall geometry in terms of the radius $r$.  On the right hand-side,  we describe the behavior of the S$^3$ and the three T$^2$ inside T$^6$ at $r=0$ and as a function of $\theta$,  the S$^3$ coordinate.  The spacetime ends smoothly at $r=0$ as a chain of $n$ smooth bolts.}
\label{fig:AdS2+T6spic}
\end{figure}

The bolt induces a compact bubble which is localized on a specific region of the S$^3$ at $r=0$,  given by the critical angles, $\theta_c^{(i)}$.  We thus find that  the solutions correspond to \textit{asymptotically-AdS$_2$ smooth bubbling geometries} without horizons.  Using the same argument as in section \ref{sec:SUSYBreaking},  we can argue that they are non-supersymmetric smooth T$^6$ deformations of AdS$_2$ in M-theory.  We have depicted the profile of the geometries in Fig.\ref{fig:AdS2+T6spic} using the same convention as the previous example.

\subsection{Analysis of the bubble equations}

Unfortunately, the regularity constraints \eqref{eq:BuEqT6s} are not analytically solvable for generic rod configurations. They only have solutions for a small number of rods $n$ and approximations can be derived at large $n$.  However, some generic properties can be discussed and illustrative examples derived.

First, notice that the bubble equations are invariant under a global scaling of rod lengths $\ell_i^2 \to \lambda \ell_i^2$.  This is due to the conformal invariance of AdS$_2$ geometries \cite{Bena:2018bbd}.  As a consequence, one rod length can be considered as a free parameter and only the other $n-1$ can possibly be constrained.  Since there are $n$ equations, at least one total M2-brane charge is fixed in terms of the torus radii: a given solution cannot exist for arbitrary asymptotic quantities and at any point in the moduli space.  However, the constraint on the asymptotic quantities depends strongly on the rod configuration and in particular on the number of rods, $n$, which can be taken as a parameter.  Therefore, while the geometries associated to any given rod configuration can only be smooth in a small subset of the moduli space of asymptotic quantities, we will argue that the set of solutions \textit{including all possible rod configurations} certainly covers the whole moduli space.


Second, consider some simple configurations, for which all the rods are of different nature. Such configurations have necessarily less rods than the total number of rod species, six. One can see using \eqref{eq:AlphaSimple} that $\bar{\alpha}_{pq}=0$ for all $p,q=1,\ldots,n$, and this trivializes  the aspect ratios $d_i$ involved in the regularity constraints \eqref{eq:BuEqT6s}: $d_i=1$. The bubble equations of these specific configurations involve only the local M2-brane charges at the rod, \eqref{eq:localcharges}, that depend only on the rods where the associated T$^2$ shrinks. We thus have three decoupled channels of constraints.  More precisely, the regularity constraints are invariant under three rescaling parameters $\lambda_I$, $I=1,2,3$,  given by $\ell_i^2 \to \lambda_I \ell_i^2$ for $i \in U_{y_{2I-1}} \cup U_{2I}$. 
As a result, we do not have $n-1$ independent length parameters but $n-3$, and the three total charges are fixed in terms of torus radii accordingly.  The three-rod solution of the previous section was an example of such a geometry and we will discuss the six-rod generalization in a moment.

As soon as we consider configurations that have several rods of the same kind, some of the aspect ratios $d_i$ are non-trivial and induce couplings between the three T$^2$ channels.  These couplings will break the scale invariants associated with the different T$^2$.  Thus, generic solutions with a large number of rods of the same type end up having a single scale invariant, and a single total M2 charge is fixed.

To illustrate these properties,  we derive illustrative examples:
\begin{itemize}
\item[•] We construct the six-rod solutions where all rods are of different types so that the solutions have three scale invariants associated with each T$^2$.  The three total M2-brane charges are all fixed in terms of the radii of the T$^6$ circles, similarly to the three-rod solutions of section \ref{sec:BuAdS2T2}.
\item[•] We consider a four-rod configuration where two rods force the same circle to shrink.  This breaks a scale invariance and releases a constraint on a  total charge.
\item[•] We go further by considering five rods where two pairs are of the same type.  These solutions are more generic in the sense that they have a single scale invariant, as expected for asymptotically-AdS$_2$ geometries. One of the total charges is fixed in terms of the radii of the T$^6$ circles.  We show that this constraint implies that at least one quantized M2-brane charge is  given by the size of the T$^6$ as in \eqref{eq:QuantChargCons3rods}.  
\item[•] We show that one can construct solutions where the three quantized M2-brane charges are much larger than the size of the T$^6$ using configurations with a large number of rods.  We consider a configurations made of a specific pattern repeated many times; these configurations are simple enough that the bubble equations can be solved approximately at large $n$.  We find that the quantized charges scale as $n^\frac{5}{6}\, \cV_6 / (l_p)^6$ for this configuration.  This suggests that for arbitrary numbers of  M2 branes and for an arbitrary size of T$^6$,  an asymptotically-AdS$_2$ smooth geometry can always be found by stacking appropriately rod sources in the spacetime.
\end{itemize}

\subsection{Chain of six different bolts}

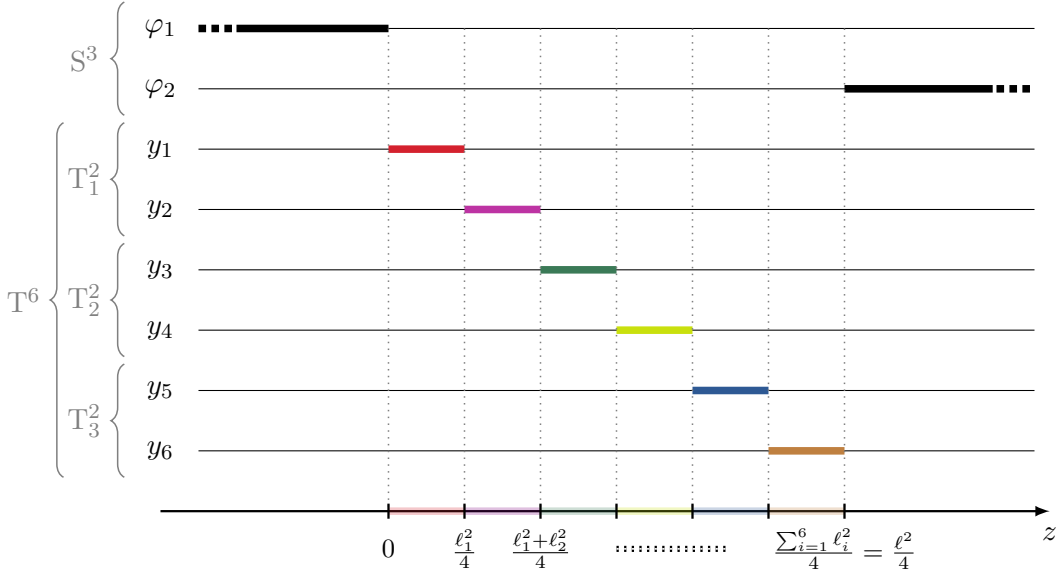
\begin{figure}[ht]
\centering
    \begin{tikzpicture}

\def\deb{-10} 
\def\inter{0.8} 
\def\ha{2.8} 
\def\zaxisline{8} 
\def\rodsize{2} 
\def\numrod{3} 

\def\fin{\deb+1+2*\rodsize+\numrod*\rodsize} 




\draw[black,thin] (\deb+1,\ha) -- (\fin,\ha);
\draw[black,thin] (\deb,\ha-\inter) -- (\fin-1,\ha-\inter);
\draw[black,thin] (\deb,\ha-2*\inter) -- (\fin,\ha-2*\inter);
\draw[black,thin] (\deb,\ha-3*\inter) -- (\fin,\ha-3*\inter);
\draw[black,thin] (\deb,\ha-4*\inter) -- (\fin,\ha-4*\inter);
\draw[black,thin] (\deb,\ha-5*\inter) -- (\fin,\ha-5*\inter);
\draw[black,thin] (\deb,\ha-6*\inter) -- (\fin,\ha-6*\inter);
\draw[black,thin] (\deb,\ha-7*\inter) -- (\fin,\ha-7*\inter);
\draw[black,->, line width=0.3mm] (\deb-0.5,\ha-\zaxisline*\inter) -- (\fin+0.2,\ha-\zaxisline*\inter);

\draw [decorate, 
    decoration = {brace,
        raise=5pt,
        amplitude=5pt},line width=0.2mm,gray] (\deb-0.8,\ha-1.5*\inter+0.05) --  (\deb-0.8,\ha+0.5*\inter-0.05);
\draw [decorate, 
    decoration = {brace,
        raise=5pt,
        amplitude=5pt},line width=0.2mm,gray] (\deb-0.8,\ha-3.5*\inter+0.05) --  (\deb-0.8,\ha-1.5*\inter-0.05);
\draw [decorate, 
    decoration = {brace,
        raise=5pt,
        amplitude=5pt},line width=0.2mm,gray] (\deb-0.8,\ha-5.5*\inter+0.05) --  (\deb-0.8,\ha-3.5*\inter-0.05);
\draw [decorate, 
    decoration = {brace,
        raise=5pt,
        amplitude=5pt},line width=0.2mm,gray] (\deb-0.8,\ha-7.5*\inter+0.05) --  (\deb-0.8,\ha-5.5*\inter-0.05);
\draw [decorate, 
    decoration = {brace,
        raise=5pt,
        amplitude=5pt},line width=0.2mm,gray] (\deb-1.6,\ha-7.5*\inter+0.05) --  (\deb-1.6,\ha-1.5*\inter-0.05);

\draw[gray] (\deb-1.5,\ha-0.5*\inter) node{S$^3$};
\draw[gray] (\deb-1.5,\ha-2.5*\inter) node{T$^2_1$};
\draw[gray] (\deb-1.5,\ha-4.5*\inter) node{T$^2_2$};
\draw[gray] (\deb-1.5,\ha-6.5*\inter) node{T$^2_3$};
\draw[gray] (\deb-2.3,\ha-4.5*\inter) node{T$^6$};

\draw (\deb-0.5,\ha) node{$\varphi_1$};
\draw (\deb-0.5,\ha-\inter) node{$\varphi_2$};
\draw (\deb-0.5,\ha-2*\inter) node{$y_1$};
\draw (\deb-0.5,\ha-3*\inter) node{$y_2$};
\draw (\deb-0.5,\ha-4*\inter) node{$y_3$};
\draw (\deb-0.5,\ha-5*\inter) node{$y_4$};
\draw (\deb-0.5,\ha-6*\inter) node{$y_5$};
\draw (\deb-0.5,\ha-7*\inter) node{$y_6$};

\draw (\fin+0.2,\ha-\zaxisline*\inter-0.3) node{$z$};


\draw[black, dotted, line width=1mm] (\deb,\ha) -- (\deb+0.5,\ha);
\draw[black,line width=1mm] (\deb+0.5,\ha) -- (\deb+0.5+\rodsize,\ha);
\draw[black,line width=1mm] (\fin-0.5-\rodsize,\ha-\inter) -- (\fin-0.55,\ha-\inter);
\draw[black, dotted,line width=1mm] (\fin-0.5,\ha-\inter) -- (\fin,\ha-\inter);


\draw[amaranthred,line width=1mm] (\deb+0.5+\rodsize,\ha-2*\inter) -- (\deb+0.5+1.5*\rodsize,\ha-2*\inter);
\draw[byzantine,line width=1mm] (\deb+0.5+1.5*\rodsize,\ha-3*\inter) -- (\deb+0.5+2*\rodsize,\ha-3*\inter);
\draw[amazon,line width=1mm] (\deb+0.5+2*\rodsize,\ha-4*\inter) -- (\deb+0.5+2.5*\rodsize,\ha-4*\inter);
\draw[bitterlemon,line width=1mm] (\deb+0.5+2.5*\rodsize,\ha-5*\inter) -- (\deb+0.5+3*\rodsize,\ha-5*\inter);
\draw[bdazzledblue,line width=1mm] (\deb+0.5+3*\rodsize,\ha-6*\inter) -- (\deb+0.5+3.5*\rodsize,\ha-6*\inter);
\draw[brown,line width=1mm] (\deb+0.5+3.5*\rodsize,\ha-7*\inter) -- (\deb+0.5+4*\rodsize,\ha-7*\inter);

\draw[amaranthred,line width=1mm,opacity=0.25] (\deb+0.5+\rodsize,\ha-\zaxisline*\inter) -- (\deb+0.5+1.5*\rodsize,\ha-\zaxisline*\inter);
\draw[purple,line width=1mm,opacity=0.25] (\deb+0.5+1.5*\rodsize,\ha-\zaxisline*\inter) -- (\deb+0.5+2*\rodsize,\ha-\zaxisline*\inter);
\draw[amazon,line width=1mm,opacity=0.25] (\deb+0.5+2*\rodsize,\ha-\zaxisline*\inter) -- (\deb+0.5+2.5*\rodsize,\ha-\zaxisline*\inter);
\draw[bitterlemon,line width=1mm,opacity=0.25] (\deb+0.5+2.5*\rodsize,\ha-\zaxisline*\inter) -- (\deb+0.5+3*\rodsize,\ha-\zaxisline*\inter);
\draw[bdazzledblue,line width=1mm,opacity=0.25] (\deb+0.5+3*\rodsize,\ha-\zaxisline*\inter) -- (\deb+0.5+3.5*\rodsize,\ha-\zaxisline*\inter);
\draw[brown,line width=1mm,opacity=0.25] (\deb+0.5+3.5*\rodsize,\ha-\zaxisline*\inter) -- (\deb+0.5+4*\rodsize,\ha-\zaxisline*\inter);


\draw[gray,dotted,line width=0.2mm] (\deb+0.5+\rodsize,\ha) -- (\deb+0.5+\rodsize,\ha-\zaxisline*\inter);
\draw[gray,dotted,line width=0.2mm] (\deb+0.5+1.5*\rodsize,\ha) -- (\deb+0.5+1.5*\rodsize,\ha-\zaxisline*\inter);
\draw[gray,dotted,line width=0.2mm] (\deb+0.5+2*\rodsize,\ha) -- (\deb+0.5+2*\rodsize,\ha-\zaxisline*\inter);
\draw[gray,dotted,line width=0.2mm] (\deb+0.5+2.5*\rodsize,\ha) -- (\deb+0.5+2.5*\rodsize,\ha-\zaxisline*\inter);
\draw[gray,dotted,line width=0.2mm] (\deb+0.5+3*\rodsize,\ha) -- (\deb+0.5+3*\rodsize,\ha-\zaxisline*\inter);
\draw[gray,dotted,line width=0.2mm] (\deb+0.5+3.5*\rodsize,\ha) -- (\deb+0.5+3.5*\rodsize,\ha-\zaxisline*\inter);
\draw[gray,dotted,line width=0.2mm] (\deb+0.5+4*\rodsize,\ha) -- (\deb+0.5+4*\rodsize,\ha-\zaxisline*\inter);

\draw[line width=0.3mm] (\deb+0.5+1.5*\rodsize,\ha-\zaxisline*\inter+0.1) -- (\deb+0.5+1.5*\rodsize,\ha-\zaxisline*\inter-0.1);
\draw[line width=0.3mm] (\deb+0.5+\rodsize,\ha-\zaxisline*\inter+0.1) -- (\deb+0.5+\rodsize,\ha-\zaxisline*\inter-0.1);
\draw[line width=0.3mm] (\deb+0.5+2*\rodsize,\ha-\zaxisline*\inter+0.1) -- (\deb+0.5+2*\rodsize,\ha-\zaxisline*\inter-0.1);
\draw[line width=0.3mm] (\deb+0.5+2.5*\rodsize,\ha-\zaxisline*\inter+0.1) -- (\deb+0.5+2.5*\rodsize,\ha-\zaxisline*\inter-0.1);
\draw[line width=0.3mm] (\deb+0.5+3*\rodsize,\ha-\zaxisline*\inter+0.1) -- (\deb+0.5+3*\rodsize,\ha-\zaxisline*\inter-0.1);
\draw[line width=0.3mm] (\deb+0.5+3.5*\rodsize,\ha-\zaxisline*\inter+0.1) -- (\deb+0.5+3.5*\rodsize,\ha-\zaxisline*\inter-0.1);
\draw[line width=0.3mm] (\deb+0.5+4*\rodsize,\ha-\zaxisline*\inter+0.1) -- (\deb+0.5+4*\rodsize,\ha-\zaxisline*\inter-0.1);

\draw (\deb+0.5+\rodsize,\ha-\zaxisline*\inter-0.5) node{{\small $0$}};
\draw (\deb+0.5+1.5*\rodsize,\ha-\zaxisline*\inter-0.5) node{{\small $\frac{\ell_1^2}{4}$}};
\draw (\deb+0.5+2*\rodsize,\ha-\zaxisline*\inter-0.5) node{{\small $\frac{\ell_1^2+\ell_2^2}{4}$}};

\draw[black,line width=0.3mm,dotted,double] (\deb+0.5+2.5*\rodsize,\ha-\zaxisline*\inter-0.5) -- (\deb+0.5+3.25*\rodsize,\ha-\zaxisline*\inter-0.5);

\draw (\deb+0.5+4*\rodsize,\ha-\zaxisline*\inter-0.5) node{{\small $\frac{\sum_{i=1}^6\ell_i^2}{4}=\frac{\ell^2}{4}$}};

\end{tikzpicture}
\caption{Rod diagram of the shrinking directions on the $z$-axis after sourcing the solutions with six connected rods that force the degeneracy of each T$^6$ circle.}
\label{fig:6rodsources}
\end{figure}  

We consider solutions induced by six rods, each of which forces the degeneracy of a different T$^6$ direction at the center of the spacetime.  Since their order does not matter in the regularity constraints, we assume for simplicity that they are correctly ordered (see Fig.\ref{fig:6rodsources}) $$U_{y_1} \= \{ 1\} \,,\qquad U_{y_2} \= \{ 2\}\,,\quad \ldots \quad \,,\qquad  U_{y_6} \= \{ 6\}\,.$$

The M-theory metric and fields are given by \eqref{eq:metAdS2+T6s}, and $\bar{\alpha}_{ij}=0$ for all $i,j=1,..,6$.  The regularity constraints \eqref{eq:BuEqT6s} give
\begin{equation}
Q_I \= R_{y_{2I-1}}^2+R_{y_{2I}}^2\,,\qquad  \ell_{2I} \= \frac{R_{y_{2I}}}{R_{y_{2I-1}}}\,\ell_{2I-1}\,,\qquad I=1,2,3\,.
\end{equation}
Thus,  the M2-brane charges are entirely fixed as expected, while the lengths of the rods that force the $y_1$,  $y_3$ and $y_5$ circles to degenerate are free parameters.  Moreover,  the lengths of the rods corresponding to the degeneracy of the $y_2$,  $y_4$ and $y_6$ circles are proportional to their T$^2$ partner.  As an illustration,  we focus on the first T$^2$,  parametrized by $y_1$ and $y_2$.  The sizes of the bolts that force the degeneracy of the $y_1$ and $y_2$ circles can be arbitrarily rescaled, but  their ratio is fixed by the ratio of the radii $R_{y_1}$ and $R_{y_2}$.  Both bolts carry M2-brane charges, equal to respectively $R_{y_1}^2$ and $R_{y_2}^2$, and the total charge $Q_1$ is the sum of the square of the radii.  

As for the three-rod solutions of section \ref{sec:BuAdS2T2},  the three scale invariants give an infinite family of smooth asymptotically-AdS$_2$ solutions.  While the UV of the solutions are fixed to be AdS$_2\times$S$^3\times$T$^6$,  they have a variety of different topological structure in the IR when dialing the three scale invariants, as discussed in section \ref{sec:InterestingLim}.  However,  the regular solutions only exist at a specific point of the moduli space where the number of M2 branes are fixed in terms of the T$^6$ size
\begin{equation}
N_1 \= N_2 \= N_3 \= \frac{2 \cV_6}{(l_p)^6}\,,
\end{equation}
where we have assumed for simplicity that each two-torus have equal sides, $R_{y_{2I-1}}=R_{y_{2I}}$.

\subsection{Solution with bolts of the same nature}

We now consider configurations that have rod sources of the same type.  We show that this releases some of the constraints on the total M2-brane charges, by breaking some scaling invariances. 

\begin{itemize}
\item[•] \underline{Four-rod solutions with two rods of the same nature:}

\begin{figure}[ht]
\centering
    \begin{tikzpicture}

\def\deb{-10} 
\def\inter{0.8} 
\def\ha{2.8} 
\def\zaxisline{8} 
\def\rodsize{2} 
\def\numrod{2} 

\def\fin{\deb+1+2*\rodsize+\numrod*\rodsize} 





\draw[black,thin] (\deb+1,\ha) -- (\fin,\ha);
\draw[black,thin] (\deb,\ha-\inter) -- (\fin-1,\ha-\inter);
\draw[black,thin] (\deb,\ha-2*\inter) -- (\fin,\ha-2*\inter);
\draw[black,thin] (\deb,\ha-3*\inter) -- (\fin,\ha-3*\inter);
\draw[black,thin] (\deb,\ha-4*\inter) -- (\fin,\ha-4*\inter);
\draw[black,thin] (\deb,\ha-5*\inter) -- (\fin,\ha-5*\inter);
\draw[black,thin] (\deb,\ha-6*\inter) -- (\fin,\ha-6*\inter);
\draw[black,thin] (\deb,\ha-7*\inter) -- (\fin,\ha-7*\inter);
\draw[black,->, line width=0.3mm] (\deb-0.5,\ha-\zaxisline*\inter) -- (\fin+0.2,\ha-\zaxisline*\inter);

\draw [decorate, 
    decoration = {brace,
        raise=5pt,
        amplitude=5pt},line width=0.2mm,gray] (\deb-0.8,\ha-1.5*\inter+0.05) --  (\deb-0.8,\ha+0.5*\inter-0.05);
\draw [decorate, 
    decoration = {brace,
        raise=5pt,
        amplitude=5pt},line width=0.2mm,gray] (\deb-0.8,\ha-3.5*\inter+0.05) --  (\deb-0.8,\ha-1.5*\inter-0.05);
\draw [decorate, 
    decoration = {brace,
        raise=5pt,
        amplitude=5pt},line width=0.2mm,gray] (\deb-0.8,\ha-5.5*\inter+0.05) --  (\deb-0.8,\ha-3.5*\inter-0.05);
\draw [decorate, 
    decoration = {brace,
        raise=5pt,
        amplitude=5pt},line width=0.2mm,gray] (\deb-0.8,\ha-7.5*\inter+0.05) --  (\deb-0.8,\ha-5.5*\inter-0.05);
\draw [decorate, 
    decoration = {brace,
        raise=5pt,
        amplitude=5pt},line width=0.2mm,gray] (\deb-1.6,\ha-7.5*\inter+0.05) --  (\deb-1.6,\ha-1.5*\inter-0.05);

\draw[gray] (\deb-1.5,\ha-0.5*\inter) node{S$^3$};
\draw[gray] (\deb-1.5,\ha-2.5*\inter) node{T$^2_1$};
\draw[gray] (\deb-1.5,\ha-4.5*\inter) node{T$^2_2$};
\draw[gray] (\deb-1.5,\ha-6.5*\inter) node{T$^2_3$};
\draw[gray] (\deb-2.3,\ha-4.5*\inter) node{T$^6$};

\draw (\deb-0.5,\ha) node{$\varphi_1$};
\draw (\deb-0.5,\ha-\inter) node{$\varphi_2$};
\draw (\deb-0.5,\ha-2*\inter) node{$y_1$};
\draw (\deb-0.5,\ha-3*\inter) node{$y_2$};
\draw (\deb-0.5,\ha-4*\inter) node{$y_3$};
\draw (\deb-0.5,\ha-5*\inter) node{$y_4$};
\draw (\deb-0.5,\ha-6*\inter) node{$y_5$};
\draw (\deb-0.5,\ha-7*\inter) node{$y_6$};

\draw (\fin+0.2,\ha-\zaxisline*\inter-0.3) node{$z$};


\draw[black, dotted, line width=1mm] (\deb,\ha) -- (\deb+0.5,\ha);
\draw[black,line width=1mm] (\deb+0.5,\ha) -- (\deb+0.5+\rodsize,\ha);
\draw[black,line width=1mm] (\fin-0.5-\rodsize,\ha-\inter) -- (\fin-0.55,\ha-\inter);
\draw[black, dotted,line width=1mm] (\fin-0.5,\ha-\inter) -- (\fin,\ha-\inter);


\draw[amaranthred,line width=1mm] (\deb+0.5+\rodsize,\ha-2*\inter) -- (\deb+0.5+1.5*\rodsize,\ha-2*\inter);
\draw[amazon,line width=1mm] (\deb+0.5+1.5*\rodsize,\ha-4*\inter) -- (\deb+0.5+2*\rodsize,\ha-4*\inter);
\draw[bdazzledblue,line width=1mm] (\deb+0.5+2*\rodsize,\ha-6*\inter) -- (\deb+0.5+2.5*\rodsize,\ha-6*\inter);
\draw[amaranthred,line width=1mm] (\deb+0.5+2.5*\rodsize,\ha-2*\inter) -- (\deb+0.5+3*\rodsize,\ha-2*\inter);

\draw[amaranthred,line width=1mm,opacity=0.25] (\deb+0.5+\rodsize,\ha-\zaxisline*\inter) -- (\deb+0.5+1.5*\rodsize,\ha-\zaxisline*\inter);
\draw[amazon,line width=1mm,opacity=0.25] (\deb+0.5+1.5*\rodsize,\ha-\zaxisline*\inter) -- (\deb+0.5+2*\rodsize,\ha-\zaxisline*\inter);
\draw[bdazzledblue,line width=1mm,opacity=0.25] (\deb+0.5+2*\rodsize,\ha-\zaxisline*\inter) -- (\deb+0.5+2.5*\rodsize,\ha-\zaxisline*\inter);
\draw[amaranthred,line width=1mm,opacity=0.25] (\deb+0.5+2.5*\rodsize,\ha-\zaxisline*\inter) -- (\deb+0.5+3*\rodsize,\ha-\zaxisline*\inter);


\draw[gray,dotted,line width=0.2mm] (\deb+0.5+\rodsize,\ha) -- (\deb+0.5+\rodsize,\ha-\zaxisline*\inter);
\draw[gray,dotted,line width=0.2mm] (\deb+0.5+1.5*\rodsize,\ha) -- (\deb+0.5+1.5*\rodsize,\ha-\zaxisline*\inter);
\draw[gray,dotted,line width=0.2mm] (\deb+0.5+2*\rodsize,\ha) -- (\deb+0.5+2*\rodsize,\ha-\zaxisline*\inter);
\draw[gray,dotted,line width=0.2mm] (\deb+0.5+2.5*\rodsize,\ha) -- (\deb+0.5+2.5*\rodsize,\ha-\zaxisline*\inter);
\draw[gray,dotted,line width=0.2mm] (\deb+0.5+3*\rodsize,\ha) -- (\deb+0.5+3*\rodsize,\ha-\zaxisline*\inter);

\draw[line width=0.3mm] (\deb+0.5+1.5*\rodsize,\ha-\zaxisline*\inter+0.1) -- (\deb+0.5+1.5*\rodsize,\ha-\zaxisline*\inter-0.1);
\draw[line width=0.3mm] (\deb+0.5+\rodsize,\ha-\zaxisline*\inter+0.1) -- (\deb+0.5+\rodsize,\ha-\zaxisline*\inter-0.1);
\draw[line width=0.3mm] (\deb+0.5+2*\rodsize,\ha-\zaxisline*\inter+0.1) -- (\deb+0.5+2*\rodsize,\ha-\zaxisline*\inter-0.1);
\draw[line width=0.3mm] (\deb+0.5+2.5*\rodsize,\ha-\zaxisline*\inter+0.1) -- (\deb+0.5+2.5*\rodsize,\ha-\zaxisline*\inter-0.1);
\draw[line width=0.3mm] (\deb+0.5+3*\rodsize,\ha-\zaxisline*\inter+0.1) -- (\deb+0.5+3*\rodsize,\ha-\zaxisline*\inter-0.1);

\draw (\deb+0.5+\rodsize,\ha-\zaxisline*\inter-0.5) node{{\small $0$}};
\draw (\deb+0.5+1.5*\rodsize,\ha-\zaxisline*\inter-0.5) node{{\small $\frac{\ell_1^2}{4}$}};
\draw (\deb+0.5+2*\rodsize,\ha-\zaxisline*\inter-0.5) node{{\small $\frac{\ell_1^2+\ell_2^2}{4}$}};
\draw (\deb+0.5+2.5*\rodsize,\ha-\zaxisline*\inter-0.5) node{{\small $\frac{\sum_{i=1}^3\ell_i^2}{4}$}};

\draw (\deb+0.5+3*\rodsize,\ha-\zaxisline*\inter-0.5) node{{\small $\frac{\ell^2}{4}$}};

\end{tikzpicture}
\caption{Rod diagram of the shrinking directions on the $z$-axis after sourcing the solutions with four connected rods that force the degeneracy of the each $y_3$ and $y_5$ once and $y_1$ twice.}
\label{fig:4rodsources}
\end{figure}
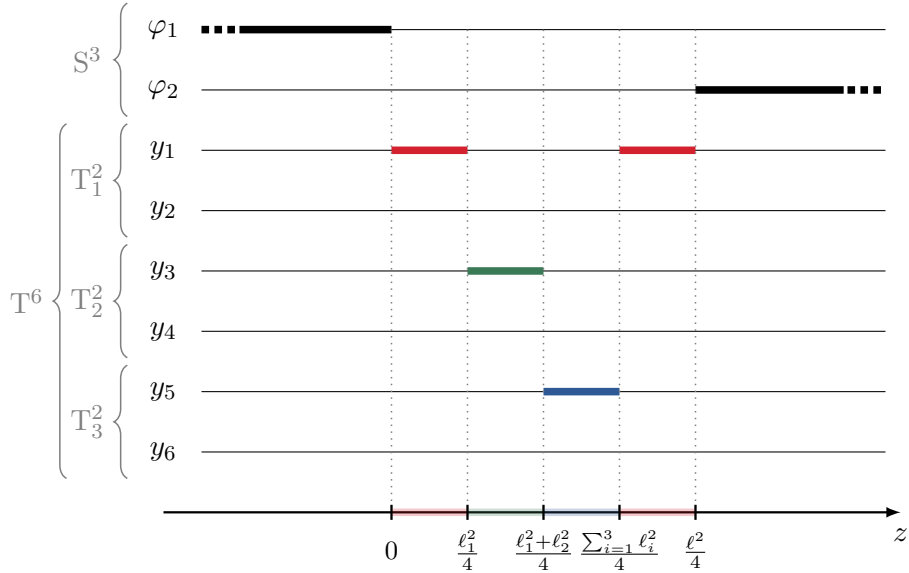  

We first consider the following four-rod configuration (see Fig.\ref{fig:4rodsources})
$$U_{y_1} \= \{ 1,4\} \,,\quad U_{y_3} \= \{ 2\}\,,\quad U_{y_5} \= \{ 3\}\,,\qquad U_{y_2}=U_{y_4}=U_{y_6}=\emptyset \,,$$
which, compared to the solution in section \ref{sec:3rod}, contains an additional rod that forces the $y_1$-circle to degenerate.

The M-theory solution can be directly derived from \eqref{eq:metAdS2+T6s}. One has $\bar{\alpha}_{ij}=0$ for all $i,j$ except $\bar{\alpha}_{14}=\bar{\alpha}_{41}=1$. The  bubble equations \eqref{eq:BuEqT6s} give
\begin{equation}
\begin{split}
Q_1& \= \frac{2R_{y_1}^2 \left(1-\sqrt{1-\frac{R_{y_3}^2}{Q_2}} \right)}{R_{y_3}^2} Q_2\,,\qquad Q_3 \= \frac{R_{y_5}^2}{R_{y_3}^2}\,Q_2\,, \\
\ell_1^2 &\= \ell_4^2 \= \frac{Q_2}{R_{y_3}^2}\,\sqrt{1-\frac{R_{y_3}^2}{Q_2}}\left(1+\sqrt{1-\frac{R_{y_3}^2}{Q_2}} \right) \,(\ell_2^2+\ell_3^2)
\end{split}
\end{equation}

One of the M2-brane charges is now unfixed: we have chosen it to be $Q_2$. It is still slightly constrained by the fact that the other charges must be real and positive: $Q_2>R_{y_3}^2$.  As a consequence,  the lengths of the two rods that force the first T$^2$ to degenerate are now fixed in terms of the others and cannot be dialed freely as in the previous solutions.  However,  we still have two scale invariances associated to the two other T$^2$ directions.  Indeed,  one can vary $\ell_2$ and $\ell_3$ arbitrarily.  The solution still describes a large variety of smooth IR topology in AdS$_2$ by tuning the sizes of these two rods.  

Furthermore, note that $Q_1$ is necessarily bounded between $R_{y_1}^2$ and $2R_{y_1}^2$, whereas $Q_2$ and $Q_3$ can be arbitrarily large with respect to the radii $R_{y_3}^2$ and $R_{y_5}^2$.  We can also convert these constraints in terms of moduli space by expressing the quantized M2-brane charge \eqref{eq:quantized_charges} and volumes of the internal spaces \eqref{eq:AsymVol}.  In one word,  the number of M2-branes along the second two-torus, $N_2$,  is now arbitrary,  and decouples from the torus size.   However,  $N_1$ and $N_3$ are fixed in terms of $N_2$ and the T$^6$ volume, and in particular, the number of M2 branes $N_1$ cannot be larger than $2\cV_6/(l_p)^6$.  Therefore,  while the constraints on the available moduli space is less restrictive than for the three-rod and six-rod configurations, solutions still only exists in a region of the moduli space where one of the quantized brane charges scales with the torus size.

Finally, notice that while the order of the rods did not matter for the three-rod solution of section \ref{sec:BuAdS2T2}, or the six-rod solution,  it does matter here, as we consider rod sources of the same nature.  Indeed, the aspect ratios $d_i$ \eqref{eq:DefdiAspect} are sensitive to the order and we would have obtained slightly different regularity conditions if we had considered $U_{y_1}=\{1,3\}$,  $U_{y_3}=\{2\}$ and $U_{y_5}=\{4\}$ for instance.

\item[•] \underline{Five-rod solutions with four rods of the same nature:}

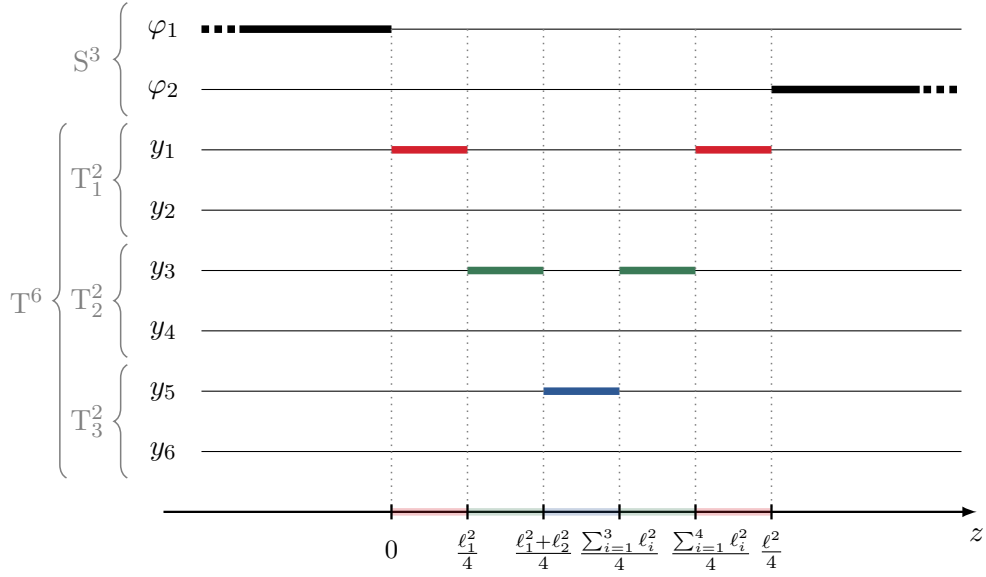
\begin{figure}[ht]
\centering
    \begin{tikzpicture}

\def\deb{-10} 
\def\inter{0.8} 
\def\ha{2.8} 
\def\zaxisline{8} 
\def\rodsize{2} 
\def\numrod{2.5} 

\def\fin{\deb+1+2*\rodsize+\numrod*\rodsize}


\draw[black,thin] (\deb+1,\ha) -- (\fin,\ha);
\draw[black,thin] (\deb,\ha-\inter) -- (\fin-1,\ha-\inter);
\draw[black,thin] (\deb,\ha-2*\inter) -- (\fin,\ha-2*\inter);
\draw[black,thin] (\deb,\ha-3*\inter) -- (\fin,\ha-3*\inter);
\draw[black,thin] (\deb,\ha-4*\inter) -- (\fin,\ha-4*\inter);
\draw[black,thin] (\deb,\ha-5*\inter) -- (\fin,\ha-5*\inter);
\draw[black,thin] (\deb,\ha-6*\inter) -- (\fin,\ha-6*\inter);
\draw[black,thin] (\deb,\ha-7*\inter) -- (\fin,\ha-7*\inter);
\draw[black,->, line width=0.3mm] (\deb-0.5,\ha-\zaxisline*\inter) -- (\fin+0.2,\ha-\zaxisline*\inter);

\draw [decorate, 
    decoration = {brace,
        raise=5pt,
        amplitude=5pt},line width=0.2mm,gray] (\deb-0.8,\ha-1.5*\inter+0.05) --  (\deb-0.8,\ha+0.5*\inter-0.05);
\draw [decorate, 
    decoration = {brace,
        raise=5pt,
        amplitude=5pt},line width=0.2mm,gray] (\deb-0.8,\ha-3.5*\inter+0.05) --  (\deb-0.8,\ha-1.5*\inter-0.05);
\draw [decorate, 
    decoration = {brace,
        raise=5pt,
        amplitude=5pt},line width=0.2mm,gray] (\deb-0.8,\ha-5.5*\inter+0.05) --  (\deb-0.8,\ha-3.5*\inter-0.05);
\draw [decorate, 
    decoration = {brace,
        raise=5pt,
        amplitude=5pt},line width=0.2mm,gray] (\deb-0.8,\ha-7.5*\inter+0.05) --  (\deb-0.8,\ha-5.5*\inter-0.05);
\draw [decorate, 
    decoration = {brace,
        raise=5pt,
        amplitude=5pt},line width=0.2mm,gray] (\deb-1.6,\ha-7.5*\inter+0.05) --  (\deb-1.6,\ha-1.5*\inter-0.05);

\draw[gray] (\deb-1.5,\ha-0.5*\inter) node{S$^3$};
\draw[gray] (\deb-1.5,\ha-2.5*\inter) node{T$^2_1$};
\draw[gray] (\deb-1.5,\ha-4.5*\inter) node{T$^2_2$};
\draw[gray] (\deb-1.5,\ha-6.5*\inter) node{T$^2_3$};
\draw[gray] (\deb-2.3,\ha-4.5*\inter) node{T$^6$};

\draw (\deb-0.5,\ha) node{$\varphi_1$};
\draw (\deb-0.5,\ha-\inter) node{$\varphi_2$};
\draw (\deb-0.5,\ha-2*\inter) node{$y_1$};
\draw (\deb-0.5,\ha-3*\inter) node{$y_2$};
\draw (\deb-0.5,\ha-4*\inter) node{$y_3$};
\draw (\deb-0.5,\ha-5*\inter) node{$y_4$};
\draw (\deb-0.5,\ha-6*\inter) node{$y_5$};
\draw (\deb-0.5,\ha-7*\inter) node{$y_6$};

\draw (\fin+0.2,\ha-\zaxisline*\inter-0.3) node{$z$};


\draw[black, dotted, line width=1mm] (\deb,\ha) -- (\deb+0.5,\ha);
\draw[black,line width=1mm] (\deb+0.5,\ha) -- (\deb+0.5+\rodsize,\ha);
\draw[black,line width=1mm] (\fin-0.5-\rodsize,\ha-\inter) -- (\fin-0.55,\ha-\inter);
\draw[black, dotted,line width=1mm] (\fin-0.5,\ha-\inter) -- (\fin,\ha-\inter);


\draw[amaranthred,line width=1mm] (\deb+0.5+\rodsize,\ha-2*\inter) -- (\deb+0.5+1.5*\rodsize,\ha-2*\inter);
\draw[amazon,line width=1mm] (\deb+0.5+1.5*\rodsize,\ha-4*\inter) -- (\deb+0.5+2*\rodsize,\ha-4*\inter);
\draw[bdazzledblue,line width=1mm] (\deb+0.5+2*\rodsize,\ha-6*\inter) -- (\deb+0.5+2.5*\rodsize,\ha-6*\inter);
\draw[amazon,line width=1mm] (\deb+0.5+2.5*\rodsize,\ha-4*\inter) -- (\deb+0.5+3*\rodsize,\ha-4*\inter);
\draw[amaranthred,line width=1mm] (\deb+0.5+3*\rodsize,\ha-2*\inter) -- (\deb+0.5+3.5*\rodsize,\ha-2*\inter);

\draw[amaranthred,line width=1mm,opacity=0.25] (\deb+0.5+\rodsize,\ha-\zaxisline*\inter) -- (\deb+0.5+1.5*\rodsize,\ha-\zaxisline*\inter);
\draw[amazon,line width=1mm,opacity=0.25] (\deb+0.5+1.5*\rodsize,\ha-\zaxisline*\inter) -- (\deb+0.5+2*\rodsize,\ha-\zaxisline*\inter);
\draw[bdazzledblue,line width=1mm,opacity=0.25] (\deb+0.5+2*\rodsize,\ha-\zaxisline*\inter) -- (\deb+0.5+2.5*\rodsize,\ha-\zaxisline*\inter);
\draw[amazon,line width=1mm,opacity=0.25] (\deb+0.5+2.5*\rodsize,\ha-\zaxisline*\inter) -- (\deb+0.5+3*\rodsize,\ha-\zaxisline*\inter);
\draw[amaranthred,line width=1mm,opacity=0.25] (\deb+0.5+3*\rodsize,\ha-\zaxisline*\inter) -- (\deb+0.5+3.5*\rodsize,\ha-\zaxisline*\inter);


\draw[gray,dotted,line width=0.2mm] (\deb+0.5+\rodsize,\ha) -- (\deb+0.5+\rodsize,\ha-\zaxisline*\inter);
\draw[gray,dotted,line width=0.2mm] (\deb+0.5+1.5*\rodsize,\ha) -- (\deb+0.5+1.5*\rodsize,\ha-\zaxisline*\inter);
\draw[gray,dotted,line width=0.2mm] (\deb+0.5+2*\rodsize,\ha) -- (\deb+0.5+2*\rodsize,\ha-\zaxisline*\inter);
\draw[gray,dotted,line width=0.2mm] (\deb+0.5+2.5*\rodsize,\ha) -- (\deb+0.5+2.5*\rodsize,\ha-\zaxisline*\inter);
\draw[gray,dotted,line width=0.2mm] (\deb+0.5+3*\rodsize,\ha) -- (\deb+0.5+3*\rodsize,\ha-\zaxisline*\inter);
\draw[gray,dotted,line width=0.2mm] (\deb+0.5+3.5*\rodsize,\ha) -- (\deb+0.5+3.5*\rodsize,\ha-\zaxisline*\inter);

\draw[line width=0.3mm] (\deb+0.5+1.5*\rodsize,\ha-\zaxisline*\inter+0.1) -- (\deb+0.5+1.5*\rodsize,\ha-\zaxisline*\inter-0.1);
\draw[line width=0.3mm] (\deb+0.5+\rodsize,\ha-\zaxisline*\inter+0.1) -- (\deb+0.5+\rodsize,\ha-\zaxisline*\inter-0.1);
\draw[line width=0.3mm] (\deb+0.5+2*\rodsize,\ha-\zaxisline*\inter+0.1) -- (\deb+0.5+2*\rodsize,\ha-\zaxisline*\inter-0.1);
\draw[line width=0.3mm] (\deb+0.5+2.5*\rodsize,\ha-\zaxisline*\inter+0.1) -- (\deb+0.5+2.5*\rodsize,\ha-\zaxisline*\inter-0.1);
\draw[line width=0.3mm] (\deb+0.5+3*\rodsize,\ha-\zaxisline*\inter+0.1) -- (\deb+0.5+3*\rodsize,\ha-\zaxisline*\inter-0.1);
\draw[line width=0.3mm] (\deb+0.5+3.5*\rodsize,\ha-\zaxisline*\inter+0.1) -- (\deb+0.5+3.5*\rodsize,\ha-\zaxisline*\inter-0.1);

\draw (\deb+0.5+\rodsize,\ha-\zaxisline*\inter-0.5) node{{\small $0$}};
\draw (\deb+0.5+1.5*\rodsize,\ha-\zaxisline*\inter-0.5) node{{\small $\frac{\ell_1^2}{4}$}};
\draw (\deb+0.5+2*\rodsize,\ha-\zaxisline*\inter-0.5) node{{\small $\frac{\ell_1^2+\ell_2^2}{4}$}};
\draw (\deb+0.5+2.5*\rodsize,\ha-\zaxisline*\inter-0.5) node{{\small $\frac{\sum_{i=1}^3\ell_i^2}{4}$}};
\draw (\deb+0.5+3.1*\rodsize,\ha-\zaxisline*\inter-0.5) node{{\small $\frac{\sum_{i=1}^4\ell_i^2}{4}$}};

\draw (\deb+0.5+3.5*\rodsize,\ha-\zaxisline*\inter-0.5) node{{\small $\frac{\ell^2}{4}$}};

\end{tikzpicture}
\caption{Rod diagram of the shrinking directions on the $z$-axis after sourcing the solutions with five connected rods that force the degeneracy of $y_5$  once and $y_1$ and $y_3$ twice.}
\label{fig:5rodsources}
\end{figure}  

We can take a step further by allowing two pairs of rods to be of the same kind.  We consider (see Fig.\ref{fig:5rodsources})
$$U_{y_1} \= \{ 1,5\} \,,\quad U_{y_3} \= \{ 2,4\}\,,\quad U_{y_5} \= \{ 3\}\,,\qquad U_{y_2}=U_{y_4}=U_{y_6}=\emptyset \,,$$

The bubble equations give
\begin{equation}
\begin{split}
Q_3& \= \frac{R_{y_5}^2}{R_{y_3}^2}\, \frac{Q_2 R_{y_1}^2(Q_1-R_{y_1}^2)}{4Q_2 R_{y_1}^2(Q_1-R_{y_1}^2)-Q_1^2 R_{y_3}^2}\,Q_2\,, \\
\ell_1^2 &\= \ell_5^2 \= \frac{2 R_{y_1}^2-Q_1}{2(Q_1-R_{y_1}^2)}\,\frac{Q_1^2 R_{y_3}^2}{4Q_2 R_{y_1}^2(Q_1-R_{y_1}^2)-Q_1^2 R_{y_3}^2}\,\ell_3^2\,,\\
\ell_2^2 &\= \ell_4^2 \= \frac{1}{2}\,\left( \frac{Q_1^2 R_{y_3}^2}{4Q_2 R_{y_1}^2(Q_1-R_{y_1}^2)-Q_1^2 R_{y_3}^2}-1\right)\,\ell_3^2\,.
\end{split}
\end{equation}

Two total M2 charges, $Q_1$ and $Q_2$,  are unconstrained but the validity of the solutions require
\begin{equation}
R_{y_1}^2 < Q_1 < 2R_{y_1}^2\,,\qquad \frac{R_{y_3}^2}{4 R_{y_1}^2} < \frac{Q_2(Q_1-R_{y_1}^2)}{Q_1^2} < \frac{R_{y_3}^2}{2R_{y_1}^2} \,.
\end{equation}
Moreover,  one rod length is free,  which we took to be $\ell_3^2$, while all others are fixed.  Thus,  one cannot dial separately different rod lengths associated to different T$^2$ anymore and one has a unique IR topology with a single scale invariance, as expected for AdS$_2$ geometries.

However, even if the total charges are independent parameters,  some are restricted to be of order the radius of the extra dimension,  and their associated quantized charges are of order the torus volume.  Once again, our solutions exist only in regions of the moduli space where brane numbers and torus size are of the same order.  This is due to the fact that a rod can be considered as a ``quantum bit'' for which the local quantized charge will scale with the internal space volume.  Thus, to construct solutions that have total charges that decouple from the internal space sizes,  one needs to turn the last parameter at hand: the number of rods.   

\end{itemize}

\subsection{Solutions with a large number of bolts}
\label{sec:large_n}

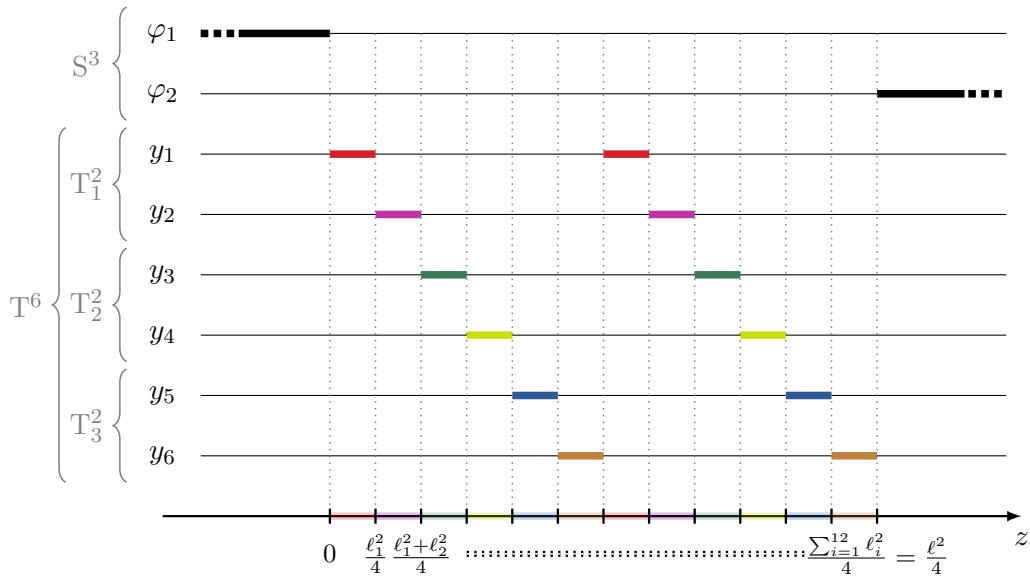
\begin{figure}[ht]
  \centering
      \begin{tikzpicture}

  \def\deb{-10} 
  \def\inter{0.8} 
  \def\ha{2.8} 
  \def\zaxisline{8} 
  \def\rodsize{1.2} 
  \def\numrod{6} 

  \def\fin{\deb+1+2*\rodsize+\numrod*\rodsize} 




  \draw[black,thin] (\deb+1,\ha) -- (\fin,\ha);
  \draw[black,thin] (\deb,\ha-\inter) -- (\fin-1,\ha-\inter);
  \draw[black,thin] (\deb,\ha-2*\inter) -- (\fin,\ha-2*\inter);
  \draw[black,thin] (\deb,\ha-3*\inter) -- (\fin,\ha-3*\inter);
  \draw[black,thin] (\deb,\ha-4*\inter) -- (\fin,\ha-4*\inter);
  \draw[black,thin] (\deb,\ha-5*\inter) -- (\fin,\ha-5*\inter);
  \draw[black,thin] (\deb,\ha-6*\inter) -- (\fin,\ha-6*\inter);
  \draw[black,thin] (\deb,\ha-7*\inter) -- (\fin,\ha-7*\inter);
  \draw[black,->, line width=0.3mm] (\deb-0.5,\ha-\zaxisline*\inter) -- (\fin+0.2,\ha-\zaxisline*\inter);

  \draw [decorate, 
      decoration = {brace,
          raise=5pt,
          amplitude=5pt},line width=0.2mm,gray] (\deb-0.8,\ha-1.5*\inter+0.05) --  (\deb-0.8,\ha+0.5*\inter-0.05);
  \draw [decorate, 
      decoration = {brace,
          raise=5pt,
          amplitude=5pt},line width=0.2mm,gray] (\deb-0.8,\ha-3.5*\inter+0.05) --  (\deb-0.8,\ha-1.5*\inter-0.05);
  \draw [decorate, 
      decoration = {brace,
          raise=5pt,
          amplitude=5pt},line width=0.2mm,gray] (\deb-0.8,\ha-5.5*\inter+0.05) --  (\deb-0.8,\ha-3.5*\inter-0.05);
  \draw [decorate, 
      decoration = {brace,
          raise=5pt,
          amplitude=5pt},line width=0.2mm,gray] (\deb-0.8,\ha-7.5*\inter+0.05) --  (\deb-0.8,\ha-5.5*\inter-0.05);
  \draw [decorate, 
      decoration = {brace,
          raise=5pt,
          amplitude=5pt},line width=0.2mm,gray] (\deb-1.6,\ha-7.5*\inter+0.05) --  (\deb-1.6,\ha-1.5*\inter-0.05);

  \draw[gray] (\deb-1.5,\ha-0.5*\inter) node{S$^3$};
  \draw[gray] (\deb-1.5,\ha-2.5*\inter) node{T$^2_1$};
  \draw[gray] (\deb-1.5,\ha-4.5*\inter) node{T$^2_2$};
  \draw[gray] (\deb-1.5,\ha-6.5*\inter) node{T$^2_3$};
  \draw[gray] (\deb-2.3,\ha-4.5*\inter) node{T$^6$};

  \draw (\deb-0.5,\ha) node{$\varphi_1$};
  \draw (\deb-0.5,\ha-\inter) node{$\varphi_2$};
  \draw (\deb-0.5,\ha-2*\inter) node{$y_1$};
  \draw (\deb-0.5,\ha-3*\inter) node{$y_2$};
  \draw (\deb-0.5,\ha-4*\inter) node{$y_3$};
  \draw (\deb-0.5,\ha-5*\inter) node{$y_4$};
  \draw (\deb-0.5,\ha-6*\inter) node{$y_5$};
  \draw (\deb-0.5,\ha-7*\inter) node{$y_6$};

  \draw (\fin+0.2,\ha-\zaxisline*\inter-0.3) node{$z$};


  \draw[black, dotted, line width=1mm] (\deb,\ha) -- (\deb+0.5,\ha);
  \draw[black,line width=1mm] (\deb+0.5,\ha) -- (\deb+0.5+\rodsize,\ha);
  \draw[black,line width=1mm] (\fin-0.5-\rodsize,\ha-\inter) -- (\fin-0.55,\ha-\inter);
  \draw[black, dotted,line width=1mm] (\fin-0.5,\ha-\inter) -- (\fin,\ha-\inter);


  \draw[amaranthred,line width=1mm] (\deb+0.5+\rodsize,\ha-2*\inter) -- (\deb+0.5+1.5*\rodsize,\ha-2*\inter);
  \draw[byzantine,line width=1mm] (\deb+0.5+1.5*\rodsize,\ha-3*\inter) -- (\deb+0.5+2*\rodsize,\ha-3*\inter);
  \draw[amazon,line width=1mm] (\deb+0.5+2*\rodsize,\ha-4*\inter) -- (\deb+0.5+2.5*\rodsize,\ha-4*\inter);
  \draw[bitterlemon,line width=1mm] (\deb+0.5+2.5*\rodsize,\ha-5*\inter) -- (\deb+0.5+3*\rodsize,\ha-5*\inter);
  \draw[bdazzledblue,line width=1mm] (\deb+0.5+3*\rodsize,\ha-6*\inter) -- (\deb+0.5+3.5*\rodsize,\ha-6*\inter);
  \draw[brown,line width=1mm] (\deb+0.5+3.5*\rodsize,\ha-7*\inter) -- (\deb+0.5+4*\rodsize,\ha-7*\inter);

  \draw[amaranthred,line width=1mm] (\deb+0.5+4*\rodsize,\ha-2*\inter) -- (\deb+0.5+4.5*\rodsize,\ha-2*\inter);
  \draw[byzantine,line width=1mm] (\deb+0.5+4.5*\rodsize,\ha-3*\inter) -- (\deb+0.5+5*\rodsize,\ha-3*\inter);
  \draw[amazon,line width=1mm] (\deb+0.5+5*\rodsize,\ha-4*\inter) -- (\deb+0.5+5.5*\rodsize,\ha-4*\inter);
  \draw[bitterlemon,line width=1mm] (\deb+0.5+5.5*\rodsize,\ha-5*\inter) -- (\deb+0.5+6*\rodsize,\ha-5*\inter);
  \draw[bdazzledblue,line width=1mm] (\deb+0.5+6*\rodsize,\ha-6*\inter) -- (\deb+0.5+6.5*\rodsize,\ha-6*\inter);
  \draw[brown,line width=1mm] (\deb+0.5+6.5*\rodsize,\ha-7*\inter) -- (\deb+0.5+7*\rodsize,\ha-7*\inter);

  \draw[amaranthred,line width=1mm,opacity=0.25] (\deb+0.5+\rodsize,\ha-\zaxisline*\inter) -- (\deb+0.5+1.5*\rodsize,\ha-\zaxisline*\inter);
  \draw[purple,line width=1mm,opacity=0.25] (\deb+0.5+1.5*\rodsize,\ha-\zaxisline*\inter) -- (\deb+0.5+2*\rodsize,\ha-\zaxisline*\inter);
  \draw[amazon,line width=1mm,opacity=0.25] (\deb+0.5+2*\rodsize,\ha-\zaxisline*\inter) -- (\deb+0.5+2.5*\rodsize,\ha-\zaxisline*\inter);
  \draw[bitterlemon,line width=1mm,opacity=0.25] (\deb+0.5+2.5*\rodsize,\ha-\zaxisline*\inter) -- (\deb+0.5+3*\rodsize,\ha-\zaxisline*\inter);
  \draw[bdazzledblue,line width=1mm,opacity=0.25] (\deb+0.5+3*\rodsize,\ha-\zaxisline*\inter) -- (\deb+0.5+3.5*\rodsize,\ha-\zaxisline*\inter);
  \draw[brown,line width=1mm,opacity=0.25] (\deb+0.5+3.5*\rodsize,\ha-\zaxisline*\inter) -- (\deb+0.5+4*\rodsize,\ha-\zaxisline*\inter);

  \draw[amaranthred,line width=1mm,opacity=0.25] (\deb+0.5+4*\rodsize,\ha-\zaxisline*\inter) -- (\deb+0.5+4.5*\rodsize,\ha-\zaxisline*\inter);
  \draw[purple,line width=1mm,opacity=0.25] (\deb+0.5+4.5*\rodsize,\ha-\zaxisline*\inter) -- (\deb+0.5+5*\rodsize,\ha-\zaxisline*\inter);
  \draw[amazon,line width=1mm,opacity=0.25] (\deb+0.5+5*\rodsize,\ha-\zaxisline*\inter) -- (\deb+0.5+5.5*\rodsize,\ha-\zaxisline*\inter);
  \draw[bitterlemon,line width=1mm,opacity=0.25] (\deb+0.5+5.5*\rodsize,\ha-\zaxisline*\inter) -- (\deb+0.5+6*\rodsize,\ha-\zaxisline*\inter);
  \draw[bdazzledblue,line width=1mm,opacity=0.25] (\deb+0.5+6*\rodsize,\ha-\zaxisline*\inter) -- (\deb+0.5+6.5*\rodsize,\ha-\zaxisline*\inter);
  \draw[brown,line width=1mm,opacity=0.25] (\deb+0.5+6.5*\rodsize,\ha-\zaxisline*\inter) -- (\deb+0.5+7*\rodsize,\ha-\zaxisline*\inter);


  \draw[gray,dotted,line width=0.2mm] (\deb+0.5+\rodsize,\ha) -- (\deb+0.5+\rodsize,\ha-\zaxisline*\inter);
  \draw[gray,dotted,line width=0.2mm] (\deb+0.5+1.5*\rodsize,\ha) -- (\deb+0.5+1.5*\rodsize,\ha-\zaxisline*\inter);
  \draw[gray,dotted,line width=0.2mm] (\deb+0.5+2*\rodsize,\ha) -- (\deb+0.5+2*\rodsize,\ha-\zaxisline*\inter);
  \draw[gray,dotted,line width=0.2mm] (\deb+0.5+2.5*\rodsize,\ha) -- (\deb+0.5+2.5*\rodsize,\ha-\zaxisline*\inter);
  \draw[gray,dotted,line width=0.2mm] (\deb+0.5+3*\rodsize,\ha) -- (\deb+0.5+3*\rodsize,\ha-\zaxisline*\inter);
  \draw[gray,dotted,line width=0.2mm] (\deb+0.5+3.5*\rodsize,\ha) -- (\deb+0.5+3.5*\rodsize,\ha-\zaxisline*\inter);
  \draw[gray,dotted,line width=0.2mm] (\deb+0.5+4*\rodsize,\ha) -- (\deb+0.5+4*\rodsize,\ha-\zaxisline*\inter);
  \draw[gray,dotted,line width=0.2mm] (\deb+0.5+4.5*\rodsize,\ha) -- (\deb+0.5+4.5*\rodsize,\ha-\zaxisline*\inter);
  \draw[gray,dotted,line width=0.2mm] (\deb+0.5+5*\rodsize,\ha) -- (\deb+0.5+5*\rodsize,\ha-\zaxisline*\inter);
  \draw[gray,dotted,line width=0.2mm] (\deb+0.5+5.5*\rodsize,\ha) -- (\deb+0.5+5.5*\rodsize,\ha-\zaxisline*\inter);
  \draw[gray,dotted,line width=0.2mm] (\deb+0.5+6*\rodsize,\ha) -- (\deb+0.5+6*\rodsize,\ha-\zaxisline*\inter);
  \draw[gray,dotted,line width=0.2mm] (\deb+0.5+6.5*\rodsize,\ha) -- (\deb+0.5+6.5*\rodsize,\ha-\zaxisline*\inter);
  \draw[gray,dotted,line width=0.2mm] (\deb+0.5+7*\rodsize,\ha) -- (\deb+0.5+7*\rodsize,\ha-\zaxisline*\inter);

  \draw[line width=0.3mm] (\deb+0.5+1.5*\rodsize,\ha-\zaxisline*\inter+0.1) -- (\deb+0.5+1.5*\rodsize,\ha-\zaxisline*\inter-0.1);
  \draw[line width=0.3mm] (\deb+0.5+\rodsize,\ha-\zaxisline*\inter+0.1) -- (\deb+0.5+\rodsize,\ha-\zaxisline*\inter-0.1);
  \draw[line width=0.3mm] (\deb+0.5+2*\rodsize,\ha-\zaxisline*\inter+0.1) -- (\deb+0.5+2*\rodsize,\ha-\zaxisline*\inter-0.1);
  \draw[line width=0.3mm] (\deb+0.5+2.5*\rodsize,\ha-\zaxisline*\inter+0.1) -- (\deb+0.5+2.5*\rodsize,\ha-\zaxisline*\inter-0.1);
  \draw[line width=0.3mm] (\deb+0.5+3*\rodsize,\ha-\zaxisline*\inter+0.1) -- (\deb+0.5+3*\rodsize,\ha-\zaxisline*\inter-0.1);
  \draw[line width=0.3mm] (\deb+0.5+3.5*\rodsize,\ha-\zaxisline*\inter+0.1) -- (\deb+0.5+3.5*\rodsize,\ha-\zaxisline*\inter-0.1);
  \draw[line width=0.3mm] (\deb+0.5+4*\rodsize,\ha-\zaxisline*\inter+0.1) -- (\deb+0.5+4*\rodsize,\ha-\zaxisline*\inter-0.1);
  \draw[line width=0.3mm] (\deb+0.5+4.5*\rodsize,\ha-\zaxisline*\inter+0.1) -- (\deb+0.5+4.5*\rodsize,\ha-\zaxisline*\inter-0.1);
  \draw[line width=0.3mm] (\deb+0.5+5*\rodsize,\ha-\zaxisline*\inter+0.1) -- (\deb+0.5+5*\rodsize,\ha-\zaxisline*\inter-0.1);
  \draw[line width=0.3mm] (\deb+0.5+5.5*\rodsize,\ha-\zaxisline*\inter+0.1) -- (\deb+0.5+5.5*\rodsize,\ha-\zaxisline*\inter-0.1);
  \draw[line width=0.3mm] (\deb+0.5+6*\rodsize,\ha-\zaxisline*\inter+0.1) -- (\deb+0.5+6*\rodsize,\ha-\zaxisline*\inter-0.1);
  \draw[line width=0.3mm] (\deb+0.5+6.5*\rodsize,\ha-\zaxisline*\inter+0.1) -- (\deb+0.5+6.5*\rodsize,\ha-\zaxisline*\inter-0.1);
  \draw[line width=0.3mm] (\deb+0.5+7*\rodsize,\ha-\zaxisline*\inter+0.1) -- (\deb+0.5+7*\rodsize,\ha-\zaxisline*\inter-0.1);

  \draw (\deb+0.5+\rodsize,\ha-\zaxisline*\inter-0.5) node{{\small $0$}};
  \draw (\deb+0.5+1.5*\rodsize,\ha-\zaxisline*\inter-0.5) node{{\small $\frac{\ell_1^2}{4}$}};
  \draw (\deb+0.5+2*\rodsize,\ha-\zaxisline*\inter-0.5) node{{\small $\frac{\ell_1^2+\ell_2^2}{4}$}};

  \draw[black,line width=0.3mm,dotted,double] (\deb+0.5+2.5*\rodsize,\ha-\zaxisline*\inter-0.5) -- (\deb+0.5+6.25*\rodsize,\ha-\zaxisline*\inter-0.5);

  \draw (\deb+0.5+7*\rodsize,\ha-\zaxisline*\inter-0.5) node{{\small $\frac{\sum_{i=1}^{12}\ell_i^2}{4}=\frac{\ell^2}{4}$}};

  \end{tikzpicture}
  \caption{Rod diagram of the shrinking directions on the $z$-axis after sourcing the solutions with $\hat n=2$ patterns of six connected rods that force the degeneracy of each T$^6$ circle.}
  \label{fig:manyrodsources}
\end{figure}  

We consider solutions induced by $n = 6 \hat n$ rods, ordered in repeating patterns of six different rods (see Fig.\ref{fig:manyrodsources} for $\hat n=2$), $$U_{y_1} \= \{ 1, \,7, \dots, \,6\hat n -5\} \,,\quad U_{y_2} \= \{ 2,\,8,\dots,\,6 \hat n-4\}\,,\ \ldots \ \,,\quad  U_{y_6} \= \{ 6,\,12,\dots,\,6\hat n\}\,.$$

Solving the bubble equations \eqref{eq:BuEqT6s} with a large number of rods requires some simplifications and approximations as discussed in \cite{Bah:2021rki}.  We consider that the three charges are equal $Q_I = Q$ and all radii are equal $R_{y_i} = R_y$. We also assume that, away from the rods at the edges,  the rods are all of the same size, $\ell_i = \bar{\ell}$. This assumption can be verified numerically (see Fig.\ref{fig:sizes_rods_numeric}), and the analytic computation of the aspect ratios also provides a non-trivial consistency check.

The resolution of the bubble equations is detailed in Appendix \ref{app:large_n}.  We find that the average length $\bar{\ell}$ is unconstrained and gives a scale to our capped asymptotically-AdS$_2$ geometries.  However,  the equal total charges are fixed to be
\begin{align}
  Q ~&=~ c\,\,\hat n^\frac{5}{6} \, R_y^2\,,\qquad c = \frac{2^{1/6} \cdot A^{10}}{e^{10/12}\cdot 3^{11/12}} \,\approx\, 2.144,
\end{align}
where $A$ is the Glaisher constant.  This approximation has been compared with numerical derivations in Fig.\ref{fig:comparison_approx_numeric} in the Appendix, and they  match remarkably,  even for relatively small values of $n$.

In previous solutions,  one of the M2 charges at least was restricted to be of order the size of the extra dimensions.  This was a consequence of considering a small number of rod sources.  With the present configuration,  we see that the total charges scale like $n^\frac{5}{6}$ with respect to the number of rods.    In terms of quantized charges,  we have
\begin{equation}
N_1 = N_2 =N_3 \,\propto\, n^\frac{5}{6} \, \frac{\cV_6}{(l_p)^6}\,.
\end{equation}
We can therefore construct smooth non-supersymmetric asymptotically-AdS$_2$ solutions in M-theory where the total number of M2-brane flux decouples from the internal space volume by just piling up more and more rod sources.  The geometries will have a fixed AdS$_2\times$S$^3\times$T$^6$ in the UV while they consist of a large chain of bolts of identical size where the T$^6$ directions smoothly degenerate alternatively.

Note that the charges look all equal in the present solution which can seem as another restriction on the moduli space.  However,  more general solutions,  with arbitrary patterns of rod sources and where the radii, $R_{y_i}$,  are not equal,  should not have this feature.  We do believe that one can construct non-supersymmetric asymptotically-AdS$_2$ solitons from our family of solutions in large region of the moduli space.

Interestingly,  the local charge carried by each bolt and the quantized number of M2 branes wrapping each bubbles are given by
$$q_\text{M2} ~\equiv~ \frac Q{2\hat n} ~\approx~ \hat n^{-1/6} R_y^2\,,\qquad n_\text{M2} ~\approx~ n^{-1/6} \,   \frac{\cV_6}{(l_p)^6}$$
and they decrease with the number of rods. Because $n_\text{M2}$ must be an integer,  this puts a bound on the maximum number of rods,  $n_\text{max}$ and total number of M2 branes,  $N_\text{max}$,  in our solutions in terms of the T$^6$ volume in Planck unit:
\begin{equation}
n_\text{max} \,\sim\,  N_\text{max} \,\sim\,  \qty(\frac{\cV_6}{(l_p)^6})^{6} \,.
\end{equation}
Since the supergravity descriptions of the brane bound states require $\cV_{6} \gg (l_p)^6$ to neglect quantum effects in the internal space,  this bound still allows for a huge number of physical asymptotically-AdS$_2$ non-supersymmetric solitons.

\subsection{\texorpdfstring{Generic T$^2$ and S$^3$ deformations}{Generic T2 and S3 deformations}}
\label{sec:generic_deformations}

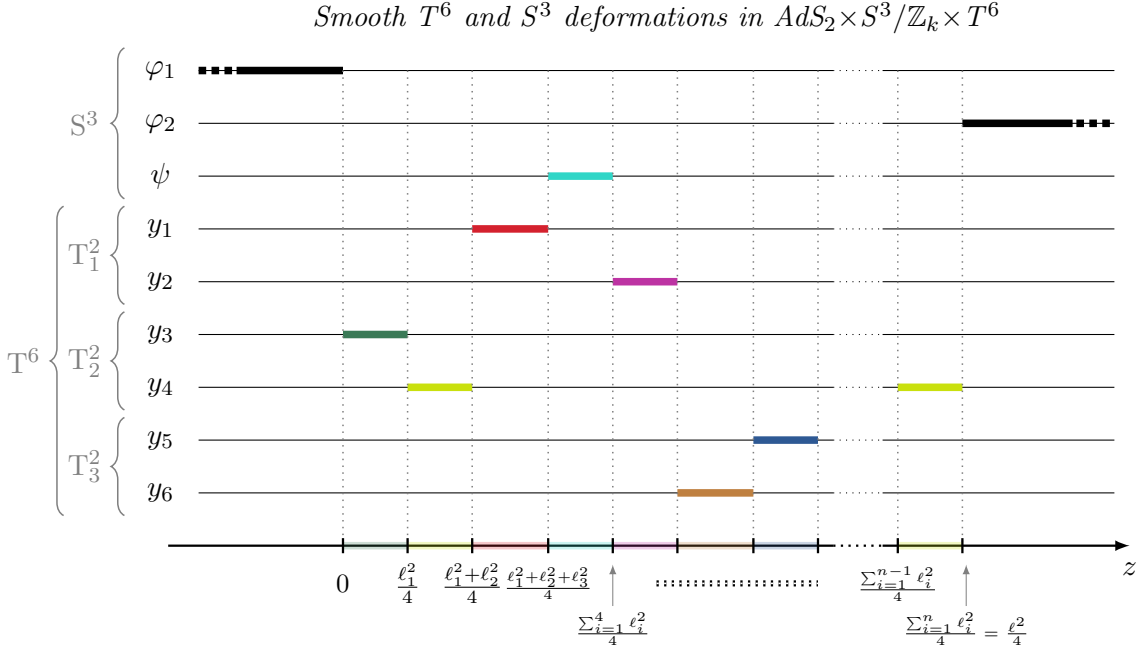
\begin{figure}[ht]
\centering
    \begin{tikzpicture}

\def\deb{-10} 
\def\inter{0.7} 
\def\ha{2.8} 
\def\zaxisline{9} 
\def\rodsize{1.7} 
\def\numrod{4.5} 

\def\fin{\deb+1+2*\rodsize+\numrod*\rodsize} 




\draw (\deb+0.5+\rodsize+0.5*\numrod*\rodsize,\ha+\inter) node{{{\it Smooth T$^{\,6}$ and S$^{\,3}$ deformations in AdS$_2\times$S$^{\,3}/\mathbb{Z}_k\times$T$^{\,6}$}}}; 

\draw [decorate, 
    decoration = {brace,
        raise=5pt,
        amplitude=5pt},line width=0.2mm,gray] (\deb-0.8,\ha-2.5*\inter+0.05) --  (\deb-0.8,\ha+0.5*\inter-0.05);
\draw [decorate, 
    decoration = {brace,
        raise=5pt,
        amplitude=5pt},line width=0.2mm,gray] (\deb-0.8,\ha-4.5*\inter+0.05) --  (\deb-0.8,\ha-2.5*\inter-0.05);
\draw [decorate, 
    decoration = {brace,
        raise=5pt,
        amplitude=5pt},line width=0.2mm,gray] (\deb-0.8,\ha-6.5*\inter+0.05) --  (\deb-0.8,\ha-4.5*\inter-0.05);
\draw [decorate, 
    decoration = {brace,
        raise=5pt,
        amplitude=5pt},line width=0.2mm,gray] (\deb-0.8,\ha-8.5*\inter+0.05) --  (\deb-0.8,\ha-6.5*\inter-0.05);
\draw [decorate, 
    decoration = {brace,
        raise=5pt,
        amplitude=5pt},line width=0.2mm,gray] (\deb-1.6,\ha-8.5*\inter+0.05) --  (\deb-1.6,\ha-2.5*\inter-0.05);

\draw[gray] (\deb-1.5,\ha-1*\inter) node{S$^3$};
\draw[gray] (\deb-1.5,\ha-3.5*\inter) node{T$^2_1$};
\draw[gray] (\deb-1.5,\ha-5.5*\inter) node{T$^2_2$};
\draw[gray] (\deb-1.5,\ha-7.5*\inter) node{T$^2_3$};
\draw[gray] (\deb-2.3,\ha-5.5*\inter) node{T$^6$};

\draw (\deb-0.5,\ha) node{$\varphi_1$};
\draw (\deb-0.5,\ha-1*\inter) node{$\varphi_2$};
\draw (\deb-0.5,\ha-2*\inter) node{$\psi$};
\draw (\deb-0.5,\ha-3*\inter) node{$y_1$};
\draw (\deb-0.5,\ha-4*\inter) node{$y_2$};
\draw (\deb-0.5,\ha-5*\inter) node{$y_3$};
\draw (\deb-0.5,\ha-6*\inter) node{$y_4$};
\draw (\deb-0.5,\ha-7*\inter) node{$y_5$};
\draw (\deb-0.5,\ha-8*\inter) node{$y_6$};

\draw (\fin+0.2,\ha-\zaxisline*\inter-0.3) node{$z$};


\draw[black,thin] (\deb,\ha) -- (\deb+0.5+4.5*\rodsize+0.2,\ha);\draw[black,thin,dotted] (\deb+0.5+4.5*\rodsize+0.2,\ha) -- (\deb+0.5+5*\rodsize,\ha);
\draw[black,thin] (\deb+0.5+5*\rodsize,\ha) -- (\fin,\ha);

\draw[black,thin] (\deb,\ha-\inter) -- (\deb+0.5+4.5*\rodsize+0.2,\ha-\inter);\draw[black,thin,dotted] (\deb+0.5+4.5*\rodsize+0.2,\ha-\inter) -- (\deb+0.5+5*\rodsize,\ha-\inter);
\draw[black,thin] (\deb+0.5+5*\rodsize,\ha-\inter) -- (\fin,\ha-\inter);

\draw[black,thin] (\deb,\ha-2*\inter) -- (\deb+0.5+4.5*\rodsize+0.2,\ha-2*\inter);\draw[black,thin,dotted] (\deb+0.5+4.5*\rodsize+0.2,\ha-2*\inter) -- (\deb+0.5+5*\rodsize,\ha-2*\inter);
\draw[black,thin] (\deb+0.5+5*\rodsize,\ha-2*\inter) -- (\fin,\ha-2*\inter);

\draw[black,thin] (\deb,\ha-3*\inter) -- (\deb+0.5+4.5*\rodsize+0.2,\ha-3*\inter);\draw[black,thin,dotted] (\deb+0.5+4.5*\rodsize+0.2,\ha-3*\inter) -- (\deb+0.5+5*\rodsize,\ha-3*\inter);
\draw[black,thin] (\deb+0.5+5*\rodsize,\ha-3*\inter) -- (\fin,\ha-3*\inter);

\draw[black,thin] (\deb,\ha-4*\inter) -- (\deb+0.5+4.5*\rodsize+0.2,\ha-4*\inter);\draw[black,thin,dotted] (\deb+0.5+4.5*\rodsize+0.2,\ha-4*\inter) -- (\deb+0.5+5*\rodsize,\ha-4*\inter);
\draw[black,thin] (\deb+0.5+5*\rodsize,\ha-4*\inter) -- (\fin,\ha-4*\inter);

\draw[black,thin] (\deb,\ha-5*\inter) -- (\deb+0.5+4.5*\rodsize+0.2,\ha-5*\inter);\draw[black,thin,dotted] (\deb+0.5+4.5*\rodsize+0.2,\ha-5*\inter) -- (\deb+0.5+5*\rodsize,\ha-5*\inter);
\draw[black,thin] (\deb+0.5+5*\rodsize,\ha-5*\inter) -- (\fin,\ha-5*\inter);

\draw[black,thin] (\deb,\ha-6*\inter) -- (\deb+0.5+4.5*\rodsize+0.2,\ha-6*\inter);\draw[black,thin,dotted] (\deb+0.5+4.5*\rodsize+0.2,\ha-6*\inter) -- (\deb+0.5+5*\rodsize,\ha-6*\inter);
\draw[black,thin] (\deb+0.5+5*\rodsize,\ha-6*\inter) -- (\fin,\ha-6*\inter);

\draw[black,thin] (\deb,\ha-7*\inter) -- (\deb+0.5+4.5*\rodsize+0.2,\ha-7*\inter);\draw[black,thin,dotted] (\deb+0.5+4.5*\rodsize+0.2,\ha-7*\inter) -- (\deb+0.5+5*\rodsize,\ha-7*\inter);
\draw[black,thin] (\deb+0.5+5*\rodsize,\ha-7*\inter) -- (\fin,\ha-7*\inter);

\draw[black,thin] (\deb,\ha-8*\inter) -- (\deb+0.5+4.5*\rodsize+0.2,\ha-8*\inter);\draw[black,thin,dotted] (\deb+0.5+4.5*\rodsize+0.2,\ha-8*\inter) -- (\deb+0.5+5*\rodsize,\ha-8*\inter);
\draw[black,thin] (\deb+0.5+5*\rodsize,\ha-8*\inter) -- (\fin,\ha-8*\inter);

\draw[black,line width=0.3mm] (\deb-0.4,\ha-\zaxisline*\inter) -- (\deb+0.5+4.5*\rodsize+0.2,\ha-\zaxisline*\inter);\draw[black,line width=0.3mm,dotted] (\deb+0.5+4.5*\rodsize+0.2,\ha-\zaxisline*\inter) -- (\deb+0.5+5*\rodsize,\ha-\zaxisline*\inter);
\draw[black,->, line width=0.3mm] (\deb+0.5+5*\rodsize,\ha-\zaxisline*\inter) -- (\fin+0.2,\ha-\zaxisline*\inter);


\draw[black, dotted, line width=1mm] (\deb,\ha) -- (\deb+0.5,\ha);
\draw[black,line width=1mm] (\deb+0.5,\ha) -- (\deb+0.5+\rodsize-0.3,\ha);
\draw[black,line width=1mm] (\fin-0.5-\rodsize+0.2,\ha-1*\inter) -- (\fin-0.55,\ha-1*\inter);
\draw[black, dotted,line width=1mm] (\fin-0.5,\ha-1*\inter) -- (\fin,\ha-1*\inter);


\draw[amazon,line width=1mm] (\deb+0.5+\rodsize-0.3,\ha-5*\inter) -- (\deb+0.5+1.5*\rodsize-0.3,\ha-5*\inter);

\draw[bitterlemon,line width=1mm] (\deb+0.5+1.5*\rodsize-0.3,\ha-6*\inter) -- (\deb+0.5+2*\rodsize-0.3,\ha-6*\inter);

\draw[amaranthred,line width=1mm] (\deb+0.5+2*\rodsize-0.3,\ha-3*\inter) -- (\deb+0.5+2.5*\rodsize-0.15,\ha-3*\inter);

\draw[turquoise,line width=1mm] (\deb+0.5+2.5*\rodsize-0.15,\ha-2*\inter) -- (\deb+0.5+3*\rodsize-0.15,\ha-2*\inter);

\draw[byzantine,line width=1mm] (\deb+0.5+3*\rodsize-0.15,\ha-4*\inter) -- (\deb+0.5+3.5*\rodsize-0.15,\ha-4*\inter);

\draw[brown,line width=1mm] (\deb+0.5+3.5*\rodsize-0.15,\ha-8*\inter) -- (\deb+0.5+4*\rodsize,\ha-8*\inter);

\draw[bdazzledblue,line width=1mm] (\deb+0.5+4*\rodsize,\ha-7*\inter) -- (\deb+0.5+4.5*\rodsize,\ha-7*\inter);

\draw[bitterlemon,line width=1mm] (\deb+0.5+5*\rodsize+0.2,\ha-6*\inter) -- (\deb+0.5+5.5*\rodsize+0.2,\ha-6*\inter);


\draw[amazon,line width=1mm,opacity=0.25] (\deb+0.5+\rodsize-0.3,\ha-\zaxisline*\inter) -- (\deb+0.5+1.5*\rodsize-0.3,\ha-\zaxisline*\inter);

\draw[bitterlemon,line width=1mm,opacity=0.25] (\deb+0.5+1.5*\rodsize-0.3,\ha-\zaxisline*\inter) -- (\deb+0.5+2*\rodsize-0.3,\ha-\zaxisline*\inter);
\draw[amaranthred,line width=1mm,opacity=0.25] (\deb+0.5+2*\rodsize-0.3,\ha-\zaxisline*\inter) -- (\deb+0.5+2.5*\rodsize-0.15,\ha-\zaxisline*\inter);

\draw[turquoise,line width=1mm,opacity=0.25] (\deb+0.5+2.5*\rodsize-0.15,\ha-\zaxisline*\inter) -- (\deb+0.5+3*\rodsize-0.15,\ha-\zaxisline*\inter);

\draw[byzantine,line width=1mm,opacity=0.25] (\deb+0.5+3*\rodsize-0.15,\ha-\zaxisline*\inter) -- (\deb+0.5+3.5*\rodsize-0.15,\ha-\zaxisline*\inter);

\draw[brown,line width=1mm,opacity=0.25] (\deb+0.5+3.5*\rodsize-0.15,\ha-\zaxisline*\inter) -- (\deb+0.5+4*\rodsize,\ha-\zaxisline*\inter);
\draw[bdazzledblue,line width=1mm,opacity=0.25] (\deb+0.5+4*\rodsize,\ha-\zaxisline*\inter) -- (\deb+0.5+4.5*\rodsize,\ha-\zaxisline*\inter);

\draw[bitterlemon,line width=1mm,opacity=0.25] (\deb+0.5+5*\rodsize+0.2,\ha-\zaxisline*\inter) -- (\deb+0.5+5.5*\rodsize+0.2,\ha-\zaxisline*\inter);


\draw[gray,dotted,line width=0.2mm] (\deb+0.5+\rodsize-0.3,\ha) -- (\deb+0.5+\rodsize-0.3,\ha-\zaxisline*\inter);
\draw[gray,dotted,line width=0.2mm] (\deb+0.5+1.5*\rodsize-0.3,\ha) -- (\deb+0.5+1.5*\rodsize-0.3,\ha-\zaxisline*\inter);
\draw[gray,dotted,line width=0.2mm] (\deb+0.5+2*\rodsize-0.3,\ha) -- (\deb+0.5+2*\rodsize-0.3,\ha-\zaxisline*\inter);
\draw[gray,dotted,line width=0.2mm] (\deb+0.5+2.5*\rodsize-0.15,\ha) -- (\deb+0.5+2.5*\rodsize-0.15,\ha-\zaxisline*\inter);
\draw[gray,dotted,line width=0.2mm] (\deb+0.5+3*\rodsize-0.15,\ha) -- (\deb+0.5+3*\rodsize-0.15,\ha-\zaxisline*\inter);
\draw[gray,dotted,line width=0.2mm] (\deb+0.5+3.5*\rodsize-0.15,\ha) -- (\deb+0.5+3.5*\rodsize-0.15,\ha-\zaxisline*\inter);
\draw[gray,dotted,line width=0.2mm] (\deb+0.5+4*\rodsize,\ha) -- (\deb+0.5+4*\rodsize,\ha-\zaxisline*\inter);
\draw[gray,dotted,line width=0.2mm] (\deb+0.5+4.5*\rodsize,\ha) -- (\deb+0.5+4.5*\rodsize,\ha-\zaxisline*\inter);
\draw[gray,dotted,line width=0.2mm] (\deb+0.5+5*\rodsize+0.2,\ha) -- (\deb+0.5+5*\rodsize+0.2,\ha-\zaxisline*\inter);
\draw[gray,dotted,line width=0.2mm] (\deb+0.5+5.5*\rodsize+0.2,\ha) -- (\deb+0.5+5.5*\rodsize+0.2,\ha-\zaxisline*\inter);

\draw[line width=0.3mm] (\deb+0.5+\rodsize-0.3,\ha-\zaxisline*\inter+0.1) -- (\deb+0.5+\rodsize-0.3,\ha-\zaxisline*\inter-0.1);
\draw[line width=0.3mm] (\deb+0.5+1.5*\rodsize-0.3,\ha-\zaxisline*\inter+0.1) -- (\deb+0.5+1.5*\rodsize-0.3,\ha-\zaxisline*\inter-0.1);
\draw[line width=0.3mm] (\deb+0.5+2*\rodsize-0.3,\ha-\zaxisline*\inter+0.1) -- (\deb+0.5+2*\rodsize-0.3,\ha-\zaxisline*\inter-0.1);
\draw[line width=0.3mm] (\deb+0.5+2.5*\rodsize-0.15,\ha-\zaxisline*\inter+0.1) -- (\deb+0.5+2.5*\rodsize-0.15,\ha-\zaxisline*\inter-0.1);
\draw[line width=0.3mm] (\deb+0.5+3*\rodsize-0.15,\ha-\zaxisline*\inter+0.1) -- (\deb+0.5+3*\rodsize-0.15,\ha-\zaxisline*\inter-0.1);
\draw[line width=0.3mm] (\deb+0.5+3.5*\rodsize-0.15,\ha-\zaxisline*\inter+0.1) -- (\deb+0.5+3.5*\rodsize-0.15,\ha-\zaxisline*\inter-0.1);
\draw[line width=0.3mm] (\deb+0.5+4*\rodsize,\ha-\zaxisline*\inter+0.1) -- (\deb+0.5+4*\rodsize,\ha-\zaxisline*\inter-0.1);
\draw[line width=0.3mm] (\deb+0.5+4.5*\rodsize,\ha-\zaxisline*\inter+0.1) -- (\deb+0.5+4.5*\rodsize,\ha-\zaxisline*\inter-0.1);
\draw[line width=0.3mm] (\deb+0.5+5*\rodsize+0.2,\ha-\zaxisline*\inter+0.1) -- (\deb+0.5+5*\rodsize+0.2,\ha-\zaxisline*\inter-0.1);
\draw[line width=0.3mm] (\deb+0.5+5.5*\rodsize+0.2,\ha-\zaxisline*\inter+0.1) -- (\deb+0.5+5.5*\rodsize+0.2,\ha-\zaxisline*\inter-0.1);

\draw (\deb+0.5+1*\rodsize-0.3,\ha-\zaxisline*\inter-0.5) node{{\small $0$}};
\draw (\deb+0.5+1.5*\rodsize-0.3,\ha-\zaxisline*\inter-0.5) node{{\small $\frac{\ell_1^2}{4}$}};
\draw (\deb+0.5+2*\rodsize-0.3,\ha-\zaxisline*\inter-0.5) node{{\small $\frac{\ell_1^2+\ell_2^2}{4}$}};

\draw (\deb+0.5+2.5*\rodsize-0.15,\ha-\zaxisline*\inter-0.5) node{{\tiny $\frac{\ell_1^2+\ell_{2}^2+\ell_{3}^2}{4}$}};
\draw[gray,->,line width=0.1mm] (\deb+0.5+3*\rodsize-0.15,\ha-\zaxisline*\inter-0.8) -- (\deb+0.5+3*\rodsize-0.15,\ha-\zaxisline*\inter-0.25);
\draw (\deb+0.5+3*\rodsize-0.15,\ha-\zaxisline*\inter-1.1) node{{\tiny $\frac{\sum_{i=1}^4\ell_{i}^2}{4}$}};

\draw[black,line width=0.3mm,dotted,double] (\deb+0.5+3.25*\rodsize,\ha-\zaxisline*\inter-0.5) -- (\deb+0.5+4.5*\rodsize,\ha-\zaxisline*\inter-0.5);

\draw (\deb+0.5+5*\rodsize+0.2,\ha-\zaxisline*\inter-0.5) node{{\tiny $\frac{\sum_{i=1}^{n-1}\ell_{i}^2}{4}$}};
\draw[gray,->,line width=0.1mm] (\deb+0.5+5.5*\rodsize+0.2+0.05,\ha-\zaxisline*\inter-0.8) -- (\deb+0.5+5.5*\rodsize+0.2+0.05,\ha-\zaxisline*\inter-0.25);
\draw (\deb+0.5+5.5*\rodsize+0.2+0.05,\ha-\zaxisline*\inter-1.1) node{{\tiny $\frac{\sum_{i=1}^n\ell_{i}^2}{4}=\frac{\ell^2}{4}$}};

\end{tikzpicture}
\caption{Rod diagram of the shrinking directions on the $z$-axis after sourcing the solutions with $n$ connected rods that force the degeneracy of either a T$^6$ direction or the Hopf angle of the S$^3$, $\psi=k(\varphi_1-\varphi_2)$.}
\label{fig:rodsourceAdS2+T6+S3s}
\end{figure}

We now consider the most generic configurations of $n$ connected rod sources,   forcing either the degeneracy of a T$^6$ direction or the degeneracy of the Hopf angle of the S$^3$,  $\psi$ \eqref{eq:DefHyperspher}.  We have depicted a generic configuration in Fig.\ref{fig:rodsourceAdS2+T6+S3s} where we have summarized the conventions on the $z$-axis.

In addition to the six sets of labels,  $U_{y_i}$ $i=1,\ldots,6$,  which indicate the rods that force the degeneracy of a T$^6$ direction we define similarly $U_\psi$ for the Hopf angle as defined in \eqref{eq:DefUx}.  Moreover,  we also consider that the S$^3$ has conical defect asymptotically,  $k\neq 1$ which will be necessary for having smooth geometries.

We derive the fields from the linear branch of solutions \eqref{eq:LinearAdS2} and use the identities \eqref{eq:SimplRelations2} to simplify their form.   We find that the metric and gauge field in M-theory are generically given by\footnote{The metric in the Weyl cylindrical coordinate system is obtained by replacing $$ \frac{dr^2}{r^2+\ell^2}+d\theta^2= \frac{4}{\left( r^2+\ell^2\cos^2\theta\right)\left( r^2+\ell^2\sin^2\theta\right)} \left(d\rho^2+dz^2\right).$$}
\begin{align}
ds_{11}^2  \= & \frac{1}{\cZ^2} \,\left[  -\frac{(r^2+\ell^2)^2}{Q^2}\,dt^2+ \frac{kQ \cZ^3\cF\cH}{r^2+\ell^2}\, dr^2 \right]+Q\, \cZ\,\sum_{I=1}^3\left( \frac{\cK_{y_{2I-1}} \,dy_{2I-1}^2+\cK_{y_{2I}}\, dy_{2I}^2}{Q_I\,\cZ_I}\right)\nn \\
&\+ kQ \cZ\cH \left[\cF\, d\theta^2+\frac{\cK_\psi}{4k^2 \cH^2} \,(d\psi+k\,A \,d\phi)^2 + \cos^2 \theta \sin^2 \theta\, d\phi^2 \right]    \,, \label{eq:metAdS2+T6sSpecial}\\ 
A_3 \= &-\sum_{I=1}^3 \frac{\cK_{y_{2I-1}} \cK_{y_{2I}}\,(r^2+\ell^2)}{Q_I\,\cZ_I} \,dt\wedge dy_{2I-1} \wedge dy_{2I}\,, \nn 
\end{align}
where we have introduced,  in addition to the deformation factors \eqref{eq:DefDefFactorsGen},
\begin{align}
\cK_{\psi} &\equi \prod_{i\in U_{\psi}} \frac{r_i^2}{r_i^2+\ell_i^2}\,,\qquad A \equi \frac{1}{\ell^2+\sum_{i\in U_\psi} \ell_i^2}\left(\ell^2 \,\cos 2\theta +\sum_{i\in U_\psi} \ell_i^2 \cos 2\theta_i \right)\,,\nn\\
\cH &\equi  \frac{r^2}{\ell^2+\sum_{i\in U_\psi} \ell_i^2} \left(1+\frac{\ell^2}{r^2} - \prod_{i\in U_\psi} \left(1+\frac{\ell_i^2}{r_i^2} \right)^{-1}\right)\,.
\end{align}

We retrieve the previous family of solutions by taking $U_\psi = \emptyset$ and $k=1$.  Moreover,  since all deformation factors go to $1$ and $A \to \cos 2\theta$ at large distance.  The solutions are indeed asymptotic to AdS$_2\times$S$^{\,3}/\mathbb{Z}_k\times$T$^{\,6}$.

The regularity and topology analysis is relatively similar to the one in section \ref{sec:RegGen}.  The spacetimes are regular and have several coordinate degeneracies that must be regularized.  

First,  at the poles of the S$^3$,  $\theta=0$ and $\pi/2$ for $r>0$,  one has $A=\pm 1$ and $\cF=1$,  so the hyperspherical angles,  $\varphi_1$ and $\varphi_2$ \eqref{eq:DefHyperspher}, degenerate respectively with the same orbifold defect as the one imposes asymptotically.  

\begin{figure}[ht]
\centering
\includegraphics[width=0.95 \columnwidth]{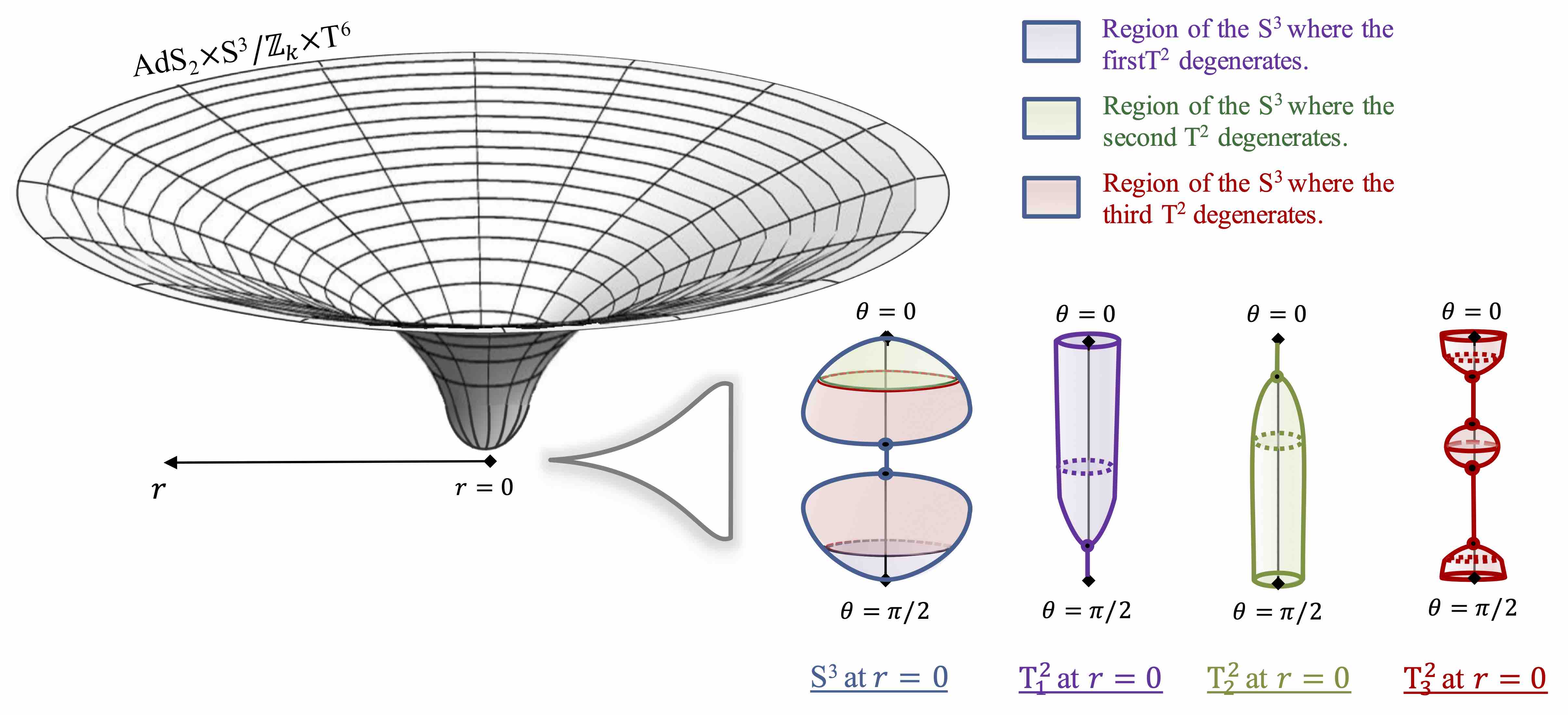}
\caption{Schematic description of the spacetime induced by an arbitrary number of connected rods, each one inducing the degeneracy of either a T$^6$ coordinate or the Hopf angle of the S$^3$.  On the left hand-side,  we depict the overall geometry in terms of the radius $r$.  On the right hand-side,  we describe the behavior of the S$^3$ and the three T$^2$ inside T$^6$ at $r=0$ and as a function of $\theta$,  the S$^3$ coordinate.  The spacetime ends smoothly at $r=0$ as a chain of $n$ smooth bolts.}
\label{fig:AdS2+T6+S3spic}
\end{figure}

Second,  the chain of rods is localized at $r=0$ where the spacetime ends, and it is  divided in section of $\theta$ as in \eqref{eq:ThetaSection}.  They form a chain of bolts,  that are origins of $\IR^2$,  but now the shrinking direction can be played by $\psi$.  The presence of those types of bolts in the chain modifies all regularity conditions.  We find that the bolts are regular if one imposes the following $n$ algebraic constraints:
\begin{equation}
\begin{split}
R_{y_{2I-1}} & \= \sqrt{\frac{\ell^2\,q_{M2\,\,i}^{(I)}}{\ell^2+\sum_{p\in U_\psi}\ell_p^2}}\,d_i\,,\qquad \text{if }i\in U_{2I-1}\,,\quad I=1,2,3\,,\\
R_{y_{2I}} &\= \sqrt{\frac{\ell^2\,q_{M2\,\,i}^{(I)}}{\ell^2+\sum_{p\in U_\psi}\ell_p^2}}\,d_i\,,\qquad \text{if }i\in U_{2I}\,,\quad I=1,2,3\,, \\
\frac{1}{k} & \= \frac{\ell \ell_i \,d_i}{\ell^2+\sum_{p\in U_\psi}\ell_p^2} ,\qquad \text{if }i\in U_{\psi}\,,
\end{split}
\label{eq:BuEqT6+S3s}
\end{equation}
where $d_i$ and $q_{M2\,\,i}$, the local M2 brane charge at the rod,  are still given by \eqref{eq:DefdiAspect} and \eqref{eq:localcharges}.  We can argue,  as in section \ref{sec:SUSYBreaking}, that the solutions break supersymmetry,  and correspond to smooth non-BPS excitations of empty AdS$_2$ in M-theory.  Finally,  we have depicted a generic smooth asymptotically-AdS$_2$ geometry obtained from this family of solutions in Fig.\ref{fig:AdS2+T6+S3spic}.

\subsection{Energy of the solutions} 
\label{sub:what_energy_for_these_solutions_}

In this section, we discuss the energy of the solutions constructed.  There is an ongoing debate about whether AdS$_2$ admits finite-energy excitations, which is one of the puzzles that makes the AdS$_2$/CFT$_1$ holography special compared to its higher-dimensional counterpart.  

Several computations point to the impossibility of such excitations. First, it has been shown in \cite{Maldacena:1998uz,Almheiri:2014cka} that the AdS$_2$ vacuum cannot support boundary-preserving dilatonic excitations at a perturbative level.  In the two-dimensional Einstein-Maxwell-Dilaton theory, it was proven using holographic renormalization that finite-energy excitations can not exist: the geometries with constant dilaton -- hence with fixed AdS$_2$ asymptotics, necessarily have zero energy \cite{Castro:2008ms,Cvetic:2016eiv}. These arguments also find resonance when studying the dual CFT$_1$, where it has been shown that if the ground states preserve conformal invariance, then the CFT is topological and there can be no time dependence: all states have zero energy.

However, other scenarios may emerge when considering all degrees of freedom in string theory.  First, in \cite{Chamon:2011xk,Jackiw:2012ur,Bena:2018bbd}, it has been argued that the CFT$_1$ dual to AdS$_2$ can be dynamical if the ground states break conformal invariance.  In the bulk, there exist geometries realized in string theory with a smooth cap at a certain scale in the AdS$_2$ throat that can support time-dependent dilatonic excitations at a perturbative level \cite{Bena:2018bbd}.  The geometries that we constructed in this paper can be viewed as backreacted realizations of these excitations, proving the existence of non-supersymmetric solitonic excitation in AdS$_2$ and giving more weight to the scenario of a dynamical CFT dual.  Note that they are much more involved than the dilatonic perturbations in AdS$_2$ considered in \cite{Maldacena:1998uz,Almheiri:2014cka,Bena:2018bbd} since they require various other fields in M-theory to be regularized.

Nevertheless,  an important question remains about the energy of these solitons in AdS$_2$.  According to the holographic renormalization computations of \cite{Cvetic:2016eiv},  excitations in AdS$_2$  have zero energy, and this could apply to the solutions of this paper despite the large variety of solitonic excitations highlighted in this paper.
Note however that the context of \cite{Cvetic:2016eiv} is different from the one in which we construct our solutions. Indeed,  these results require a specific AdS$_2$ dilaton gravity model where the dilaton is either running or constant.  These dilaton profiles arise from a specific compactification of the four-dimensional STU model that can also be uplifted in M-theory.  Clearly,  our solutions will have a much more complex dilaton profile and field contents  when reduced to two dimensions than those of \cite{Cvetic:2016eiv}, and this can greatly change the holographic renormalization procedure and the final result.  Indeed, if the dilaton would be constant asymptotically,  it will blow up in the IR, as a consequence of the coordinate singularities in higher dimensions.

To derive rigorously the energy of the solutions,  one should apply a strategy similar to \cite{Castro:2008ms,Cvetic:2016eiv}.  First,  one should compute the bulk action in the M-theory frame where the solutions are regular\footnote{The Chern-Simons contribution has been omitted since it is trivially zero for all solutions constructed in this paper.}
\begin{equation}
I_\text{bulk} = \frac{1}{16 \pi G_{11}} \int_\cM d^{11} x \sqrt{-g} \left[ R -\frac{1}{2} |F_4|^2  \right],
\end{equation}
The total action requires two extra terms defined at the boundary of $\cM$,  $\partial \cM$,  since the spacetimes are non-compact: a Gibbons-Hawking-York term for a consistent variational principle of the metric \cite{Gibbons:1976ue,York:1972sj}, as well as counter terms for the matter fields. The total action is thus $I_\text{tot} \equiv  I_\text{bulk} + I_\text{GHY}+ I_\text{c.t.}$, with
\begin{equation}
I_\text{GHY}= \frac{1}{8 \pi G_{11}} \int_{\partial \cM} d^{11} x \sqrt{-h}\,  K \,,\qquad I_\text{c.t.}= \frac{1}{8 \pi G_{11}} \int_{\partial \cM} d^{11} x \sqrt{-h} \, \cL_\text{c.t.}\,,
\end{equation}
where $h$ is the induced metric on $\partial M$ and $K$ is the extrinsic curvature.  To derive the energy of an on-shell solution,  one can compute the free energy,  $F$,  which is proportional to the Euclidean action on-shell,  $I_\text{tot}^{(E)} = - I_\text{tot}$:
\begin{equation}
\beta F \= I_\text{tot}^{(E)} \= \beta \, E -S\,.
\end{equation}
In this equation, $\beta$ is the periodicity of the thermal circle, $E$ is the energy and $S$ the entropy of the solution.  Because we are dealing with specific states,  the entropy vanishes $S=0$, and the Euclidean action is directly related to the energy.   

The main challenge in deriving the on-shell Euclidean action lies in the counterterms, $I_\text{c.t.}$.  Indeed, the bulk part and the Gibbons-Hawking-York term have an explicit formula and can be computed directly.  Deriving the counter-terms of the two-dimensional fields is much more involved and is beyond the scope of this article.  Indeed, it is first necessary to perform a holographic renormalization for each electric components in $F_4$ to have a consistent variational principle and resolve UV divergences. This can be done in the same way as in \cite{Castro:2008ms,Cvetic:2016eiv}, but the field content here is richer than a single $U(1)$ gauge field. We thus postpone this computation to a future project.

In this section,  we simply give heuristic arguments suggesting that the solutions have non-zero energy and leave the complete derivation of the energy for a future project:
\begin{itemize}
\item[•] First, the integrable structure in AdS$_2$ highlighted in this paper allows  an immense variety of solutions that preserve the same asymptotics and have the same conserved charges.  We have found that the difference of the action minus the counterterms between two such solutions,  $\Delta \widetilde{I} = (I_\text{bulk}+I_\text{GHY})_{\text{sol}_1} - (I_\text{bulk}+I_\text{GHY})_{\text{sol}_2}$,  is generically non-zero and depends non-trivially on the internal parameters forming the IR bubbling topology.  A priori,  it is possible that this difference can be balanced by the counterterm contribution so that $\Delta I_\text{tot}= \Delta \widetilde{I} + (I_\text{c.t.})_{\text{sol}_1} - (I_\text{c.t.})_{\text{sol}_2}=0$,  indicating that all states have the same energy.  However,  the counterterms should depend only on the three-form field strength $A_3$ at the boundary as in \cite{Castro:2008ms}.  In section \ref{sec:generic_deformations},  we have seen that one can add S$^3$ deformations that do not affect the gauge field \eqref{eq:metAdS2+T6sSpecial}.  Thus,  suppose we consider two solutions given by \eqref{eq:metAdS2+T6sSpecial} with the same T$^2$ deformations, but the first has no S$^3$ deformations,  $\cK_\psi =1$ and $U_\psi = \emptyset$,  while the second has some,  $\cK_\psi \neq 1$.  Both solutions have the same three-form gauge field so that one should expect $(I_\text{c.t.})_{\text{sol}_1} = (I_\text{c.t.})_{\text{sol}_2}$, and thus generically their total Euclidean action will differ.  Therefore,  they are expected to have two different values of energy.  This argument must of course be confirmed by an explicit derivation of $I_\text{c.t.}$.  
\item[•] In this paper,  we have been interested in solutions that are entirely smooth with rod sources that are connected to each other.  However,  as pointed out in \cite{Bah:2021owp},  when two of the rod sources are disconnected,  a string with negative tension is needed to separate them, manifesting in the geometry as a conical excess.  The so-called ``strut'' carries negative energy density \cite{Costa:2000kf}, which captures the binding energy of the chain of bubbles that hold them together from gravitational attraction. The energy of the struts for the AdS$_2$ solutions constructed in this paper can be locally derived following  \cite{Costa:2000kf,Bah:2021owp}. Having a bound state that develops struts when the individual constituents are separated is an indirect but strong argument that they have non-zero energy: if that was not the case, they would not need these struts to be held apart.
\item[•] Despite the unknown counter-terms,  the renormalized holographic stress tensor, and hence the energy of the solutions,  is expected to be read from the first terms of Fefferman-Graham expansion of the metric \cite{Castro:2008ms}.  More precisely, the AdS$_2$ metric should take the asymptotic form:
  \begin{equation}
    ds_2^2 ~=~ \frac{dz^2}{z^2} + h_0 \big(z^2 + h_1 + \mathcal{O}(z^{-2})\big) dt^2 \,,
  \end{equation}
  where $h_0$ and $h_1$ can be generically functions of $t$, but they are constants for our time-independent solutions. The constant $h_0$ controls the radius of the AdS space, and it will be fixed by the choice of charges.  On the other hand,  the constant $h_1$  is expected to encode the energy of the solutions \cite{Castro:2008ms}. Thus, finding solutions with different constants $h_1$ should indicate that the energy of these solutions differs.  \\
  The asymptotic analysis of our solutions do not only depend on the radius $r$, but also on $\theta$,  the angular variable on the sphere. As such,  their two-dimensional reduction is not trivial, and it is not clear whether they can be embedded in a consistent truncation. Regardless, by making an appropriate change of coordinates, we can expand asymptotically the five-dimensional metric \eqref{eq:ds5met},  and ensure that the first terms of the expansion are independent of $\theta$.  For the three-bubble solutions of section \ref{sec:3rod} and a four-bubble extension as given in section \ref{sec:ArbitraryT2s},  this constrains the sizes of the rods so that
  \begin{gather}
    \ell_1 = \ell_3 = 2^{-1/4} \ell_2 \qq{for the 3-rods solution,} \nn \\
    \ell_1^2 =  \ell_3^2 + \ell_4^2 \,,\ \ell_2^4 = 2 \ell_3^4 + 3 \ell_3^2 \ell_4^2 + 2 \ell_4^4 \qq{for the 4-rods solution.}
   \end{gather}
   Comparing their asymptotic expansions,  we find that $h_0 = -16 (Q_1 Q_2 Q_3)^{-2/3}$ for both solutions,  but the $h_1$ differ:
   \begin{equation}
     h_1 = \left\{
     \begin{aligned}
       &\frac{\ell_3^4}{64}  &\qq{for the 3-rods solution,} \\
       &\frac{2 \ell_3^4 + 2 \ell_3^2 \ell_4^2  + 2 \ell_4^4}{128} &\qq{for the 4-rods solution.}
     \end{aligned}
     \right.
   \end{equation}
   Assuming that the energy of these solutions is a function of $h_1$ \cite{Castro:2008ms}, this is yet another indication that they correspond to finite-energy excitations.
\end{itemize}

These three arguments suggest that the non-supersymmetric solutions constructed in this paper are regular finite-energy excitations of AdS$_2$.  A final proof must be performed by an explicit derivation of the energy following the procedure described above.

\section{Conclusion}
\label{sec:conclusion}

\subsection*{Summary}

The main result of this paper is the construction of a large class of new smooth solitonic excitations of AdS$_2$ in M-theory. It has been previously assumed that such excitations cannot exist, based on the proof of their non-existence in simple models of gravity in AdS$_2$ \cite{Maldacena:1998uz,Almheiri:2014cka}. The solutions presented in this paper rely crucially on additional string-theory degrees of freedom, avoiding the previous pitfalls. Indeed, starting with eleven-dimensional supergravity, we have used geometric and topological transitions of the compact internal space,  S$^3 \times$T$^6$, and M2-brane flux in the IR,  to generate a smooth non-supersymmetric cap deep inside AdS$_2$, while preserving the asymptotics.

This work uses the techniques developed in \cite{Heidmann:2021cms} to build non-supersymmetric M-theory solutions that are sourced by regular bolts,  and fits in the larger framework of the construction of non-supersymmetric asymptotically-AdS$_D$ solitons pioneered in \cite{Bah:2022pdn}. We first constructed explicitly and analyzed in details the simplest set of solutions, consisting of three rods. We proved that these solutions break supersymmetry,  give a large variety of smooth solitonic excitations in the IR and that they exist in a specific corner of the moduli space of brane charges in M-theory.  Moreover,  we derived more sophisticated configurations with additional rod sources and showed that generic solitons have very little constraints on the moduli space,  and form an incredibly large family of smooth non-supersymmetric solutions in AdS$_2$ from string theory.

\subsection*{Discussion}

Our results open several directions for future work. First, it would be very interesting to develop a holographic dictionary with our solutions.  Applying the usual AdS/CFT philosophy, the geometries we have constructed are dual to non-supersymmetric states of a CFT$_1$.   As mentioned in the introduction, the essence of AdS$_2$/CFT$_1$ duality is not yet known.  Our work provides examples of fully backreacted non-supersymmetric smooth bulk geometries that are asymptotically AdS$_2$ in string theory, that do not require running dilatons in lower dimensions, and, because they terminate smoothly in the IR, they have only one AdS boundary.

As a first step,  we would like to analyze the description of our geometries when reduced to two dimensions.  Ultimately,  we would like to derive an effective two-dimensional action that contains all fields required by our solitons and to compare the matter contents with respect to Jackiw-Teitelboim gravity.

A major question, that has been partially addressed in section \ref{sub:what_energy_for_these_solutions_},  is whether the solutions that we have built in supergravity have non-zero energy and therefore correspond to finite-energy excitations in the CFT$_1$. In \cite{Cvetic:2016eiv}, it was argued using holographic computations in an Einstein-Maxwell-Dilaton model of two-dimensional gravity that such excitations cannot exist. However, this model does not include the essential degrees of freedom of string theory that are necessary for our solutions,  and so the result may be different. This has  been mentioned in \cite{Bena:2018bbd}, where the authors argued that such excitations must exist. The discussion was based on the conjecture that the backreaction of certain non-supersymmetric perturbations would not break the AdS$_2$ asymptotics. The fact that we are now able to construct fully-backreacted non-supersymmetric solutions preserving the asymptotics certainly gives weight to this argument.  Arguments in section \ref{sub:what_energy_for_these_solutions_}, suggest that the solitons constructed in this paper have finite energy.  However,  one should derive exactly these energies by performing holographic renormalization and computing the holographic stress tensor in M-theory to give a definitive answer to this question.  This will be subject of a future project.

In any way, these solutions can serve as bulk data points to obtain an understanding of a potential dual CFT$_1$. Using standard holographic techniques to estimate two-point functions in our geometries, one could for example compute correlators in the CFT, similarly to what has been done in higher dimensions \cite{Kanitscheider:2006zf,Taylor:2007hs,Kanitscheider:2007wq,Giusto2015,Giusto:2019qig,Rawash:2021pik}.

We expect the precise matching of our geometries with CFT states to be challenging. A mistake would be to try to treat these geometries as small non-supersymmetric perturbations on top of a supersymmetric background.  As we saw in section \ref{sec:SUSYBreaking}, the moduli space of our geometries do possess a specific supersymmetric limit, when the sizes of all the rods are identically zero: empty AdS$_2$. However, because of the conformal nature of AdS$_2$, all the solutions are invariant under a global rescaling of the rod sizes,  and it is not possible to give a precise idea of an ``expansion in small rod sizes'': this limit is pathological and our solitons emerge from inherently non-perturbative back-reactions.

Another interesting question concerns the stability. Their construction heavily relies on a large number of spacetime symmetries and Kaluza-Klein bubbles.  On general grounds, we expect the solutions to be generically unstable under perturbations that break one of these symmetries.  Moreover,  vacuum Kaluza-Klein bubbles in flat space suffered from an instability that make them eat up the full spacetime \cite{Witten:1981gj}.  However,  it has been shown that they become meta-stable gravitational states when electromagnetic fluxes are added \cite{Bah:2021irr}.  Moreover,  the AdS asymptotics drastically change the dynamics since it has a ``box'' effect that can prevent from instability.  It would be interesting to study the stability properties of our solutions,  elucidate in which channel they can decay,  and if their radiation can be related to Hawking radiation of near-extremal black holes with a similar derivation as in \cite{Chowdhury:2007jx,Bena:2019azk,Bena:2020yii}. This will make them good candidates for describing coherent microstates of such black holes.

\section*{Acknowledgments}
We are grateful to Iosif Bena,  Ibou Bah,  and Nick Warner for interesting and stimulating discussions.  The work of PH is supported by NSF grant PHY-2112699.  The work of AH is supported in part by the ERC Grant 787320 - QBH Structure.

\appendix

\section{Integrable structure of the M2-M2-M2 system}
\label{App:EOM}

We consider M-theory solutions that depend on two variables, and are static and axially-symmetric. In the Weyl formalism, we can freely choose these coordinates, denoted as $(\rho,z)$, such that the induced metric on the two-dimensional space is conformally flat and that the induced metric on the remaining nine-dimensional spacetime satisfies $\det h \= -\rho^2$, where $h$ is the induced metric \cite{Weyl:book,Emparan:2001wk}. Moreover, one can consider one of the U(1) isometry, denoted as $\phi$, to have a metric coefficient proportional to $\rho^2$ such that the $(\rho,z,\phi)$ space defines a three-dimensional base in the Weyl cylindrical coordinate system, and $z$ plays the role of the axis of symmetry \cite{Emparan:2001wk,Bah:2020pdz,Heidmann:2021cms}.

The solutions are constructed on T$^6$ and are supported by M2-M2-M2 flux.  The three stacks of M2 branes are wrapping three orthogonal 2-tori inside the T$^6$ that we parametrized by $(y_1,y_2,y_3,y_4,y_5,y_6)$.  Finally,  we consider the remaining S$^1$, parametrized by an angle $\psi$,  as a Hopf fibration over the $(\rho,z,\phi)$ base with a KK vector along $\phi$.

The ansatz of metric and fields that suits the spacetime symmetries and flux is given by \eqref{eq:MtheoryAnsatz}.  We refer the reader interested in the derivation of this ansatz and the equations of motion to \cite{Heidmann:2021cms}.  We introduce the cylindrical Laplacian operator of a flat three-dimensional base for axisymmetric functions:
\begin{equation}
\Delta \equi \frac{1}{\rho}\,\partial_\rho \left( \rho \,\partial_\rho \right) \+ \partial_z^2\,.
\label{eq:Laplacian}
\end{equation}
The Einstein-Maxwell equations can be written down in a uniform way if one defines an electric dual of the magnetic KK gauge potential and decompose $\nu$ such that 
\begin{equation}
dT_0 \equi \frac{1}{\rho Z_0^2} \,\star_2 dH_0 \,, \qquad \nu =\sum_{\Lambda=0}^3( \nu_{Z_\Lambda}+ \nu_{W_\Lambda})\,,
\label{eq:ElecDual}
\end{equation}
where $\star_2$ is the Hodge star operator in the $(\rho,z)$ flat space and $\nu_X$ are the individual contributions of each warp factor in $\nu$. The equations of motion decompose into 8 sectors:
\begin{itemize}
\item[•] \underline{Four vacuum sectors for $ \Lambda=0,1,2,3$:}
\begin{align}
&\Delta \log W_\Lambda  \= 0\,,\label{eq:EOMVac}\\
&\frac{2}{\rho}\, \partial_z \nu_{W_\Lambda} \= \partial_\rho \log W_\Lambda \,\partial_z \log W_\Lambda\,, \quad \frac{4}{\rho}\, \partial_\rho \nu_{W_\Lambda} \= \left( \partial_\rho \log W_\Lambda\right)^2 \- \left(\partial_z \log W_\Lambda\right)^2. \nn
\end{align}
\item[•] \underline{Four Maxwell sectors for  $ \Lambda=0,1,2,3$:}
\begin{align}
& \Delta \log Z_\Lambda \= - Z_\Lambda^2  \left[ (\partial_\rho T_\Lambda)^2 + (\partial_z T_\Lambda)^2 \right]\,, \quad  \partial_\rho \left(\rho Z_\Lambda^2 \,\partial_\rho T_\Lambda \right)\+\partial_z \left( \rho Z_\Lambda^2 \,\partial_z T_\Lambda\right)  \=0 \,,\nn \\
&\frac{2}{\rho}\, \partial_z \nu_{Z_\Lambda} \= \partial_\rho \log Z_\Lambda \,\partial_z \log Z_\Lambda - Z_\Lambda^2 \,\partial_\rho  T_\Lambda \partial_z  T_\Lambda \,, \label{eq:EOMMaxwell}\\
& \frac{4}{\rho}\, \partial_\rho \nu_{Z_\Lambda} \=\left( \partial_\rho \log Z_\Lambda \right)^2 \- \left(\partial_z \log Z_\Lambda\right)^2 - Z_\Lambda^2 \, \left((\partial_\rho T_\Lambda)^2-(\partial_z T_\Lambda)^2 \right)\,. \nn
\end{align}
\end{itemize}
The equations for the T$^6$ deformations,  $W_\Lambda$, and their associated $\nu_W$ is a linear system of equations. They are identical to equations obtained for vacuum Weyl solutions \cite{Weyl:book,Emparan:2001wk}. The logarithms of $W_\Lambda$ are harmonic functions for which solutions sourced by segments, i.e. rods, or points on the $z$-axis are explicitly known.

The equations in the Maxwell sectors are coupled non-linear equations.  However, they are identical to equations one obtains for static axially-symmetric geometries in four dimensions.  More precisely,  four-dimensional solutions given by
\begin{equation}
ds_4^2 \= - \frac{dt^2}{Z_\Lambda^2} + Z_\Lambda^2 \left[ e^{8 \nu_{Z_\Lambda}} \left( d\rho^2 +dz^2 \right) +\rho^2 d\phi^2 \right]\,,\qquad F = -2\, dT_\Lambda \wedge dt\,, \label{eq:4dsys}
\end{equation}
leads to the exact same equations as for the three Maxwell sectors \eqref{eq:EOMMaxwell}.\footnote{We have considered the four-dimensional Einstein-Maxwell action $$(16 \pi G_4)\,S_4 =\int d^4x \sqrt{g} \left(R - \frac{1}{4} F_{\mu\nu}F^{\mu\nu} \right).$$}

As explained in \cite{Heidmann:2021cms,Bah:2022pdn},  the equations of motion of this system admit integrable structures that are well established from the Ernst formalism and inverse scattering.  These integrable structures follow the fact that the ansatz in \eqref{eq:4dsys} admit an action from the Geroch group \cite{Geroch:1970nt,Geroch:1972yt}.  Our general system above will inherit all of these structures that can allow for a large phase of solutions.  Indeed, monodromy methods and B\"acklund transformations can be used to extract solutions \cite{Belinsky:1979mh,PhysRevLett.41.1197,Alekseev:1999kj,Alekseev:1999bv,Stephani:2003tm}.  In particular they can be used as solution generating methods for non-BPS AdS solutions \cite{Bah:2022pdn}.  

In this paper,  we focus on a specific linear class of solutions of the Maxwell system which can be obtained from the integrable structure of the Ernst formalism.\footnote{See section 18.6.3 of \cite{Stephani:2003tm}.} The gravitational potential of the spacetime in \eqref{eq:4dsys} is the redshift factor $Z_\Lambda$.  We can consider an ansatz where the electric potential is a function of the gravitational potential  $T_\Lambda(Z_\Lambda)$.  By plugging into the equations of motion of $(Z_\Lambda,T_\Lambda)$ in \eqref{eq:EOMMaxwell},  we found that both potentials are expressed in terms of a function for which the logarithm is harmonic and three complex constants:
\begin{equation}\label{eq:soluX}
Z \= \frac{e^{b} L - e^{-b}L^{-1}}{2 a}\,,\qquad T \= \frac{\sqrt{1+a^2 Z^2}}{Z}+c\,, \qquad \Delta \log L \=0
\end{equation} where $(a,b,c) \in \mathbb{C}$ and we have dropped the $\Lambda$ index for clarity.   

The key ingredient is the potential $L$ for which the logarithm satisfies the three-dimensional axially-symmetric Laplace equation.  Arbitrary solutions can be obtained by considering arbitrary sources to this linear equation in a similar fashion as for vacuum Weyl solutions \cite{Weyl:book,Emparan:2001wk},  but with now non-trivial electromagnetic flux turned on.  This is the reason why this branch of solutions has been denoted as ``the charged Weyl formalism'' in \cite{Bah:2020ogh,Bah:2020pdz,Bah:2021owp,Heidmann:2021cms,Bah:2021rki}. This is an explicit realization of the integrable structure which exists for the four-dimensional system in \eqref{eq:4dsys},  and thus inherited by the M2-M2-M2 system in \eqref{eq:MtheoryAnsatz}. 

Note that the electric potential $T$ does not simply reduce to the BPS branch where it counterbalances the gravitational potential $T=\frac{1}{Z}$. Thus, the structure in \eqref{eq:soluX} can be taken as a non-BPS but still linear generalization of BPS multicenter solutions \cite{Gauntlett:2002nw,Bena:2005va,Bena:2007kg,Heidmann:2017cxt,Bena:2017fvm,Bena:2018bbd,Heidmann:2018vky}.  This has composed much of the recent progress in constructing asymptotically-flat non-BPS smooth horizonless solutions in \cite{Bah:2020ogh,Bah:2020pdz,Bah:2021owp,Heidmann:2021cms,Bah:2021rki,Bah:2022pdn}, while \cite{Bah:2022yji} exploits inverse scattering methods.

\section{Regularity conditions for a large number of bolts} 
\label{app:large_n}

In this Appendix we detail the computation of the regularity conditions of section \ref{sec:large_n}. We recall that we consider solutions induced by $n = 6\hat n$ rods, ordered in repeating pattern of six different rods (see Fig.\ref{fig:manyrodsources} for $\hat n=2$)$$U_{y_1} \= \{ 1, \,7, \dots, \,6\hat n-5\} \,,\quad U_{y_2} \= \{ 2,\,8,\dots,\,6\hat n-4\}\,,\ \ldots \ \,,\quad  U_{y_6} \= \{ 6,\,12,\dots,\,6\hat n\}\,.$$
For simplicity, we only consider solutions where all the global charges and all the radii are equal: $Q_I \equiv Q$, $R_{y_i} \equiv R_y$. We assume that the sizes of the rods are all equal $\ell_j \equiv \bar{\ell}$, which is confirmed to be true numerically away from the extremities of the configuration (when $1 \ll j \ll n$).

To solve the regularity conditions, one first needs to determine the large $n$ (or equivalently large $\hat n$) limit of the aspect ratios $d_i$ \eqref{eq:DefdiAspect}. We recall its expression and decompose it in three terms:
\begin{equation}
\begin{aligned}
d_i &\equi d_i^{(1)}\, d_i^{(2)} \, d_i^{(3)} \,,\qquad\quad
 d_i^{(1)} \equi \prod_{p=1}^{i-1} \prod_{q=i+1}^n \left[\frac{1+ \frac{\ell_q^2}{\sum_{r=p}^{q-1} \ell_r^2}}{1+ \frac{\ell_q^2}{\sum_{r=p+1}^{q-1} \ell_r^2}} \right]^{\frac{\bar{\alpha}_{pq}}{2}} \,,\\
 d_i^{(2)} &\equi \prod_{p=1}^{i-1}\left(1+ \frac{\ell_p^2}{\sum_{q=p+1}^{i} \ell_q^2} \right)^{\frac{\bar{\alpha}_{ip}}{2}} \,,\qquad
 d_i^{(3)} \equi \prod_{p=i+1}^{n}\left(1+ \frac{\ell_p^2}{\sum_{q=i}^{p-1} \ell_q^2} \right)^{\frac{\bar{\alpha}_{ip}}{2}}\,. \
\end{aligned}
\end{equation}

Under the assumption that all rods sizes are equal, one can compute the second term exactly:
\begin{equation}
 d_i^{(2)} ~=~ \prod_{p=1}^{\lfloor i/6 \rfloor}\left(1+ \frac{1}{6p} \right)^{1/2}~=~ \qty(\frac{\Gamma(7/6 + \lfloor i/6 \rfloor)}{\Gamma(7/6)\Gamma(1 + \lfloor i/6 \rfloor)})^{1/2} \,,
\end{equation}
which in the large $n$ limit becomes
\begin{equation}
  d_i^{(2)} ~\sim~ \frac{\qty(x\hat n)^{1/12}}{\sqrt{\Gamma(7/6)}} \,, \qquad x \equiv i/n \,.
\end{equation}
Similarly, the third term in the aspect ratio becomes, in the large $n$ limit
\begin{equation}
  d_i^{(3)} ~\sim~  \frac{\qty((1-x)\hat n)^{1/12}}{\sqrt{\Gamma(7/6)}} \,.
\end{equation}

Remains to analyze the first term $d_i^{(1)}$. To compute this term we need to perform a shuffling of the way the products are performed. First note that, because of the patterns of the configuration we consider, the power appearing in the products, $\bar\alpha_{pq}$, is non-zero only when $p$ and $q$ are separated by a multiple of $6$. This motivates the following change of variable: one can replace the product over rods at positions $(p,q)$, by a product over the pair $(s=i-p,d=\frac{q-p}6)$. Here $s$ can be viewed as the shift of the first rod with respect to the position $i$, while $6d$ is the distance between the two rods, and $d$ is necessarily an integer. Next, we need to determine the bounds on these new variables:
\begin{equation}
  \begin{aligned}
    p \geq 1 &\implies s \leq i-1 \,,\qquad  p \leq i-1 \implies s \geq 1 \,,
    \\
    q \geq i+1 &\implies s \leq 6d - 1 \,,\qquad q \leq n \implies s \geq 6d-(n-i) \,.
  \end{aligned}
\end{equation}
We thus find that the shift $s$ is constrained to lie between $s_{\text{min}}(d)=\max(1,6d-(n-i))$ and $s_{\text{max}}(d)\equiv\min(i-1,6d-1)$.

Finally, note that the multiplicands in $d_i^{(1)}$ depend only on $d$, not $s$, so that
\begin{equation}
  d_i^{(1)} ~=~ \prod_{d=1}^{\hat n} \qty[\frac{(6d+1)(6d-1)}{(6d)^2}]^{(s_{\text{max}}(d)-s_{\text{min}}(d)+1)/2} \,.
  \label{eq:rewriting_product}
\end{equation}
 This is an exact statement, under the assumption that rods are all of the same size. We now want to determine the large $N$ limit of this term. Because of the symmetry between $i$ and $n-i$,  we will assume for simplicity that $i<n/2$. Define $\alpha \equiv (s_{\text{max}}(d)-s_{\text{min}}(d)+1)/2$. The product \eqref{eq:rewriting_product} has three regimes. 
\begin{itemize}
  \item When $6d < i$, one has $\alpha = 3d-1/2$, and the product of the terms below this bound converges in the large $n$ limit to
  \begin{equation}
    \frac{e^{1/2}\Gamma(1/6)}{2^{7/12}3^{13/24}A^5} (x\hat n)^{-1/12} \comm{\ \qty(1+\mathcal{O}\qty(\hat n^{-1}))}\,,
  \end{equation}
  where $A \equiv \exp(1/12 - \zeta'(1))$ is the Glaisher constant.
  \item When $i \leq 6d \leq n+1-i$, one has $\alpha = (i-1)/2$, and the product of the terms between these bounds converges in the large $n$ limit to
  \begin{equation}
    \exp(-\frac{1-2x}{12(1-x)}) \comm{\ \qty(1+\mathcal{O}\qty(n^{-1/6}))} \,.
  \end{equation}
  \item When $6d > n+1-i$, one has $\alpha = (n-6d)/2$, and the product of the terms above this bound converges in the large $n$ limit to
  \begin{equation}
     \exp(-\frac{x}{12(1-x)}) (1-x)^{-1/12} \comm{\ \qty(1+\mathcal{O}\qty(N^{-1/6}))} \,.
   \end{equation}
\end{itemize}

As a result, one finds
\begin{equation}
   d_i^{(1)} ~\sim~ 3^{11/24}\,(2e)^{5/12} \, \Gamma(7/6) \, A^{-5} \ (x(1-x)\hat n)^{-1/12}
\end{equation}
and, multiplying all the terms, the large $n$ limit of the aspect ratio is
\begin{equation}
  d_i ~\sim~ 3^{11/24}\,(2e)^{5/12} \, A^{-5} \ \hat n^{1/12} \,.
\end{equation}

Note that the dependence in $x$ disappeared completely from the final expression of the aspect ratios. This is a non-trivial check of the validity of the hypothesis that the rods are all of the same size.

The regularity condition then gives
\begin{align}
  Q ~&=~ \frac{2\hat n}{d_i^2}\, R_y^2 = \frac{2^{1/6} \cdot A^{10}}{e^{10/12}\cdot 3^{11/12}} \,\hat n^{5/6} \, R_y^2 \approx~ 2.144 \,\hat n^{5/6} \, R_y^2 \,.
\end{align}

\begin{figure}[ht]
  \centering
  \includegraphics[width=.7\textwidth]{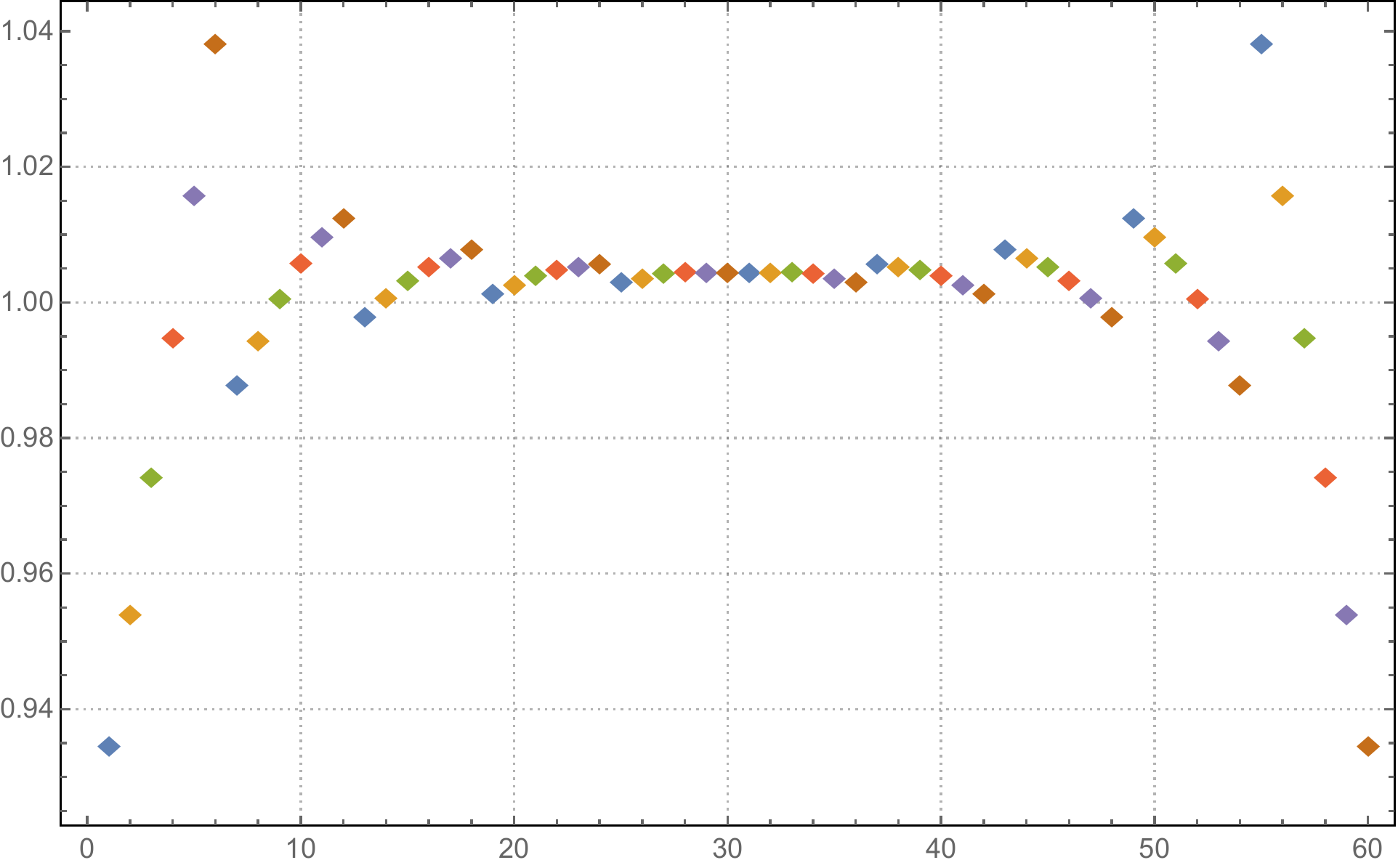}
  \caption{Numerical resolution of the bubble equations for $\hat n=10$ six-rods patterns. Aside from the rods located at the edge of the solution, a constant rod size is a good approximation.}
  \label{fig:sizes_rods_numeric}
\end{figure}

\begin{figure}[ht]
  \centering
  \vspace{2ex}

  \includegraphics[width=.7\textwidth]{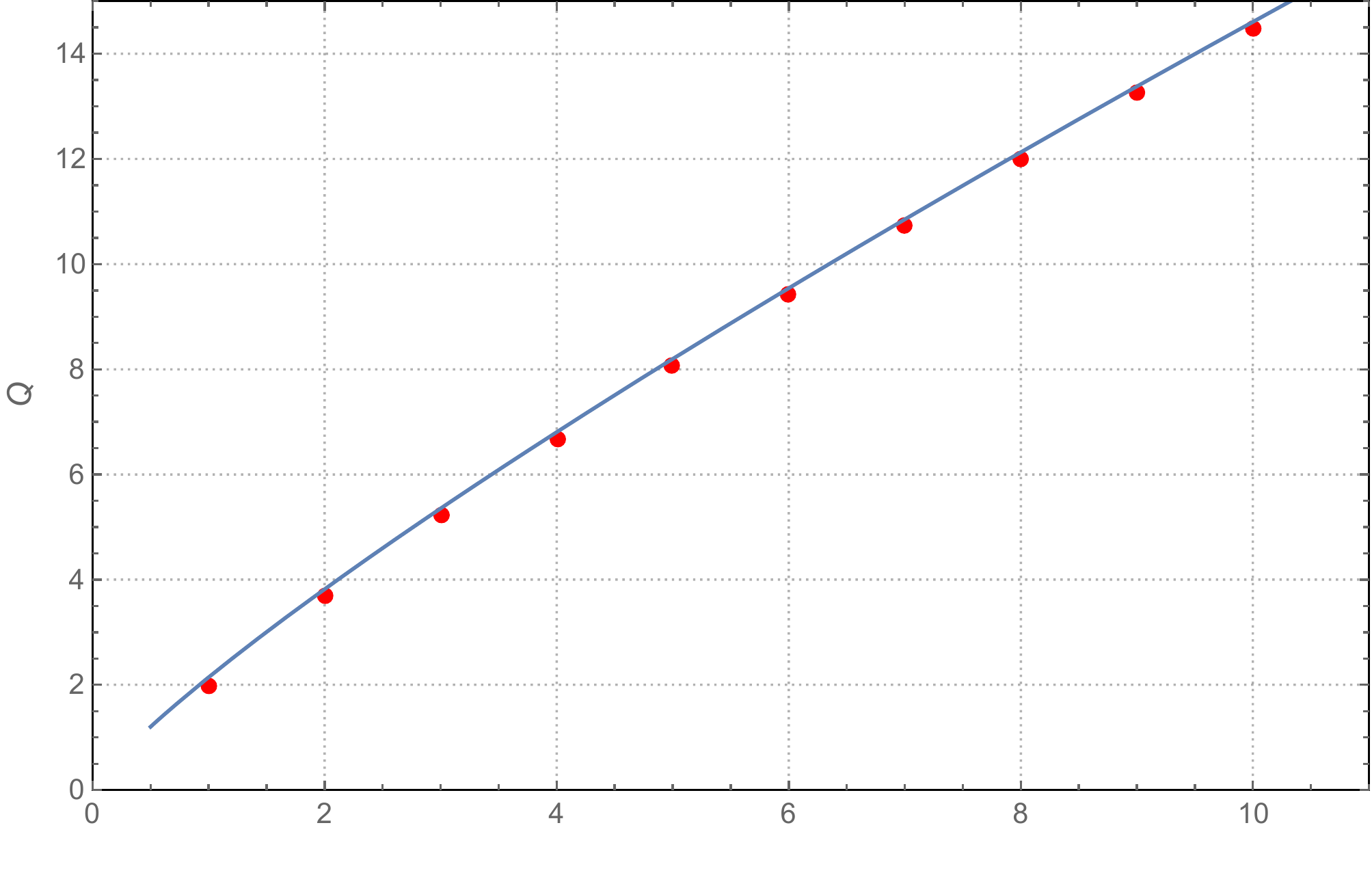}
  \caption{Global M2 charge of the solution as a function of the number $\hat n$ of 6-rod patterns. The red dots are the numerical resolution of the bubble equations, while the blue line correspond to the large-$n$ approximation.}
  \label{fig:comparison_approx_numeric}
\end{figure}

\section{Numerical results for a large number of bolts} 
\label{app:numerics}

In this Appendix we gather the Figures referred to in Section \ref{sec:large_n}. The numerical resolution of the bubble equations for $\hat n=10$ six-rods patterns is presented in Figure~\ref{fig:sizes_rods_numeric}, while a comparison between the large-$\hat n$ analytical results and the numerics is given in Figure~\ref{fig:comparison_approx_numeric}.

\bibliographystyle{utphys}      

\bibliography{microstates}       


\end{document}